\documentclass[10pt]{article}
\usepackage{amsfonts}
\usepackage{indentfirst}
\usepackage{booktabs}
\usepackage{cite}
\usepackage{color}
\usepackage{amssymb,amsmath,mathrsfs}
\usepackage{caption,subfigure,graphicx}
\usepackage{subfigure}
\usepackage{extarrows}

\begin{document}

\title{ A nonlocal coupled modified complex integrable dispersionless equation: Darboux transformation, soliton-type solutions and its asymptotic
behavior }
\author{  Hong-Qian Sun$^{a}$, Shou-Feng Shen$^{a}$
\footnote{Author to whom correspondence should be addressed: zjutssf@163.com(SFS), sunhq0118@163.com(HQS)}, Zuo-Nong Zhu$^{b,c}$\\
\small{$^a$ School of Mathematical Sciences, Zhejiang University of Technology, Hangzhou, 310023,
P. R. China}\\
\small{$^b$ School of Mathematical Sciences, Shanghai Jiao Tong University, Shanghai, 200240, P. R. China}\\
\small{$^c$ Fuyao University of Science and Technology, Fuzhou, 350122, P. R. China}}
\date{ }
\maketitle
\begin{abstract}
In this paper, we primarily construct Darboux transformation(DT) of the nonlocal coupled modified complex integrable dispersionless (cm-CID)
equation, which is first proposed by the connection with a nonlocal coupled modified complex short pulse(cm-CSP) equation.  Utilizing DT, we
present soliton-type solutions for the nonlocal cm-CID equation under vanishing and non-vanishing boundary conditions. Soliton-type solutions
include periodic wave, growing-, decaying-periodic wave, periodic-like wave (which consists of a mixture of periodic wave and breather wave, a
combination of periodic wave and background plane), breather-like wave and rational solution. Furthermore, we have also analyzed asymptotic
behavior and properties of these solutions theoretically and graphically. We must emphasis that soliton solutions of the nonlocal cm-CID equation
possess novel properties that are distinct from those of the cm-CID equation, such as the nonlocal cm-CID equation has the growing-,
decaying-periodic solution and periodic-like solution. The implications of these findings could potentially contribute to the description of
optical pulse behavior during propagation in optical fibers. \\\\
{\bf Keywords:} Soliton-type solutions,~Asymptotic behavior,~The nonlocal coupled modified complex integrable dispersionless equation,~Darboux transformation
\end{abstract}

\section{Introduction}

Nonlinear Schr\"{o}dinger(NLS) equation
\begin{equation}
\text{i}u_t+u_{xx}+|u|^2u=0
\end{equation}
is an fundamental equation in integrable systems, which has many physical applications\cite{Hasegawa1973a,Hasegawa1973b,Benney1967,Zakharov1972}, for
instance, nonlinear optical fibers, deep water wave, and plasma physics.
Recently, multi-component form of the nonlinear integrable equation has been paid attention of researchers for their vital physical applications, for
instance the coupled NLS(CNLS) equation
\begin{equation}
\text{i}u_t+u_{xx}+2(|u|^2+|v|^2)u=0,~~\text{i}v_t+v_{xx}+2(|u|^2+|v|^2)v=0
\end{equation}
was demonstrated to be integrable by Manakov\cite{Manakov1974}. The CNLS equation possess abundant solution structures
\cite{Kang1996,Chen1997,Evangelides1992,Hoefer2011}, which governs an extensive variety of physical phenomena, such as the interaction of two
incoherent light beams in crystals, the transmission of light in a randomly birefringent optical fiber, and the evolution of two-component
Bose-Einstein condensates.

Very recently, Ablowitz and Musslimani\cite{Ablowitz2013} proposed the nonlocal NLS equation
\begin{equation}
\text{i}u_t+u_{xx}+u^2u^*(-x,t)=0,
\end{equation}
and investigated its soliton solutions through the inverse scattering transform method. Note that the nonlocal NLS equation is an integrable
equation, which has Lax pair, infinite conservation laws and $N$-soliton solutions\cite{Ablowitz2013,Ablowitz2016,Li2015}. Meanwhile, the nonlocal
NLS equation are gauge equivalent to the Heisenberg-like equation and a coupled Landau-Lifshitz equation, its physical and geometrical aspects are
studied\cite{Ma2016,Gadzhimuradov2016}. Research shows that the nonlocal NLS equation has different properties from the NLS equation, such as the
nonlocal NLS equation simultaneously has both bright and dark soliton\cite{Ablowitz2013}, the nonlocal NLS equation still exists periodic singular
solution\cite{Li2015}. The coupled nonlocal NLS equation
\begin{equation}
\begin{aligned}
&\text{i}u_t+u_{xx}+(auu^*(-x,t)+bvv^*(-x,t))u=0,~~\\
&\text{i}v_t+v_{xx}+(auu^*(-x,t)+bvv^*(-x,t))v=0
\end{aligned}
\end{equation}
was proposed and studied for its soliton\cite{Khare2015}, where $a,b$ are real numbers.
Due to the rich properties of nonlocal integrable equations, it has been widely concerned by many
researchers\cite{Sarma2014,Ji2017a,Ma2017,Fokas2016}.

When the width of optical pulse is of the order $10^{-15}$s, NLS equation becomes less accurate for describing the propagation of ultra short pulses
in nonlinear media. While the short pulse equation provides an increasingly better approximation to the corresponding solution for the Maxwell
equations. The short-pulse (SP) equation
\begin{equation}\label{SP}
u_{xt}=u+\frac{1}{6}(u^{3})_{xx}
\end{equation}
was derived originally for describing the pseudo-spherical surfaces\cite{Rabelo1989,Beals1989}, and later introduced for displaying the propagation
of ultra-short optical pulses in silicon fiber\cite{Schafer2004,Chung2005}. Note that SP equation also was proposed in the negative WKI
hierarchy\cite{Qiao2003}. Research indicates that Eq.\eqref{SP} is complete integrable\cite{Sakovich2005,Brunelli2005,Brunelli2006}, and its soliton
solutions\cite{Sakovich2006,Kuetche2007,Matsuno2007,Matsuno2008}, discretization form\cite{Feng2010}, multi-component
form\cite{Matsuno2011,Feng2012}, and high order correction\cite{Kurt2013} have been studied. More recently, from Maxwell's equation, Feng proposed a
complex short-pulse(CSP) equation\cite{Feng2015}
\begin{equation}\label{CSP}
u_{xt}=u+\frac{1}{2}(|u|^{2}u_x)_{x}
\end{equation}
and a coupled complex short-pulse(CCSP) equation
\begin{equation}\label{CCSP}
u_{xt}=u+\frac{1}{2}((|u|^{2}+|v|^{2})u_x)_{x},~~v_{xt}=v+\frac{1}{2}((|u|^{2}+|v|^{2})v_x)_{x}
\end{equation}
to describe the ultra-short pulse propagation in optical fibers. Numerous properties of the CSP equation and the CCSP equation have been studied, for
instance soliton solution, rogue wave solution\cite{Feng2015,Feng2016,Shen2016,Ling2016,Gkogkou2022,Feng2022}, long-time asymptotic
behavior\cite{Xu2020}, and its spatial discretization\cite{Feng2021,Sun2023}.

Matsuno\cite{Matsuno2016} introduced the multi-component modified SP equation
\begin{equation}\label{multiMSP}
u_{i,xt}=u_i+\frac{1}{2}(\sum_{1\leq j,k\leq n}(c_{jk}u_ju_k)u_{i,x})_{x}-\frac{1}{2}(\sum_{1\leq j,k\leq n}c_{jk}u_{j,x}u_{k,x})u_i.
\end{equation}
Note that Eq.\eqref{multiMSP} can be reduced to the coupled modified SP equation
\begin{equation}\label{cmSP}
u_{xt}=u+\frac{1}{2}v(u^2)_{xx},~~v_{xt}=v+\frac{1}{2}u(v^2)_{xx}.
\end{equation}
For the equation \eqref{cmSP}, its integrability (eg. Lax pair, infinite conservation laws, and multi-soliton solution) have been
constructed\cite{Matsuno2016,Lv2022b}.
When $n=4$, Eq.\eqref{multiMSP} can be
reduced to a coupled modified complex SP(cm-CSP) equation
\begin{equation}\label{cmCSP}
\begin{aligned}
&u_{xt}=u+((|u|^2+|v|^2)u_x)_x-(|u_x|^2+|v_x|^2)u,\\
&v_{xt}=v+((|u|^2+|v|^2)v_x)_x-(|u_x|^2+|v_x|^2)v,
\end{aligned}
\end{equation}
Soliton solution, semi-rational solution and periodic solution of Eq.\eqref{cmCSP} were obtained by DT method and RH approach
\cite{Li2022,Lv2024,Sun2025}.
With a hodograph transformation
\begin{equation}\label{tr0}
dy=\rho^{-1}dx+\rho^{-1}(|u|^2+|v|^2)dt,~~~ds=-dt,
\end{equation}
where $\rho=(1+|u_x|^2+|v_x|^2)^{-1}$, Eq.\eqref{cmCSP} transforms to a coupled modified CID(cm-CID) equation\cite{Sun2025}
\begin{equation}\label{cmCID}
\begin{aligned}
&\rho_s-(|u|^2+|v|^2)_y=0,\\
&u_{ys}+\frac{u}{\rho}(\rho^2-|u_y|^2-|v_y|^2)=0,\\
&v_{ys}+\frac{v}{\rho}(\rho^2-|u_y|^2-|v_y|^2)=0.
\end{aligned}
\end{equation}
One can obtain the soliton solution and periodic solution of Eq.\eqref{cmCID} with the transformation\eqref{tr0}.

Inspired by the above research work, in this paper, we mainly investigate a nonlocal cm-CID equation
\begin{equation}\label{ncmCID}
\begin{aligned}
&\rho_s-(\sigma_1u\widetilde{u}^*+\sigma_2v\widetilde{v}^*)_y=0,\\
&u_{ys}+\frac{u}{\rho}(\rho^2-\sigma_1u_y\widetilde{u}^*_y-\sigma_2v_y\widetilde{v}^*_y)=0,\\
&v_{ys}+\frac{v}{\rho}(\rho^2-\sigma_1u_y\widetilde{u}^*_y-\sigma_2v_y\widetilde{v}^*_y)=0,
\end{aligned}
\end{equation}
where $\widetilde{u}=u(-y,-s)$, $\sigma_1=\sigma_2=1$ represents the focusing-focusing(f-f) case, $\sigma_1=-\sigma_2=1$ stands for the focusing-defocusing(f-def) case, $\sigma_1=\sigma_2=-1$ symbolizes the defocusing-defocusing (def-def) case.
Note that under the transformation
\begin{equation}\label{tr}
dx=\rho dy-(\sigma_1u\widetilde{u}^*+\sigma_2v\widetilde{v}^*)ds,ds=-dt,
\end{equation}
where $x$ is real, $\rho=(1+\sigma_1u_xu^*(-x,-t)_x+\sigma_2v_xv^*(-x,-t)_x)^{-1}$,
the nonlocal cm-CID equation\eqref{ncmCID} converts to the nonlocal cm-CSP equation
\begin{equation}\label{ncmCSP}
\begin{aligned}
&u_{xt}=u+((\sigma_1uu^*(-x,-t)+\sigma_2vv^*(-x,-t))u_x)_x-(\sigma_1u_xu^*(-x,-t)_x+\sigma_2v_xv^*(-x,-t)_x)u,\\
&v_{xt}=v+((\sigma_1uu^*(-x,-t)+\sigma_2vv^*(-x,-t))v_x)_x-(\sigma_1u_xu^*(-x,-t)_x+\sigma_2v_xv^*(-x,-t)_x)v.
\end{aligned}
\end{equation}
To our knowledge, Eqs.\eqref{ncmCID}\eqref{ncmCSP} and their integrability have not been reported in the literature. What distinctions exist between
the properties of the nonlocal cm-CID equation\eqref{ncmCID} and those of the cm-CID equation\eqref{cmCID}? What are the differences in the solutions
among the f-f equation, f-def equation and f-def equation? These are the questions that attract our interest.

Darboux transformation\cite{Gu1987,Nimmo1997,Bian2014,Ling2015,Nimmo2015} is a significant and effective approach for seeking various exact solutions
to nonlinear integrable equations, without the necessity for inverse spectral analysis. One key aim of the present paper is the construction of DT
for the nonlocal cm-CID equation. Another challenge in this paper lies in choosing an appropriate non-zero seed solution so that the obtained
solution has interesting properties.

The organization of this paper is presented as follows. In Section 2, we construct DT of the nonlocal cm-CID equation, and provide a rigorous proof.
In Section 3, with the vanishing boundary condition(VBC), we obtain classical periodic wave, double periodic solution, growing-, decaying-periodic
wave, periodic-like solution, breather-like solution and interaction of soliton and breather-like waves. Furthermore, properties and asymptotic
behavior of these solution also are analyzed theoretically. In Section 4, with the non-vanishing boundary condition(NVBC), various of exact solution
of the nonlocal cm-CID equation are derived, including bright-, dark-periodic solution, M-periodic solution, growing-, decaying- periodic wave,
breather-like solution and rational solution. Note that the properties of exact solution for the nonlocal cm-CID equation are different from those
for the cm-CID equation, such as the nonlocal cm-CID equation simultaneously has both bright-, dark-periodic wave, and growing-, decaying-periodic
wave. The solutions of the f-f equation, f-def equation, and def-def equation possess different properties, for instance, the f-def equation exists
dark periodic wave and rational solution; the def-def equation has M-periodic wave solution.

\section{ \bf Construction of Darboux transformation}

In this section, we mainly construct Darboux transformation and give a rigorous proof. By utilizing the DT, we obtain the explicit expressions of
one-, two-soliton solutions and their properties.

Eq.\eqref{ncmCID} has the following linear spectral problem
\begin{equation}\label{ncmCID-lax}
\begin{aligned}
&\Psi_y=U(\rho,u,v;\lambda)\Psi,\Psi_s=V(u,v;\lambda)\Psi,\\
&U(\rho,u,v;\lambda)=\lambda(\rho I_4+\frac{1}{\rho}G_y^2-2G_y)\Lambda,V(u,v;\lambda)=-\frac{1}{4\lambda}\Lambda+G,
\end{aligned}
\end{equation}
where $I_4$ denotes the $4\times4$ identity matrix, $\Lambda=\text{diag}(1,1,-1,-1)$ and
\begin{eqnarray*}
G=\left(
\begin{array}{cccc}
0&0&u&v\\
0&0&-\sigma_2\widetilde{v}^{*}&\sigma_1\widetilde{u}^{*}\\
-\sigma_1\widetilde{u}^{*}&v&0&0\\
-\sigma_2\widetilde{v}^{*}&-u&0&0
\end{array}
\right).
\end{eqnarray*}\\
{\bf Proposition 1.} Under the condition $\widetilde{\rho}^*=\rho$, the matrices $U(\lambda)$ and $V(\lambda)$ satisfy the symmetric properties as
\begin{equation}
\begin{aligned}
&(\widetilde{U}(\lambda))^*=M_1^{-1}U(-\lambda^*)M_1,~~(\widetilde{V}(\lambda))^*=M_1^{-1}V(-\lambda^*)M_1,~\\
&(\widetilde{U}(\lambda))^{\dagger}=-M_2U(\lambda^*)M_2,~~(\widetilde{V}(\lambda))^{\dagger}=-M_2V(\lambda^*)M_2,~
\end{aligned}
\end{equation}
where ``$\dagger$'' denotes conjugate transpose,  and
\begin{equation*}
M_1=\left(
\begin{array}{cccc}
0&1&0&0\\
-\sigma_1\sigma_2&0&0&0\\
0&0&0&-\sigma_1\\
0&0&\sigma_2&0
\end{array}
\right),~M_2=\text{diag}(1,\sigma_1\sigma_2,-\sigma_1,-\sigma_2).
\end{equation*}
Proposition 1 can be proved by direct verification. It is noted that in this paper, it is required that the linear spectral problem\eqref{ncmCID-lax}
satisfies $\widetilde{\rho}^*=\rho$.

Assume that the column vector
\begin{equation}\label{y}
|\zeta(\lambda_k)\rangle=(\psi_{1}^{(k)},\psi_{2}^{(k)},\psi_{3}^{(k)},\psi_{4}^{(k)})^T,~~k=1,2,\cdots,N,
\end{equation}
is an eigenfunction of the linear spectral problem \eqref{ncmCID-lax} at $\lambda=\lambda_k$. According to {\bf Proposition 1}, naturally one can
conclude that
\begin{equation}\label{eta}
|\eta(\lambda_k)\rangle=M_1|y(\lambda_k)\rangle=(\widetilde{\psi}_{2}^{(k)^{*}},-\sigma_1\sigma_2\widetilde{\psi}_{1}^{(k)^{*}},-\sigma_1\widetilde{\psi}_{4}^{(k)^{*}},\sigma_2\widetilde{\psi}_{3}^{(k)^{*}})^{T}
\end{equation}
is an eigenfunction of the linear spectral problem \eqref{ncmCID-lax} at $\lambda=-\lambda_k^{*}$, and $\langle\vartheta_k|=\langle
\widetilde{\zeta}_k|M_2=(|\widetilde{\zeta}_k\rangle)^{\dagger}M_2$ is an eigenfunction of the adjoint problem for Eq.\eqref{ncmCID-lax}
\begin{eqnarray}\label{adjiont}
\langle\vartheta|_y=-\langle\vartheta|U(\lambda),\quad\langle\vartheta|_s=-\langle\vartheta|V(\lambda),
\end{eqnarray}
at $\lambda=\lambda_k^{*}$.

According to the above facts, we can deduce the main theorem of this paper.\\\\
{\bf Theorem 2.}
The gauge transformation
\begin{equation}\label{tran}
\Psi^{(N)}=T^{(N)}\Psi,
\end{equation}
where
\begin{equation}\label{DT}
\begin{aligned}
&T^{N}=I-\mathbf{K}_NW_N^{-1}\Gamma(K_N),\\
&\mathbf{K}_N=(|\zeta_{1}\rangle,|\eta_{1}\rangle,|\zeta_{2}\rangle,|\eta_{2}\rangle,\cdots|\zeta_{N}\rangle,|\eta_{N}\rangle)\triangleq(K_{1},K_{2},\cdots,K_{N}),\\
&W_N=\left(
\begin{array}{cc}
\Omega(K_{1},K_{1})& \Omega(K_{1},K_{2}) \quad\cdots \quad\Omega(K_{1},K_{N})\\
\Omega(K_{2},K_{1})& \Omega(K_{2},K_{2}) \quad\cdots\quad \Omega(K_{2},K_{N})\\
\vdots &\vdots\qquad \ddots\qquad \vdots \\
\Omega(K_{N},K_{1})& \Omega(K_{N},K_{2}) \quad\cdots \quad\Omega(K_{N},K_{N})
\end{array}
\right),\\
&\Gamma(\mathbf{K}_N)=\left(
\begin{array}{cc}
\Gamma(K_{1})\\ \Gamma(K_{2})\\ \vdots\\ \Gamma(K_{N})
\end{array}
\right),\Gamma(K_{k})=\left(
\begin{array}{cc}
\frac{4\lambda_k^*\lambda}{\lambda_k^*-\lambda}\langle\widetilde{\zeta}_{k}|M\\
\frac{4\lambda_k\lambda}{\lambda_k+\lambda}\langle\widetilde{\eta}_{k}|M
\end{array}
\right), \\
&\Omega(K_{k},K_{j})=\left(
\begin{array}{cc}
\frac{4\lambda_k^*\lambda_j}{\lambda_k^*-\lambda_j}\langle \widetilde{\zeta}_{k}|M|\zeta_{j}\rangle&
\frac{4\lambda_k^*\lambda_j^*}{-\lambda_k^*-\lambda_j^*}\langle \widetilde{\zeta}_{k}|M|\eta_{j}\rangle\\
\frac{4\lambda_k\lambda_j}{\lambda_k+\lambda_j}\langle \widetilde{\eta}_{k}|M|\zeta_{j}\rangle&
\frac{4\lambda_k\lambda_j^*}{-\lambda_k+\lambda_j^*}\langle \widetilde{\eta}_{k}|M|\eta_{j}\rangle
\end{array}
\right),1\leq k,j\leq N,
\end{aligned}
\end{equation}
converts the linear spectral problem \eqref{ncmCID-lax} to a new linear spectral problem
\begin{equation}\label{problem}
\begin{aligned}
\Psi_{y}^{(N)}=U^{(N)}(\rho^{(N)},u^{(N)},v^{(N)};\lambda)\Psi^{(N)},\Psi_{s}^{(N)}=V^{(N)}(u^{(N)},v^{(N)};\lambda)\Psi^{(N)}.
\end{aligned}
\end{equation}
by replacing the old potential function $(\rho,u,v)$ with the new potential function $(\rho^{(N)},u^{(N)},v^{(N)})$ as
\begin{equation}\label{relation1}
\begin{aligned}
&\rho^{(N)}=\rho+2(\left|\begin{array}{cc}
W_N&  \widetilde{\mathbf{h}}_{1}^{(N)\dag}\\
\mathbf{h}_{1}^{(N)}& 0
\end{array}
\right|/|W_N |)_y=\rho-2(\mathbf{h}_{1}^{(N)}W_N^{-1}\widetilde{\mathbf{h}}_{1}^{(N)\dag})_y,\\
&u^{(N)}=u+2\left|\begin{array}{cc}
W_N& \widetilde{\mathbf{h}}_{3}^{(N)\dag}\\
\mathbf{h}_{1}^{(N)}& 0
\end{array}
\right|/|W_N |=u-2\mathbf{h}_{1}^{(N)}W_N^{-1} \widetilde{\mathbf{h}}_{3}^{(N)\dag},\\
&v^{(N)}=v+2\left|\begin{array}{cc}
W_N& \widetilde{\mathbf{h}}_{4}^{(N)\dag}\\
\mathbf{h}_{1}^{(N)}& 0
\end{array}
\right|/|W_N |=v-2\mathbf{h}_{1}^{(N)}W_N^{-1} \widetilde{\mathbf{h}}_{4}^{(N)\dag},
\end{aligned}
\end{equation}
where $\rho^{(N)}(y,s)$ satisfies $\rho^{(N)*}(-y,-s)=\rho^{(N)}(y,s)$, and
\begin{equation}\label{134}
\begin{aligned}
&\mathbf{h}_1^{(N)}=(\psi_1^{(1)},\widetilde{\psi}_2^{(1)*},\psi_1^{(2)},\widetilde{\psi}_2^{(2)*},\cdots,\psi_1^{(N)},\widetilde{\psi}_2^{(N)*}),\\
&\mathbf{h}_3^{(N)}=(\psi_3^{(1)},-\sigma_1\widetilde{\psi}_4^{(1)*},\psi_3^{(2)},-\sigma_1\widetilde{\psi}_4^{(2)*},\cdots,\psi_3^{(N)},-\sigma_1\widetilde{\psi}_4^{(N)*}),\\
&\mathbf{h}_4^{(N)}=(\psi_4^{(1)},\sigma_2\widetilde{\psi}_3^{(1)*},\psi_4^{(2)},\sigma_2\widetilde{\psi}_3^{(2)*},\cdots,\psi_4^{(N)},\sigma_2\widetilde{\psi}_3^{(N)*}).
\end{aligned}
\end{equation}
Only when $\rho^{(N)*}(y,s)=\rho^{(N)}(y,s)$, i.e. $\rho^{(N)}$ is real function, this solution can be transformed back to the nonlocal cm-CSP
equation\eqref{ncmCSP} through the transformation\eqref{tr}.

{\bf Proof.} For the temporal part, we prove that the structures of matrix $V^{(N)}(u^{(N)},v^{(N)})$ and $V(u,v)$ are the same.

With the identities $\widetilde{G}^{\dag}M=-MG$ and $G\Lambda=-\Lambda G$, we have
\begin{eqnarray*}
&&(\langle\widetilde{\zeta}_{k}|M|\zeta_{j}\rangle)_s=\frac{-\lambda_k^*+\lambda_j}{4\lambda_k^*\lambda_j}\langle
\widetilde{\zeta}_{k}|M\Lambda|\zeta_{j}\rangle,~(\langle
\widetilde{\eta}_{k}|M|\zeta_{j}\rangle)_s=\frac{-\lambda_k-\lambda_j}{4\lambda_k\lambda_j}\langle \eta_{k}|M\Lambda|\zeta_{j}\rangle,\\
&&(\langle \widetilde{\zeta}_{k}|M|\eta_{j}\rangle)_s=\frac{\lambda_k^*+\lambda_j^*}{4\lambda_k^*\lambda_j^*}\langle
\widetilde{\zeta}_{k}|M\Lambda|\eta_{j}\rangle,~(\langle
\widetilde{\eta}_{k}|M|\eta_{j}\rangle)_s=\frac{\lambda_k-\lambda_j^*}{4\lambda_k\lambda_j^*}\langle
\widetilde{\eta}_{k}|M\Lambda|\eta_{j}\rangle,~~k,j=1,2,3,...N.
\end{eqnarray*}
Through direct calculation, we obtain the following equations
\begin{align*}
(\mathbf{K}_{N})_s=-\frac{1}{4}\Lambda\mathbf{K}_ND_N+G\mathbf{K}_N,~(W_{N})_s=-\widetilde{\mathbf{K}}_N^{\dagger}M\Lambda
\mathbf{K}_N,~(\Gamma(\mathbf{K}_N)\Psi)_s=-\widetilde{\mathbf{K}}_N^{\dagger}M\Lambda\Psi,
\end{align*}
where
\begin{equation*}
D_N=\text{diag}(\frac{1}{\lambda_1},-\frac{1}{\lambda_1^*},\frac{1}{\lambda_2},-\frac{1}{\lambda_2^*},...,\frac{1}{\lambda_N},-\frac{1}{\lambda_N^*}).
\end{equation*}
Consequently, we can draw the conclusion that
\begin{align*}
\Psi_{s}^{(N)}&=\Psi_{s}-(\mathbf{K}_{N})_sW_N^{-1}\Gamma(\mathbf{K}_N)\Psi-\mathbf{K}_{N}(W_N^{-1})_s\Gamma(\mathbf{K}_N)\Psi-\mathbf{K}_{N}W_N^{-1}(\Gamma(\mathbf{K}_N)\Psi)_s\\
&=V(\lambda)\Psi+\frac{1}{4}\Lambda \mathbf{K}_ND_NW_N^{-1}\Gamma(\mathbf{K}_N)\Psi-G\mathbf{K}_NW_N^{-1}\Gamma(\mathbf{K}_N)\Psi
\\
&\quad-\mathbf{K}_{N}W_N^{-1}\widetilde{\mathbf{K}}_N^{\dagger}M\Lambda
\mathbf{K}_NW_N^{-1}\Gamma(\mathbf{K}_N)\Psi+\mathbf{K}_{N}W_N^{-1}\widetilde{\mathbf{K}}_N^{\dagger}M\Lambda\Psi\\
&=-\frac{1}{4\lambda}\Lambda\Psi+G\Psi-G\mathbf{K}_NW_N^{-1}\Gamma(\mathbf{K}_N)\Psi+\mathbf{K}_{N}W_N^{-1}\widetilde{\mathbf{K}}_N^{\dagger}M\Lambda\Psi
\\
&\quad-\mathbf{K}_{N}W_N^{-1}\widetilde{\mathbf{K}}_N^{\dagger}M\Lambda\mathbf{K}_NW_N^{-1}\Gamma(\mathbf{K}_N)\Psi+\frac{1}{4\lambda}\Lambda
\mathbf{K}_NW_N^{-1}\Gamma(\mathbf{K}_N)\Psi\\
&\quad+\Lambda \mathbf{K}_NW_N^{-1}\widetilde{\mathbf{K}}_N^{\dagger}M\mathbf{K}_NW_N^{-1}\Gamma(\mathbf{K}_N)\Psi-\Lambda
\mathbf{K}_NW_N^{-1}\widetilde{\mathbf{K}}_N^{\dagger}M\Psi\\
&=(G-\frac{1}{4\lambda}\Lambda+[\mathbf{K}_{N}W_N^{-1}\widetilde{\mathbf{K}}_N^{\dagger}M,\Lambda]_-)(\Psi-\mathbf{K}_{N}W_N^{-1}\Gamma(\mathbf{K}_N)\Psi)\\
&\triangleq(G^{(N)}-\frac{1}{4\lambda}\Lambda)\Psi^{(N)}=V^{(N)}(u^{(N)},v^{(N)};\lambda)\Psi^{(N)},
\end{align*}
where we utilize the following identities as
\begin{equation}
D_N^{\dag}W_N-W_ND_N=-4\widetilde{\mathbf{K}}_N^{\dag}M\mathbf{K}_N,~(D_N^{\dag}-\frac{1}{\lambda}I_2)\Gamma(\mathbf{K}_N)=-4\widetilde{\mathbf{K}}_N^{\dag}M.
\end{equation}

Denoting $\Theta=(\Theta_{j,k})_{1\leq j,k\leq4}=\mathbf{K}_{N}W_N^{-1}\widetilde{\mathbf{K}}_N^{\dagger}$, i.e.
\begin{equation}\label{theta}
\Theta_{jk}=-\left|\begin{array}{cc}
W_N & \widetilde{\mathbf{h}}_k^{\dag}\\
\mathbf{h}_j & 0
\end{array}
\right|/|W_N|=\mathbf{h}_jW_N^{-1}\widetilde{\mathbf{h}}_k^{\dag},~1\leq j,k\leq4,
\end{equation}
where $\mathbf{h}_j^{(N)}$ $(j=1,3,4)$ are given in Eq.\eqref{134}, and
\begin{align*}
&\mathbf{h}_2^{(N)}=(\psi_2^{(1)},-\sigma_1\sigma_2\widetilde{\psi}_1^{(1)*},\psi_2^{(2)},-\sigma_1\sigma_2\widetilde{\psi}_1^{(2)*},\cdots,\psi_2^{(N)},-\sigma_1\sigma_2\widetilde{\psi}_1^{(N)*}).
\end{align*}
Then the matrix $G^{(N)}$ can be written as
\begin{align*}
G^{(N)}=\left(\begin{array}{cccc}
0 & 0 & 2\sigma_1\Theta_{13}+u & 2\sigma_2\Theta_{14}+v\\
0 & 0 & 2\sigma_1\Theta_{23}-\sigma_2\widetilde{v}^* & 2\sigma_2\Theta_{24}+\sigma_1\widetilde{u}^*\\
2\Theta_{31}-\sigma_1\widetilde{u}^* & 2\sigma_1\sigma_2\Theta_{32}+v & 0 & 0\\
2\Theta_{41}-\sigma_2\widetilde{v}^* & 2\sigma_1\sigma_2\Theta_{42}-u & 0 & 0
\end{array}
\right).
\end{align*}
In the following, we prove the consistency of the elements of matrix $G^{(N)}$.

For proving the consistency of the matrix elements $G^{(N)}$, it is merely requisite to demonstrate the following equations
\begin{align*}
&\Theta_{23}=-\sigma_1\widetilde{\Theta}_{14}^*,~\Theta_{24}=\sigma_2\widetilde{\Theta}_{13}^*,
~\Theta_{31}=-\sigma_2\Theta_{24},\\
&\Theta_{32}=\sigma_1\Theta_{14},~\Theta_{41}=\sigma_1\widetilde{\Theta}_{23}^*,
~\Theta_{42}=-\sigma_2\Theta_{13}.
\end{align*}
Direct computation leads to the following equations
\begin{align*}
&\langle \widetilde{\zeta}_k|M|\zeta_j\rangle=(\langle \widetilde{\zeta}_j|M|\zeta_k\rangle)^*=\sigma_1\sigma_2\langle
\widetilde{\eta}_j|M|\eta_k\rangle=\sigma_1\sigma_2(\langle\widetilde{\eta}_k|M|\eta_j\rangle)^*,\\
&\langle \widetilde{\zeta}_k|M|\eta_j\rangle=-\langle \widetilde{\zeta}_j|M|\eta_k\rangle=(\langle \widetilde{\eta}_j|M|\zeta_k\rangle)^*=-(\langle
\widetilde{\eta}_k|M|\zeta_j\rangle)^*,
\end{align*}
thereby, we obtain $\Omega(K_k,K_j)=-\widetilde{\Omega}(K_j,K_k)^{\dagger}$, i.e. $\widetilde{W}_N^{\dagger}=-W_N$.
Bring in a matrix
\begin{align}
~~~~~A=\left(\begin{array}{ccccccc}
0 & -\sigma_1\sigma_2 & 0 & 0&\cdots&0&0\\
1 & 0 & 0 & 0&\cdots&0&0\\
0 & 0 & 0 & -\sigma_1\sigma_2&\cdots&0&0\\
0 & 0 & 1 & 0&\cdots&0&0\\
\vdots & \vdots & \vdots & \vdots&~&\vdots&\vdots\\
0 & 0 & 0 & 0&\cdots&0&-\sigma_1\sigma_2\\
0 & 0 & 0 & 0&\cdots&1&0
\end{array}
\right),
\end{align}
thus we obtain the relations
\begin{align*}
A^T=-\sigma_1\sigma_2A, W_N=-AW_N^TA, \mathbf{h}_2^{(N)}=\widetilde{\mathbf{h}}_1^{(N)*}A,~\mathbf{h}_4^{(N)}=-\sigma_1\widetilde{\mathbf{h}}_3^{(N)*}A.
\end{align*}
Via the aforementioned relation, we obtain
\begin{align*}
&\Theta_{31}=\left|\begin{array}{cc}
AW_NA & \sigma_1A\widetilde{\mathbf{h}}_4^{(N)\dag}\\
\sigma_1\sigma_2\mathbf{h}_2^{(N)}A & 0
\end{array}
\right|/|W_N|=-\sigma_2\mathbf{h}_2^{(N)}W_N^{-1}\widetilde{\mathbf{h}}_4^{(N)\dag}=-\sigma_2\Theta_{24},\\
&\Theta_{41}=-\left|\begin{array}{cc}
AW_NA & \sigma_2A\widetilde{\mathbf{h}}_3^{(N)\dag}\\
\sigma_1\sigma_2\mathbf{h}_2^{(N)}A & 0
\end{array}
\right|/|W_N|=\sigma_1\mathbf{h}_2^{(N)}W_N^{-1}\widetilde{\mathbf{h}}_3^{(N)\dag}=\sigma_1\Theta_{23}^*,\\
&\Theta_{32}=-\left|\begin{array}{cc}
AW_NA & \sigma_1A\widetilde{\mathbf{h}}_4^{(N)\dagger}\\
\mathbf{h}_1^{(N)}A & 0
\end{array}
\right|/|W_N|=\sigma_1\mathbf{h}_1^{(N)}W_N^{-1}\widetilde{\mathbf{h}}_4^{(N)\dag}=\sigma_1\Theta_{14},\\
&\Theta_{42}=-\left|\begin{array}{cc}
-AW_NA & \sigma_2A\widetilde{\mathbf{h}}_3^{(N){\dagger}}\\
\mathbf{h}_1^{(N)}A & 0
\end{array}
\right|/|W_N|=-\sigma_2\mathbf{h}_1^{(N)}W_N^{-1}\widetilde{\mathbf{h}}_3^{(N)^\dag}=-\sigma_2\Theta_{13},\\
&\Theta_{23}=-\sigma_1\left|\begin{array}{cc}
W_N^T & \mathbf{h}_4^{(N)^T}\\
\widetilde{\mathbf{h}}_1^{(N)^*} & 0
\end{array}
\right|/|W_N|=-\sigma_1\widetilde{\Theta}_{14}^*,\Theta_{24}=\sigma_2(\left|\begin{array}{cc}
W_N^T & \mathbf{h}_3^{(N)^T}\\
\widetilde{\mathbf{h}}_1^{(N)^*} & 0
\end{array}
\right|)^*/|W_N|=\sigma_2\widetilde{\Theta}_{13}^*.
\end{align*}
Through the proof of the compatibility of the potential matrix $G^{(N)}$, it can be conclude that the structures of potential matrices $G^{(N)}$ and
$G$ are identical.

For the spatial part, we prove that the matrix $U^{(N)}$ has the same structure with the matrix $U$. In other words, we need to verify the validity
of the equation
\begin{equation}\label{left-right}
(T^{(N)}_y+T^{(N)}U)T^{(N)^{-1}}=U^{(N)},
\end{equation}
where
\begin{align*}
&U^{(N)}=\lambda(\rho^{(N)}+\frac{1}{\rho^{(N)}}G_y^{(N)^2}-2G_y^{(N)})\Lambda,\\
&G^{(N)}=G+\mathbf{K}_{N}W_N^{-1}\widetilde{\mathbf{K}}_N^{\dagger}M\Lambda-\Lambda \mathbf{K}_{N}W_N^{-1}\widetilde{\mathbf{K}}_N^{\dagger}M,\\
&\rho^{(N)}\Lambda=\rho\Lambda-(\mathbf{K}_{N}W_N^{-1}\widetilde{\mathbf{K}}_N^{\dagger}M)_y-\Lambda(\mathbf{K}_{N}W_N^{-1}\widetilde{\mathbf{K}}_N^{\dagger}M)_y\Lambda.
\end{align*}
Through a routine calculations, we have the following equation
\begin{equation}\label{GN}
\begin{aligned}
&(\langle \widetilde{\zeta}_k|M|\zeta_j\rangle)_y=(-\lambda_k^*+\lambda_j)\langle \widetilde{\zeta}_k|M(\rho
I_4+\frac{1}{\rho}G_y^2-2G_y)\Lambda|\zeta_j\rangle,\\
&(\langle \widetilde{\eta}_k|M|\zeta_j\rangle)_y=(\lambda_k+\lambda_j)\langle \widetilde{\eta}_k|M(\rho
I_4+\frac{1}{\rho}G_y^2-2G_y)\Lambda|\zeta_j\rangle,\\
&(\langle \widetilde{\zeta}_k|M|\eta_j\rangle)_y=(-\lambda_k^*-\lambda_j^*)\langle \widetilde{\zeta}_k|M(\rho
I_4+\frac{1}{\rho}G_y^2-2G_y)\Lambda|\eta_j\rangle,\\
&(\langle \widetilde{\eta}_k|M|\eta_j\rangle)_y=(\lambda_k-\lambda_j^*)\langle \widetilde{\eta}_k|M(\rho
I_4+\frac{1}{\rho}G_y^2-2G_y)\Lambda|\eta_j\rangle,
\end{aligned}
\end{equation}
and then we get that
\begin{align*}
&(\mathbf{K}_N)_y=(\rho I_4+\frac{1}{\rho}G_y^2-2G_y)\Lambda \mathbf{K}_ND_N^{-1},\\
&(W_N)_y=-4(D_N^{-1})^{*}\widetilde{\mathbf{K}}_N^{\dagger}M(\rho I_4+\frac{1}{\rho}G_y^2-2G_y)\Lambda \mathbf{K}_ND_N^{-1},\\
&(\Gamma(\mathbf{K}_N))_y=-4\lambda\Upsilon_N\widetilde{\mathbf{K}}_N^{\dagger}\Lambda(\rho I_4+\frac{1}{\rho}G_y^2+2MG_yM)M,
\end{align*}
where
\begin{equation*}
\Upsilon_N=\text{diag}\left(\frac{\lambda_1^{*2}}{\lambda_1^*-\lambda},\frac{-\lambda_1^{2}}{\lambda_1+\lambda},\frac{\lambda_2^{*2}}{\lambda_2^*-\lambda},
\frac{-\lambda_2^{2}}{\lambda_2+\lambda},\cdots,\frac{\lambda_N^{*2}}{\lambda_N^*-\lambda},\frac{-\lambda_N^{2}}{\lambda_N+\lambda}\right).
\end{equation*}
By simplifying the equation \eqref{left-right}, we get
\begin{align*}
&-\frac{1}{\lambda}(\mathbf{K}_{N}W_N^{-1}\Gamma(\mathbf{K}_N))_yT^{(N)^{-1}}\rho^{(N)}\Lambda+(I_{2N}-\mathbf{K}_{N}W_N^{-1}\Gamma(\mathbf{K}_N))(\rho+\frac{1}{\rho}G_y^{2}-2G_y)\Lambda
T^{(N)^{-1}}\rho^{(N)}\Lambda\\
&=\rho^{(N)^2}I_4+G_y^{(N)^2}-2\rho^{(N)}G_y^{(N)},
\end{align*}
where
\begin{equation}
T^{(N)^{-1}}=I_4+\mathbf{K}_{N}\overline{\Upsilon}_NW_N^{-1}\widetilde{\mathbf{K}}_{N}^{\dagger}M,
\end{equation}
with
\begin{equation*}
\overline{\Upsilon}_N=\text{diag}(\frac{4\lambda_1\lambda}{\lambda_1-\lambda},
\frac{4\lambda_1^*\lambda}{-\lambda_1^*-\lambda},\frac{4\lambda_2\lambda}{\lambda_2-\lambda},
\frac{4\lambda_2^*\lambda}{-\lambda_2^*-\lambda},\cdots,\frac{4\lambda_N\lambda}{\lambda_N-\lambda},
\frac{4\lambda_N^*\lambda}{-\lambda_N^*-\lambda}).
\end{equation*}
Through intricate computations, we obtain
\begin{align*}
(\mathbf{K}_{N}W_N^{-1}\widetilde{\mathbf{K}}_N^{\dagger})_y=L_1\rho+L_2+L_2\rho^{-1},(\mathbf{K}_{N}W_N^{-1}\Gamma(\mathbf{K}_N))_y=L_4\rho+L_5+L_6\rho^{-1},
\end{align*}
with
\begin{align*}
&L_1=\Lambda\mathbf{K}_{1}D_1W_1^{-1}\widetilde{\mathbf{K}}_1^{\dagger}
-\mathbf{K}_{1}W_1^{-1}D_1^*\widetilde{\mathbf{K}}_1^{\dagger}\Lambda+4\mathbf{K}_{1}W_1^{-1}D_1^*\widetilde{\mathbf{K}}_1^{\dagger}M\Lambda\mathbf{K}_1D_1W_1^{-1}\widetilde{\mathbf{K}}_1^{\dagger},\\
&L_2=8\mathbf{K}_{1}W_1^{-1}D_1^*\widetilde{\mathbf{K}}_1^{\dagger}M\Lambda
G_y\mathbf{K}_{1}D_1W_1^{-1}\widetilde{\mathbf{K}}_1^{\dagger}-2G_y\Lambda\mathbf{K}_{1}D_1W_1^{-1}\widetilde{\mathbf{K}}_1^{\dagger}-2\mathbf{K}_{1}W_1^{-1}D_1^*\widetilde{\mathbf{K}}_1^{\dagger}\Lambda
MG_yM,\\
&L_3=G_y^2(\Lambda\mathbf{K}_{1}W_1^{-1}D_1\widetilde{\mathbf{K}}_1^{\dagger}-\mathbf{K}_{1}W_1^{-1}D_1^*\widetilde{\mathbf{K}}_1^{\dagger}\Lambda+4\mathbf{K}_{1}W_1^{-1}D_1^*\widetilde{\mathbf{K}}_1^{\dagger}M\Lambda\mathbf{K}_{1}D_1W_1^{-1}\widetilde{\mathbf{K}}_1^{\dagger}),\\
&L_4=\Lambda\mathbf{K}_{1}D_1W_1^{-1}\widetilde{\mathbf{K}}_1^{\dagger}+4\mathbf{K}_{1}W_1^{-1}D_1^*\widetilde{\mathbf{K}}_1^{\dagger}M\Lambda\mathbf{K}_{1}D_1W_1^{-1}\Gamma(\mathbf{K}_{1})-\mathbf{K}_{1}W_1^{-1}D_2\widetilde{\mathbf{K}}_1^{\dagger}\Lambda
M,\\
&L_5=8\mathbf{K}_{1}W_1^{-1}D_1^*\widetilde{\mathbf{K}}_1^{\dagger}M\Lambda
G_yD_1\mathbf{K}_{1}W_1^{-1}\Gamma(\mathbf{K}_1)-2G_y\Lambda\mathbf{K}_{1}D_1W_1^{-1}\Gamma(\mathbf{K}_1)-2\mathbf{K}_{1}W_1^{-1}D_2\widetilde{\mathbf{K}}_1^{\dagger}\Lambda
MG_y,\\
&L_6=G_y^2(\Lambda\mathbf{K}_{1}D_1W_1^{-1}\Gamma(\mathbf{K}_1)-\mathbf{K}_{1}W_1^{-1}D_2\widetilde{\mathbf{K}}_1^{\dagger}\Lambda
M+4\mathbf{K}_{1}W_1^{-1}D_1^*\widetilde{\mathbf{K}}_1^{\dagger}M\Lambda\mathbf{K}_{1}D_1W_1^{-1}\Gamma(\mathbf{K}_1)).
\end{align*}
Then we get the following equations
\begin{align*}
&\rho^{(1)}\Lambda=\rho(\Lambda-[L_1M\Lambda,\Lambda]_+)-[L_2M\Lambda,\Lambda]_+-\rho^{-1}[L_3M\Lambda,\Lambda]_+,\\
&G^{(1)}_y=\rho[L_1M,\Lambda]_-+G_y+[L_2M,\Lambda]_-+\rho^{-1}[L_3M,\Lambda]_-,
\end{align*}
where $[A,B]_\pm=AB\pm BA$.
Some complicated calculations yields the coefficients of $\rho^j$ $(j=0,\pm1,\pm2)$ in Eq.\eqref{left-right} are respectively
\begin{align*}
&\rho^2:(\Lambda-\mathbf{K}_{1}W_1^{-1}\Gamma(\mathbf{K}_1)\Lambda-\frac{1}{\lambda}L_4)T^{(1)^{-1}}(\Lambda-[L_1M\Lambda,\Lambda]_+)\\
&~~=(\Lambda-3L_1M+\Lambda L_1M\Lambda)(\Lambda-[L_1M\Lambda,\Lambda]_+)+[L_1M,\Lambda]_-^2,\\
&\rho:(2\Lambda G_y-2\mathbf{K}_{1}W_1^{-1}\Gamma(\mathbf{K}_1)\Lambda G_y-\frac{1}{\lambda}L_5)T^{(1)^{-1}}(\Lambda-[L_1M\Lambda,\Lambda]_+)\\
&~~~-(\Lambda-\mathbf{K}_{1}W_1^{-1}\Gamma(\mathbf{K}_1)\Lambda-\frac{1}{\lambda}L_4)T^{(1)^{-1}}[L_2M\Lambda,\Lambda]_+\\
&~~=(G_y+[L_2M,\Lambda]_-)[L_1M,\Lambda]_-+[L_1M,\Lambda]_-(G_y+[L_2M,\Lambda]_-)\\
&~~~+(3L_1M-\Lambda-\Lambda L_1M\Lambda)[L_2M\Lambda,\Lambda]_++(\Lambda L_2M\Lambda-2G_y\Lambda-3L_2M)(\Lambda-[L_1M\Lambda,\Lambda]_+),\\
&\rho^0:(G_y^2\Lambda-\mathbf{K}_{1}W_1^{-1}\Gamma(\mathbf{K}_1)G_y^2\Lambda -\frac{1}{\lambda}L_6)T^{(1)^{-1}}(\Lambda-[L_1M\Lambda,\Lambda]_+)\\
&~~~+(2\Lambda G_y-2\mathbf{K}_{1}W_1^{-1}\Gamma(\mathbf{K}_1)\Lambda G_y-\frac{1}{\lambda}L_5)T^{(1)^{-1}}[L_2M\Lambda,\Lambda]_+\\
&~~~-(\Lambda-\mathbf{K}_{1}W_1^{-1}\Gamma(\mathbf{K}_1)\Lambda-\frac{1}{\lambda}L_4)T^{(1)^{-1}}[L_3M\Lambda,\Lambda]_+\\
&~~=(G_y+[L_2M,\Lambda]_-)^2+[L_3M,\Lambda]_-[L_1M,\Lambda]_-\\
&~~~+[L_1M,\Lambda]_-[L_3M,\Lambda]_-+(3L_2M-\Lambda L_2M\Lambda+2G_y\Lambda)[L_2M\Lambda,\Lambda]_+\\
&~~~-(3L_3M-\Lambda L_3M\Lambda)(\Lambda-[L_1M\Lambda,\Lambda]_+)-(\Lambda-3L_1M+\Lambda L_3M\Lambda)[L_3M\Lambda,\Lambda]_+,\\
&\rho^{-1}:-(G_y^2\Lambda-\mathbf{K}_{1}W_1^{-1}\Gamma(\mathbf{K}_1)G_y^2\Lambda -\frac{1}{\lambda}L_6)T^{(1)^{-1}}[L_2M\Lambda,\Lambda]_+\\
&~~~-(2\Lambda G_y-2\mathbf{K}_{1}W_1^{-1}\Gamma(\mathbf{K}_1)\Lambda G_y-\frac{1}{\lambda}L_5)T^{(1)^{-1}}[L_3M\Lambda,\Lambda]_+\\
&~~=(G_y+[L_2M,\Lambda]_-)[L_3M,\Lambda]_-+[L_3M,\Lambda]_-(G_y+[L_2M,\Lambda]_-)\\
&~~~+(3L_2M-\Lambda L_2M\Lambda+2G_y\Lambda)[L_3M\Lambda,\Lambda]_+-(\Lambda L_3M\Lambda-3L_3M)[L_2M\Lambda,\Lambda]_+,
\end{align*}
\begin{align*}
&\rho^{-2}:-(G_y^2\Lambda-\mathbf{K}_{1}W_1^{-1}\Gamma(\mathbf{K}_1)G_y^2\Lambda -\frac{1}{\lambda}L_6)T^{(1)^{-1}}[L_3M\Lambda,\Lambda]_+\\
&~~=(3L_2M-\Lambda L_2M\Lambda)[L_3M\Lambda,\Lambda]_++[L_3M\Lambda,\Lambda]_-^2.
\end{align*}
Thus $T^{(1)}_y+T^{(1)}U=U^{(1)}T^{(1)}$ is true.

In the following, we proof the compatibility of the elements of the matrix $\rho^{(N)}\Lambda$.

With Eq.\eqref{GN}, we have
\begin{equation}
\rho^{(N)}\Lambda=\rho\Lambda-2\left(\begin{array}{cccc}
\Theta_{11,y} & \sigma_1\sigma_2\Theta_{12,y} & 0 & 0 \\
\Theta_{21,y} & \sigma_1\sigma_2\Theta_{22,y} & 0 & 0 \\
0 & 0 & -\sigma_1\Theta_{33,y} & -\sigma_2\Theta_{34,y}\\
0 & 0 & -\sigma_1\Theta_{43,y} & -\sigma_2\Theta_{44,y}
\end{array}
\right),
\end{equation}
where $\Theta_{jk}$ have given in \eqref{theta}.

In the way we proved in the first step, we can get the relation
\begin{align*}
\Theta_{22}=\sigma_1\sigma_2\Theta_{11},~\Theta_{44}=\sigma_1\sigma_2\Theta_{33},~
\Theta_{21}=-\widetilde{\Theta}_{12}^*,~\Theta_{43}=-\widetilde{\Theta}_{34}^*.
\end{align*}
Thus we only need to prove the following identity relations
\begin{align*}
\Theta_{33,y}=\sigma_1\Theta_{11,y},~\Theta_{12,y}=0,~\Theta_{34,y}=0.
\end{align*}
As the structure of W is extremely complicated, we only present the proof herein for $N=1$ and $N=2$.

When $N=1$, we have
\begin{align}\label{W1}
W_1^{-1}=\left(\begin{array}{cc}
\frac{-\text{i}\lambda_{1,I}}{2|\lambda_1|^2\langle \widetilde{\zeta}_1|M|\zeta_1\rangle} & 0\\
0 & \frac{-\sigma_1\sigma_2\text{i}\lambda_{1,I}}{2|\lambda_1|^2\langle \widetilde{\zeta}_1|M|\zeta_1\rangle}
\end{array}
\right),
\end{align}
where $|\zeta_{1}\rangle=(\psi_{1}^{(1)},\psi_{2}^{(1)},\psi_{3}^{(1)},\psi_{4}^{(1)})^T$, and then we get
\begin{align*}
&\Theta_{12}=0,~\Theta_{34}=0,~\Theta_{33}=\sigma_1\Theta_{1}+\frac{\text{i}\sigma_1\lambda_{1,I}}{2|\lambda_1|^2}.
\end{align*}
Therefore $\Theta_{33,y}=\sigma_1\Theta_{11,y}$ and the formula\eqref{relation1} of the function $\rho^{(1)}$ can be obtained.

When $N=2$, we have
\begin{equation}\label{W2}
\begin{aligned}
&W_2^{-1}=\Xi_1\left(\begin{array}{cccc}
w_{33} & 0 & -w_{13} & -\sigma_1\sigma_2w_{14}\\
0 & \sigma_1\sigma_2w_{33} & -\sigma_1\sigma_2\widetilde{w}_{14}^* & \sigma_1\sigma_2\widetilde{w}_{13}^*\\
\widetilde{w}_{13}^* & \sigma_1\sigma_2w_{14} & w_{11} & 0\\
\sigma_1\sigma_2\widetilde{w}_{14}^* & -\sigma_1\sigma_2w_{13} & 0 & \sigma_1\sigma_2w_{11}
\end{array}
\right)
\end{aligned}
\end{equation}
with
\begin{equation*}
\begin{aligned}
&\Xi_1=\frac{1}{w_{11}w_{33}+w_{13}\widetilde{w}_{13}^*+\sigma_1\sigma_2w_{14}\widetilde{w}_{14}^*},\\
&w_{11}=\frac{4|\lambda_1|^2}{\lambda_1^*-\lambda_1}\langle \widetilde{\zeta}_1|M|\zeta_1\rangle,~~
w_{33}=\frac{4|\lambda_2|^2}{\lambda_2^*-\lambda_2}\langle \widetilde{\zeta}_2|M|\zeta_2\rangle,~\\
&w_{13}=\frac{4\lambda_1^*\lambda_2}{\lambda_1^*-\lambda_2}\langle \widetilde{\zeta}_1|M|\zeta_2\rangle,~~
w_{14}=\frac{4\lambda_1^*\lambda_2^*}{-\lambda_1^*-\lambda_2^*}\langle \widetilde{\zeta}_1|M|\eta_2\rangle.
\end{aligned}
\end{equation*}
where $|\zeta_{k}\rangle$, $|\eta_{k}\rangle$ $(k=1,2)$ are defined by Eqs.\eqref{y}\eqref{eta}.
Then we get
\begin{align*}
&\Theta_{12}=0,~\Theta_{34}=0,~
\Theta_{33}=\sigma_1\Theta_{11}+\sigma_1(\frac{\text{i}\lambda_{1,I}}{2|\lambda_1|^2}+\frac{\text{i}\lambda_{2,I}}{2|\lambda_2|^2}).
\end{align*}
Therefore $\Theta_{33,y}=\sigma_1\Theta_{11,y}$ and we obtain the formula of $\rho^{(2)}$ as
\begin{equation}
\rho^{(2)}=\rho+(\frac{2\left|\begin{array}{cc}
W_2&  \widetilde{\mathbf{h}}_{1}^{(2)\dag}\\
\mathbf{h}_{1}^{(2)}& 0
\end{array}
\right|}{|W_2 |})_y=\rho-2(\mathbf{h}_{1}^{(2)}W_2^{-1}\widetilde{\mathbf{h}}_{1}^{(2)\dag})_y.
\end{equation}
For the case of $N=3,...$, $T^{(N)}_y+T^{(N)}U=U^{(N)}T^{(N)}$ can be proved similarly.

The proof for the compatibility of the elements of the matrix $\rho^{(N)}\Lambda$ have been given, then the formula of the solution $\rho^{(N)}$ can
be derived as
\begin{equation}
\rho^{(N)}=\rho+(\frac{2\left|\begin{array}{cc}
W_N&  \widetilde{\mathbf{h}}_{1}^{(N)\dag}\\
\mathbf{h}_{1}^{(N)}& 0
\end{array}
\right|}{|W_N |})_y=\rho-2(\mathbf{h}_{1}^{(N)}W_N^{-1}\widetilde{\mathbf{h}}_{1}^{(N)\dag})_y.
\end{equation}
Note that the matrix $W_N$ satisfies $\widetilde{W}_N^{\dagger}=-W_N$,
so that
\begin{align*}
&\widetilde{\rho}^{(N)*}=\widetilde{\rho}^*-(\frac{2\left|\begin{array}{cc}
\widetilde{W}_N^{\dagger}& \widetilde{\mathbf{h}}_{1}^{(N)\dag} \\
\mathbf{h}_{1}^{(N)}& 0
\end{array}
\right|}{|\widetilde{W}_N^{\dagger}|})_y=\rho-2(\mathbf{h}_{1}^{(N)}W_N^{-1}\widetilde{\mathbf{h}}_{1}^{(N)\dag})_y=\rho^{(N)}.
\end{align*}
This completes the proof of Theorem 2.2.~~~~~~~$\blacksquare$


\section{\bf Soliton solution and periodic wave solution with vanishing boundary condition}

In this section, under VBC, we present various solution of the nonlocal cm-CID equation \eqref{cmCID}. By the first DT, we obtain soliton solution,
periodic wave, growing-, decaying- and growing-decaying periodic wave solution for the nonlocal cm-CID equation $\eqref{ncmCID}$. Through the
quadratic DT, various solutions of the nonlocal cm-CID equation $\eqref{ncmCID}$ are derived, including double periodic wave (which are periodic in
both time and space, and there are two peaks of different values within each cycle), periodic-like solution (which is mixture of periodic wave and
breather wave), collision solution(which is collision of breather wave and soliton) of this equation. Meanwhile, the properties of these solutions
are also analyzed.

\subsection{ One-soliton solutions }

{\bf Proposition 3.} When $N=1$, we have the first Darboux matrix as
\begin{align*}
T^{(1)}=I-\frac{\lambda(\lambda_1^*-\lambda_1)}{\langle \widetilde{\zeta}_1|M|\zeta_1\rangle}\left(\frac{|\zeta_1\rangle\langle
\widetilde{\zeta}_1|M}{\lambda_1(\lambda_1^*-\lambda)}+\frac{\sigma_1\sigma_2|\eta_1\rangle\langle
\widetilde{\eta}_1|M}{\lambda_1^*(\lambda_1+\lambda)}\right),
\end{align*}
with $|\zeta_1\rangle$, $|\eta_1\rangle$ are defined by Eqs.\eqref{y}\eqref{eta}. Therefore, the relation of new potential
$(u^{(1)},v^{(1)},\rho^{(1)})$ and old potential $(u,v,\rho)$ can be written as

\begin{equation}\label{solution1}
\begin{aligned}
&u^{(1)}=u+\frac{\sigma_1(\lambda_1^*-\lambda_1)(\phi_1\widetilde{\phi}_3^*-\sigma_2\widetilde{\phi}_2^*\phi_4)}{2|\lambda_1|^2(\phi_1\widetilde{\phi}_1^*+\sigma_1\sigma_2\phi_2\widetilde{\phi}_2^*-\sigma_1\phi_3\widetilde{\phi}_3^*-\sigma_2\phi_4\widetilde{\phi}_4^*)},\\
&v^{(1)}=v+\frac{\sigma_2(\lambda_1^*-\lambda_1)(\phi_1\widetilde{\phi}_4^*+\sigma_1\widetilde{\phi}_2^*\phi_3)}{2|\lambda_1|^2(\phi_1\widetilde{\phi}_1^*+\sigma_1\sigma_2\phi_2\widetilde{\phi}_2^*-\sigma_1\phi_3\widetilde{\phi}_3^*-\sigma_2\phi_4\widetilde{\phi}_4^*)},\\
&\rho^{(1)}=\rho-\frac{(\lambda_1^*-\lambda_1)\Delta_1}{(\phi_1\widetilde{\phi}_1^*+\sigma_1\sigma_2\phi_2\widetilde{\phi}_2^*-\sigma_1\phi_3\widetilde{\phi}_3^*-\sigma_2\phi_4\widetilde{\phi}_4^*)^2},
\end{aligned}
\end{equation}
with
\begin{align*}
&\Delta_1=(\lambda_1^*(\phi_1\widetilde{\phi}_1^*+\sigma_1\sigma_2\phi_2\widetilde{\phi}_2^*)-\lambda_1(\sigma_1\phi_3\widetilde{\phi}_3^*+\sigma_2\phi_4\widetilde{\phi}_4^*))(\widetilde{\phi}_1^*(u_y\phi_3+v_y\phi_4)+\sigma_2\widetilde{\phi}_2^*(\widetilde{u}^*_y\phi_4-\sigma_1\sigma_2\widetilde{v}^*_y\phi_3))\\
&~~~~~~~~+(\lambda_1(\phi_1\widetilde{\phi}_1^*+\sigma_1\sigma_2\phi_2\widetilde{\phi}_2^*)-\lambda_1^*(\sigma_1\phi_3\widetilde{\phi}_3^*+\sigma_2\phi_4\widetilde{\phi}_4^*))(\phi_1(\widetilde{u}^*_y\widetilde{\phi}_3^*+\widetilde{v}^*_y\widetilde{\phi}_4^*)+\sigma_2\phi_2(u_y\widetilde{\phi}_4^*-\sigma_1\sigma_2v_y\widetilde{\phi}_3^*))\\
&~~~~~~~~+\frac{\lambda_1^*-\lambda_1}{\rho}(\rho^2-\sigma_1u_y\widetilde{u}^*_y-\sigma_2v_y\widetilde{v}^*_y)(\phi_1\widetilde{\phi}_1^*+\sigma_1\sigma_2\phi_2\widetilde{\phi}_2^*)(\sigma_1\phi_3\widetilde{\phi}_3^*+\sigma_2\phi_4\widetilde{\phi}_4^*).
\end{align*}

According to  Theorem 2, the conclusion of Proposition 3 can be naturally obtained.

For seeking soliton solution, we take zero seed solution $\rho=\gamma$, $u=0$, $v=0$. Solving the linear spectral problem \eqref{ncmCID-lax} at
$\lambda_1=\alpha_1+\text{i}\beta_1 (\beta_1\neq0)$ yields the eigenfunction
\begin{equation}\label{eigenfunction}
\psi_{1}^{(1)}=c_{1}\text{e}^{\xi_{1}},\psi_{2}^{(1)}=c_{2}\text{e}^{\xi_{1}},
\psi_{3}^{(1)}=c_{3}\text{e}^{-\xi_{1}},\psi_{4}^{(1)}=c_{4}\text{e}^{-\xi_{1}},
\end{equation}
where $\xi_{1}=\gamma\lambda_1y-\frac{1}{4\lambda_1}s$, and $c_{j}$ ($j=1,2,3,4$) are complex constants. Substituting the eigenfunction
\eqref{eigenfunction} to the formula \eqref{solution1}, we obtain one-soliton solution of the nonlocal cm-CID equation \eqref{ncmCID} as
\begin{equation}\label{solution11}
\begin{aligned}
&u^{(1)}=\frac{\sigma_1(\lambda_1^*-\lambda_1)(c_1c_3^*\text{e}^{2\xi_{1,R}}-\sigma_2c_2^*c_4\text{e}^{-2\xi_{1,R}})}{|\lambda_1|^2\Delta_2},\\
&v^{(1)}=\frac{\sigma_2(\lambda_1^*-\lambda_1)(c_1c_4^*\text{e}^{2\xi_{1,R}}+\sigma_1c_2^*c_3\text{e}^{-2\xi_{1,R}})}{|\lambda_1|^2\Delta_2},\\
&\rho^{(1)}=\gamma-\frac{\gamma(\lambda_1^*-\lambda_1)^2(|c_1|^2+\sigma_1\sigma_2|c_2|^2)(\sigma_1|c_3|^2+\sigma_2|c_4|^2)}{|\lambda_1|^2\Delta_2^2},
\end{aligned}
\end{equation}
where $R$, $I$ stand for represents the real and imaginary parts, i.e. $\xi_{1,R}=\gamma\alpha_1y-\frac{\alpha_1s}{4|\lambda_1|^2}$,
$\xi_{1,I}=\gamma\beta_1y+\frac{\beta_1s}{4|\lambda_1|^2}$, and
\begin{align*}
&\Delta_2=(|c_1|^2+\sigma_1\sigma_2|c_2|^2-\sigma_1|c_3|^2-\sigma_2|c_4|^2)\cos(2\xi_{1,I})\\
&~~~~+\text{i}(|c_1|^2+\sigma_1\sigma_2|c_2|^2+\sigma_1|c_3|^2+\sigma_2|c_4|^2)\sin(2\xi_{1,I}).
\end{align*}
Here condition
$(|c_1|^2+\sigma_1\sigma_2|c_2|^2-\sigma_1|c_3|^2-\sigma_2|c_4|^2)(|c_1|^2+\sigma_1\sigma_2|c_2|^2+\sigma_1|c_3|^2+\sigma_2|c_4|^2)\neq0$ is
satisfied.
Note that here $\rho^{(1)}$ is complex function and $\rho^{(1)*}(-y,-s)=\rho^{(1)}(y,s)$, there do not exist real $x$ in Eq.\eqref{tr}, so we only
obtain the solutions of the nonlocal cm-CID equation \eqref{ncmCID}.

\begin{figure*}
\centering
\includegraphics[height=2.8cm]{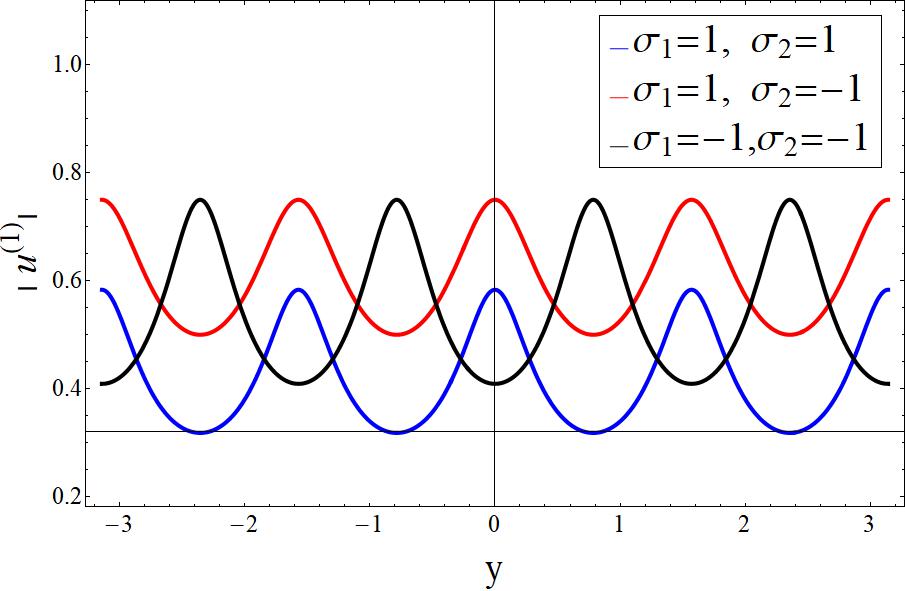}
~~\includegraphics[height=2.8cm]{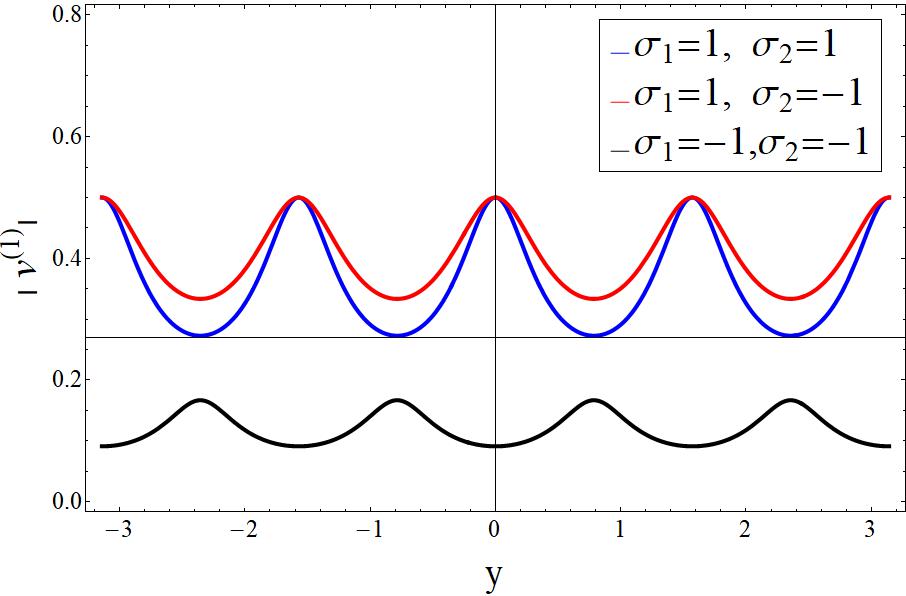}
~~\includegraphics[height=2.8cm]{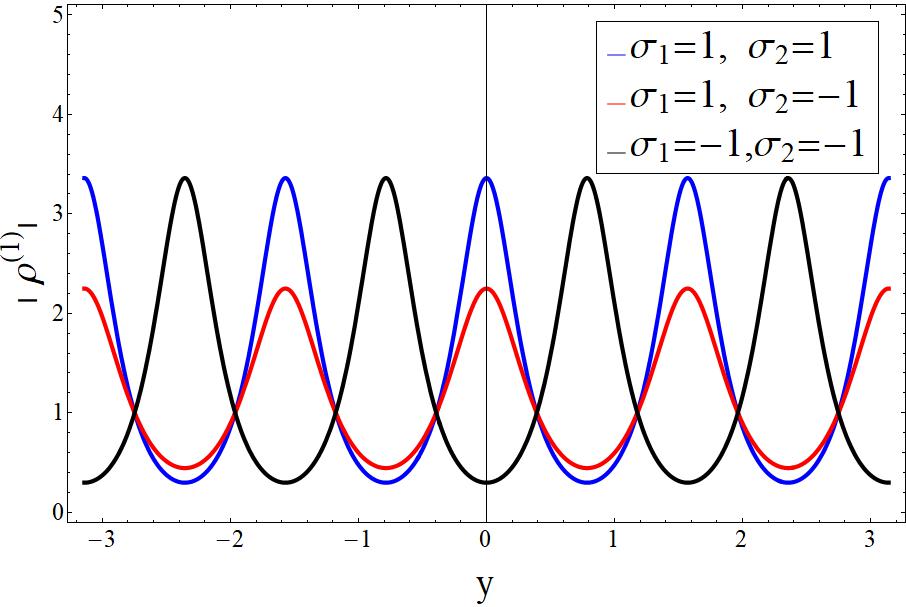}\\
{$(a)$ $|u^{(1)}|$ \hspace{3.5cm}  $(b)$ $|v^{(1)}|$ \hspace{3.5cm} $(c)$  $|\rho^{(1)}|$ }\\
\caption{Periodic wave solution for the f-f, f-def and def-def nonlocal cm-CID equation with $\lambda_1=\text{i}$.}
\label{Fig.1}
\end{figure*}

When $\alpha_1=0$, i.e. $\lambda_1$ is pure imaginary number, we have $\xi_{1,R}=0$. Thus the solution ($u^{(1)},v^{(1)},\rho^{(1)}$) are all
periodic waves. Its minimum periods in time and space are $T_{s}=2\beta_1\pi$, $T_{y}=\frac{\pi}{2\gamma\beta_1}$, respectively. Taking $\gamma=1$,
$c_1=1$, $c_2=c_4=\frac{1}{2}$, $c_3=2$, $\beta_1=1$, we give the plots of periodic wave ($|u^{(1)}|,|v^{(1)}|,|\rho^{(1)}|$) for the f-f, f-def and
def-def nonlocal cm-CID equation at $t=0$ (see FIG.~\ref{Fig.1}).

\begin{figure*}
\centering
\includegraphics[height=2.7cm]{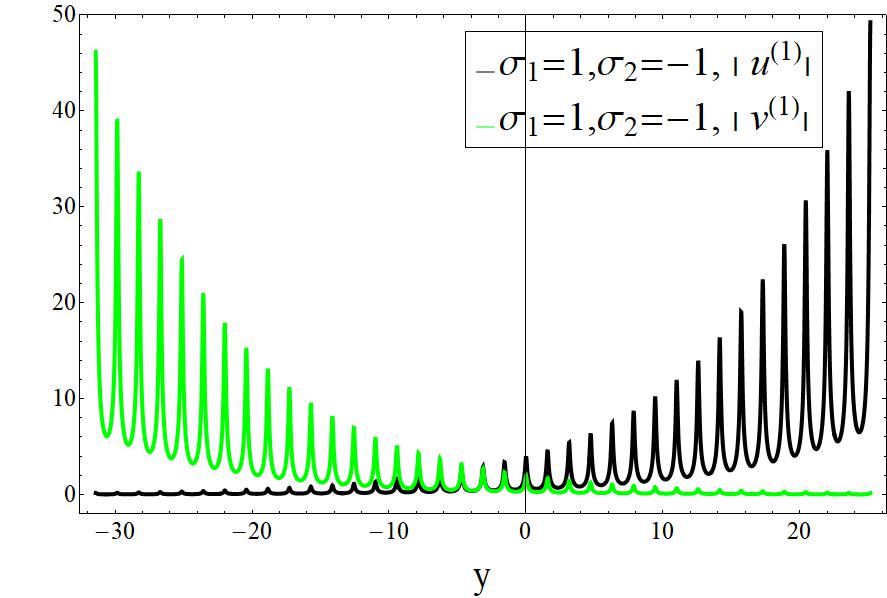}
\includegraphics[height=2.7cm]{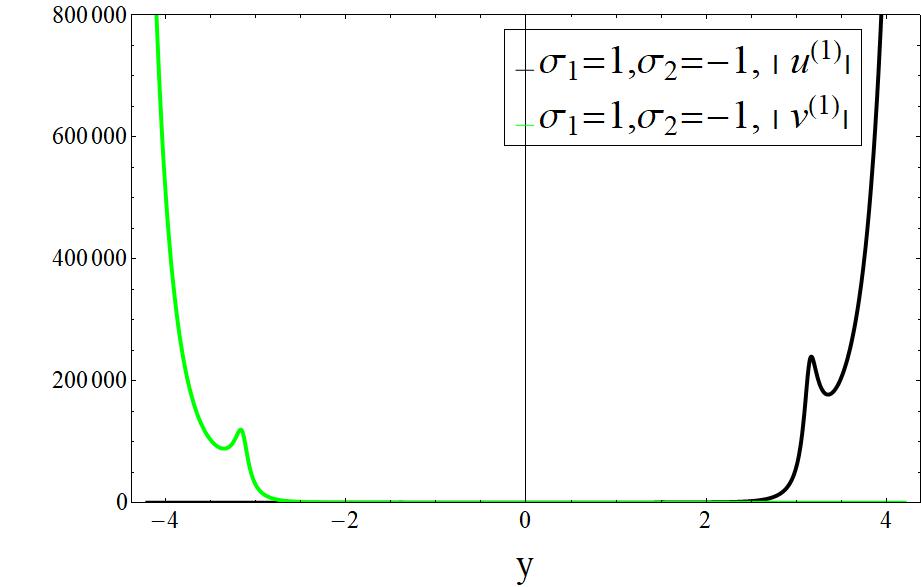}
\includegraphics[height=2.7cm]{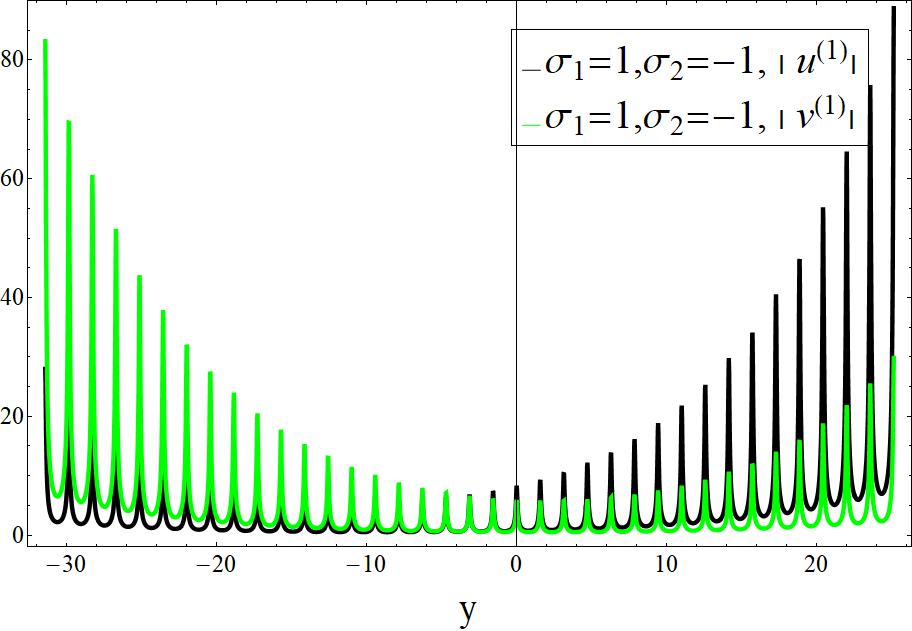}
\includegraphics[height=2.7cm]{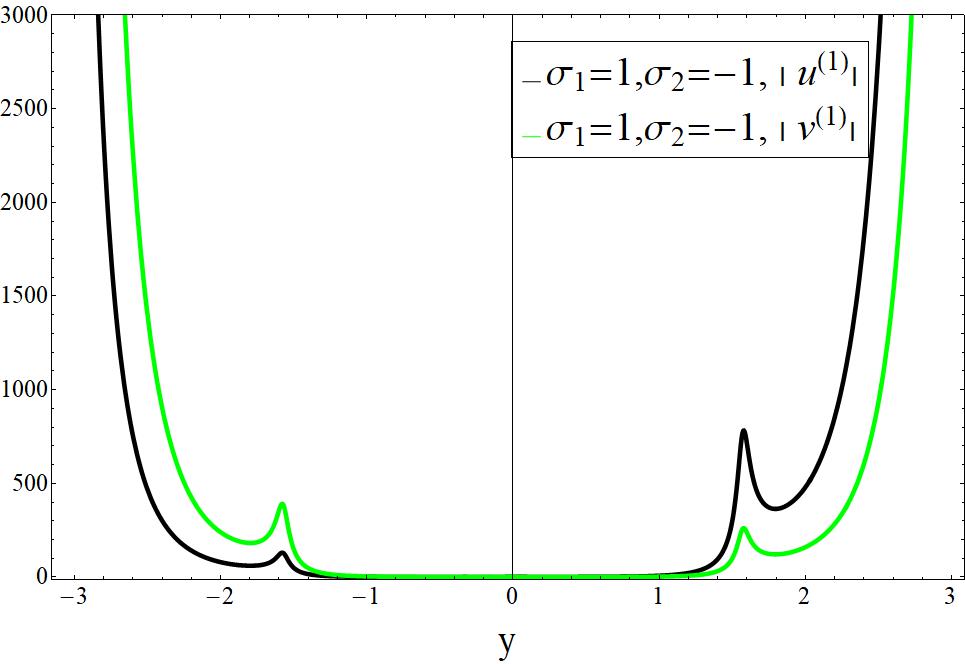}\\
{$(a)$ decaying- or growing- ~~~ $(b)$ decaying- or growing- ~~~
$(c)$ decaying-growing~~~ $(d)$ decaying-growing  }\\
\caption{Decaying-, growing- and decaying-growing periodic solution for the nonlocal cm-CID equation: $(a)$ $c_4=0$,
$\lambda_1=\frac{1}{20}+\text{i}$, $(b)$ $c_4=0$, $\lambda_1=2+\text{i}$, $(c)$ $c_4=\frac{1}{3}$, $\lambda_1=\frac{1}{20}+\text{i}$ $(d)$
$c_4=\frac{1}{3}$, $\lambda_1=2+\text{i}$.}
\label{Fig.2}
\end{figure*}

When $\alpha_1\neq0$, this solution $\rho^{(1)}$ is periodic wave, while ($u^{(1)},v^{(1)}$) is growing, decaying or growing-decaying- periodic
solution, that means the peaks and valleys of this periodic solution will increasing exponentially or decrease exponentially. As an example, taking
$\gamma=c_1=c_3=1,c_2=\frac{1}{2}$, we give the plots of the growing-, decaying- and decaying-growing periodic solution for the f-def nonlocal cm-CID
equation at $t=0$ (see FIG.~\ref{Fig.2}). It is can be seen that as the value of $\alpha_1$ increases (e.g. $\alpha_1=\frac{1}{20}$ and
$\alpha_1=2$), the speed of the periodic wave growing or decaying is accelerating with $c_4=0$(see FIG.~\ref{Fig.2}$(a)(b)$); with $c_4=\frac{1}{3}$,
the plots of the decaying-growing soliton solution for the nonlocal cm-CID equation at $t=0$ are showed in FIG.~\ref{Fig.2}$(c)(d)$. The plots of
solutions for the f-f and def-def equations are similar to that of the f-def equation; the plots of $|\rho^{(1)}|$ is similar to FIG.~\ref{Fig.1}.
\subsection{Two-soliton solutions}

{\bf Proposition 4.} When $N=2$, the quadratic Darboux matrix can be written as
\begin{align*}
T^{[2]}=I-(|\zeta_1\rangle,|\eta_1\rangle,|\zeta_2\rangle,|\eta_2\rangle)W_2^{-1}
\left(\begin{array}{cccc}
\frac{4\lambda_1^*\lambda}{\lambda_1^*-\lambda}\langle \widetilde{\zeta}_1|M\\
\frac{4\lambda_1\lambda}{\lambda_1+\lambda}\langle \widetilde{\eta}_1|M\\
\frac{4\lambda_2^*\lambda}{\lambda_2^*-\lambda}\langle \widetilde{\zeta}_2|M\\
\frac{4\lambda_2\lambda}{\lambda_2+\lambda}\langle \widetilde{\eta}_2|M
\end{array}
\right),
\end{align*}
where the matrix $W_2^{-1}$ and the eigenfunctions $|\zeta_k\rangle$, $|\eta_k\rangle$ $(k=1,2)$ are defined by Eqs.\eqref{y}\eqref{eta}\eqref{W2}.
The relation of new potential $(u^{(2)},v^{(2)},\rho^{(2)})$ and old potential $(u,v,\rho)$ can be given by
\begin{equation}\label{solution21}
\begin{aligned}
&u^{(2)}(y,s)=u-2\sigma_1(\psi_1^{(1)},\widetilde{\psi}_2^{(1)*},\psi_1^{(2)},\widetilde{\psi}_2^{(2)*})W_2^{-1}(\widetilde{\psi}_3^{(1)*},\sigma_1\psi_4^{(1)},\widetilde{\psi}_3^{(2)*},\sigma_1\psi_4^{(2)})^T,\\
&v^{(2)}(y,s)=v-2\sigma_2(\psi_1^{(1)},\widetilde{\psi}_2^{(1)*},\psi_1^{(2)},\psi_2^{(2)*})W_2^{-1}(\widetilde{\psi}_4^{(1)*},-\sigma_2\psi_3^{(1)},\widetilde{\psi}_4^{(2)*},-\sigma_2\psi_3^{(2)})^T,\\
&\rho^{(2)}=\gamma-2((\psi_1^{(1)},\widetilde{\psi}_2^{(1)*},\psi_1^{(2)},\widetilde{\psi}_2^{(2)*})W_2^{-1}(\widetilde{\psi}_1^{(1)*},\psi_2^{(1)},\widetilde{\psi}_1^{(2)*},\psi_2^{(2)})^T)_y,
\end{aligned}
\end{equation}
where $(\psi_{1}^{(j)},\psi_{2}^{(j)},\psi_{3}^{(j)},\psi_{4}^{(j)})(j=1,2)$ are eigenfunctions of the linear spectral problem \eqref{ncmCID-lax} at
$\lambda=\lambda_j$, which are defined by \eqref{eigenfunction}. Then substituting the eigenfunctions to the formula \eqref{solution1}, and
two-soliton solutions of the nonlocal cm-CID equation can be derived.

The conclusion of Proposition 4 can be naturally deduced in accordance with Theorem 2.

Next, we discuss the soliton solutions in three cases according to whether both $\alpha_1$ and $\alpha_2$ are zero, only one of $\alpha_1$ and
$\alpha_2$ is zero, and neither $\alpha_1$ nor $\alpha_2$ is zero, respectively.
\begin{figure*}
\centering
\includegraphics[height=2.8cm]{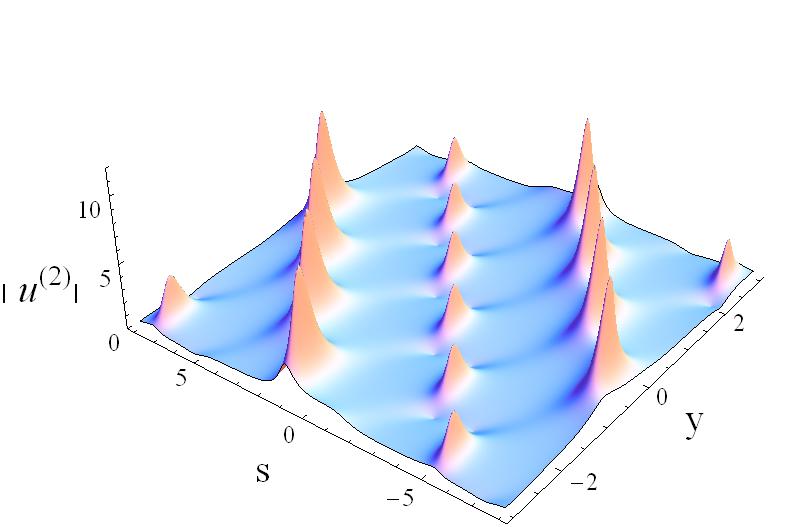}
~~\includegraphics[height=2.8cm]{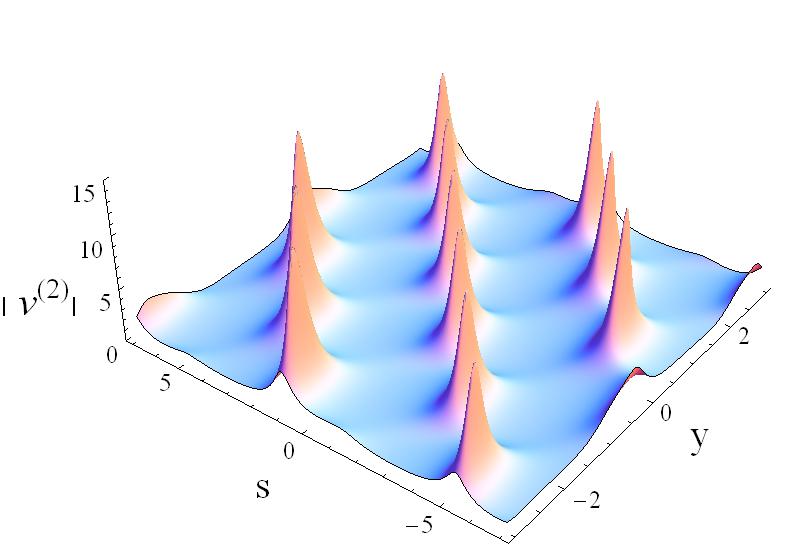}
~~\includegraphics[height=2.8cm]{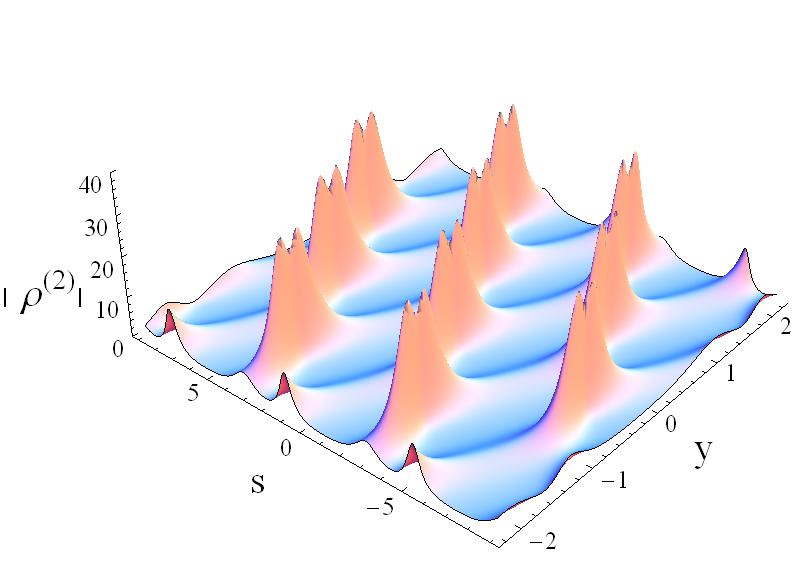}\\
{$(a)$ $|u^{(2)}|$ \hspace{3.5cm}  $(b)$ $|v^{(2)}|$ \hspace{3.5cm} $(c)$  $|\rho^{(2)}|$ }\\
\caption{Double periodic solution for the nonlocal f-def cm-CID equation with $\lambda_1=\frac{\text{i}}{2},\lambda_2=-\text{i}$.}
\label{Fig.3}
\end{figure*}

\textbf{Case 1. Double periodic wave solution}

When $\alpha_1=0$ and $\alpha_2=0$, this solution is double periodic solution. Let $c_1=c_5=1$, this double periodic solution can be given by
\begin{equation}\label{2-periodic}
\begin{aligned}
&u^{(2)}=\frac{\text{i}(\beta_1^2-\beta_2^2)(\beta_2X_1(\sigma_2c_2^*c_4-c_3^*)+\sigma_1\beta_1X_2(\sigma_2c_6^*c_8-c_7^*))}{Y_1},\\
&v^{(2)}=\frac{\text{i}\sigma_2(\beta_1^2-\beta_2^2)(\beta_2X_1(c_2^*c_3+\sigma_1c_4^*)+\beta_1X_2(\sigma_1c_6^*c_7+c_8^*))}{Y_1},\\
&\rho^{(2)}=\gamma+\frac{4\gamma\sigma_1(\beta_1^2-\beta_2^2)^2}{Y_1^2}(1+\sigma_1\sigma_2|c_2|^2)(|c_3|^2+\sigma_1\sigma_2|c_4|^2)X_4^2\\
&~~~~+\frac{4\gamma\sigma_1(\beta_1^2-\beta_2^2)^2}{Y_1^2}\left((1+\sigma_1\sigma_2|c_6|^2)(|c_7|^2+\sigma_1\sigma_2|c_8|^2)X_5^2-X_3X_4X_5\right),\\
\end{aligned}
\end{equation}
where
\begin{align*}
&X_1=(|c_7|^2+\sigma_1\sigma_2|c_8|^2)\text{e}^{-2\text{i}\xi_{2,I}}-\sigma_1(1+\sigma_1\sigma_2|c_6|^2)\text{e}^{2\text{i}\xi_{2,I}},\\
&X_2=(1+\sigma_1\sigma_2|c_2|^2)\text{e}^{2\text{i}\xi_{1,I}}-\sigma_1(|c_3|^2+\sigma_1\sigma_2|c_4|^2)\text{e}^{-2\text{i}\xi_{1,I}},\\
&X_3=\sigma_1\sigma_2(c_2-c_6)(c_3^*c_8^*-c_4^*c_7^*)+(\sigma_2c_3^*c_7+\sigma_1c_4^*c_8)(1+\sigma_1\sigma_2c_2^*c_6)\\
&~+\sigma_1\sigma_2(c_2^*-c_6^*)(c_3c_8-c_4c_7)+(\sigma_2c_3c_7^*+\sigma_1c_4c_8^*)(1+\sigma_1\sigma_2c_2c_6^*),\\
&X_4=(|c_7|^2+\sigma_1\sigma_2|c_8|^2)\text{e}^{-2\text{i}\xi_{2,I}}+\sigma_1(1+\sigma_1\sigma_2|c_6|^2)\text{e}^{2\text{i}\xi_{2,I}},\\
&X_5=(1+\sigma_1\sigma_2|c_2|^2)\text{e}^{2\text{i}\xi_{1,I}}+\sigma_1(|c_3|^2+\sigma_1\sigma_2|c_4|^2)\text{e}^{-2\text{i}\xi_{1,I}},\\
&Y_1=(\beta_1-\beta_2)^2(1+\sigma_1\sigma_2|c_2|^2)(1+\sigma_1\sigma_2|c_6|^2)\text{e}^{2\text{i}(\xi_{1,I}+\xi_{2,I})}\\
&~+(\beta_1-\beta_2)^2(|c_3|^2+\sigma_1\sigma_2|c_4|^2)(|c_7|^2+\sigma_1\sigma_2|c_8|^2)\text{e}^{-2\text{i}(\xi_{1,I}+\xi_{2,I})}\\
&~-\sigma_1(\beta_1+\beta_2)^2(1+\sigma_1\sigma_2|c_2|^2)(|c_7|^2+\sigma_1\sigma_2|c_8|^2)\text{e}^{2\text{i}(\xi_{1,I}-\xi_{2,I})}\\
&~-\sigma_1(\beta_1+\beta_2)^2(|c_3|^2+\sigma_1\sigma_2|c_4|^2)(1+\sigma_1\sigma_2|c_6|^2)\text{e}^{-2\text{i}(\xi_{1,I}-\xi_{2,I})}+4\beta_1\beta_2\omega_3,
\end{align*}
with $\xi_{k,I}=\gamma\beta_ky+\frac{s}{4\beta_k}$ $(k=1,2)$.
In Figure~\ref{Fig.3}, we give plots of double-periodic wave for the nonlocal f-def cm-CID equation with parameters
$\gamma=1,\beta_1=\frac{1}{2},\beta_2=-1$, $c_2=c_6=\frac{1}{2},c_3=c_8=0,c_4=\frac{1}{2},c_7=2$. It is a periodic wave in both the $y-$ and $s-$
directions, and there are two different peak values in each period. The plots of solutions for the f-f and def-def equations are similar to that of
the f-def equation.
\begin{figure*}
\centering
\includegraphics[height=2.8cm]{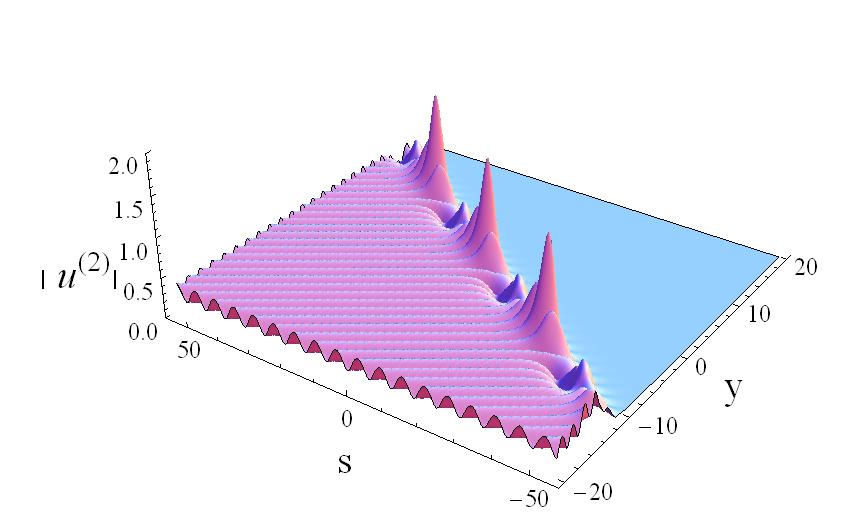}
~~\includegraphics[height=2.8cm]{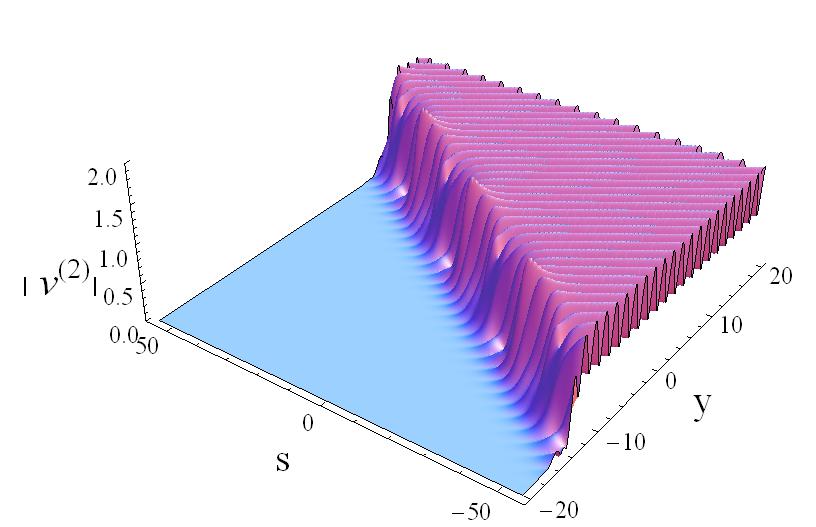}
~~\includegraphics[height=2.8cm]{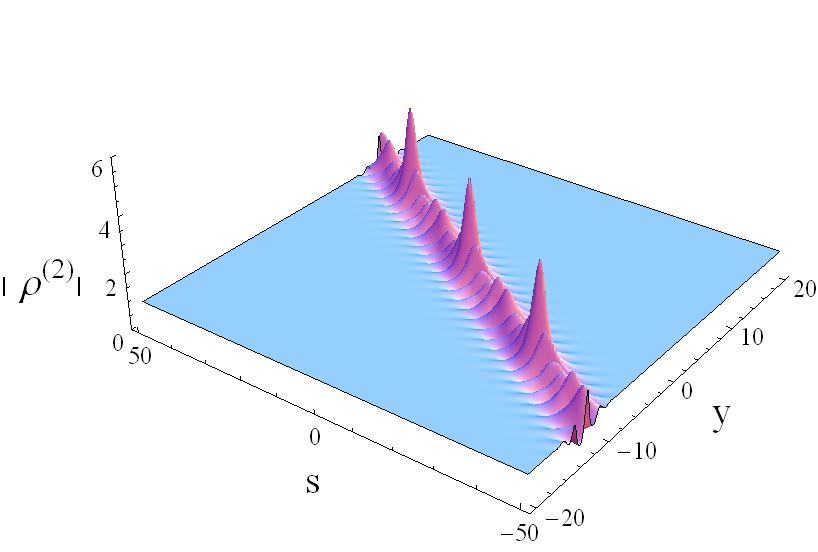}\\
{$(a)$ $|u^{(2)}|$ \hspace{4cm}  $(b)$ $|v^{(2)}|$ \hspace{4cm} $(c)$  $|\rho^{(2)}|$ }\\
\includegraphics[height=2.8cm]{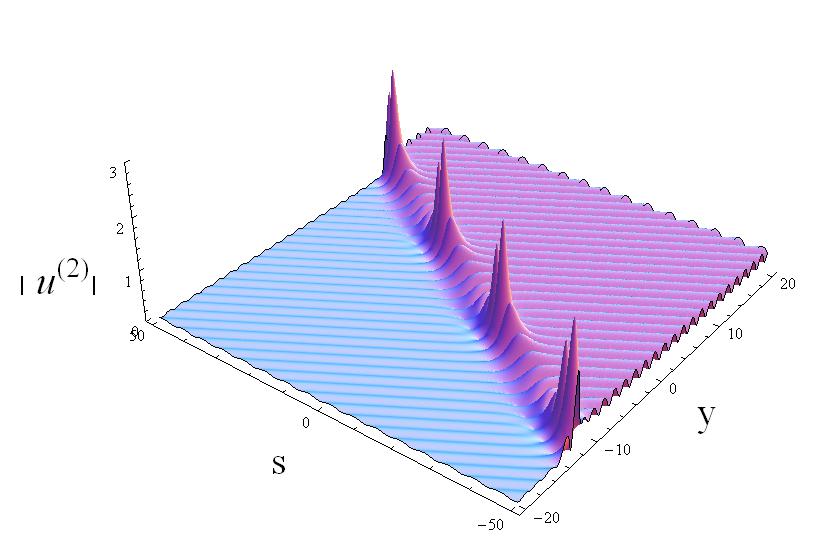}
~~\includegraphics[height=2.8cm]{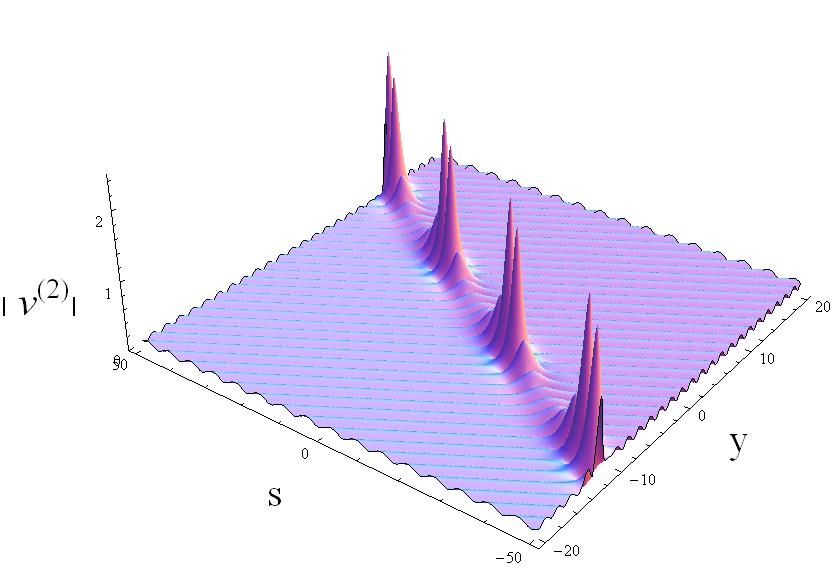}
~~\includegraphics[height=2.8cm]{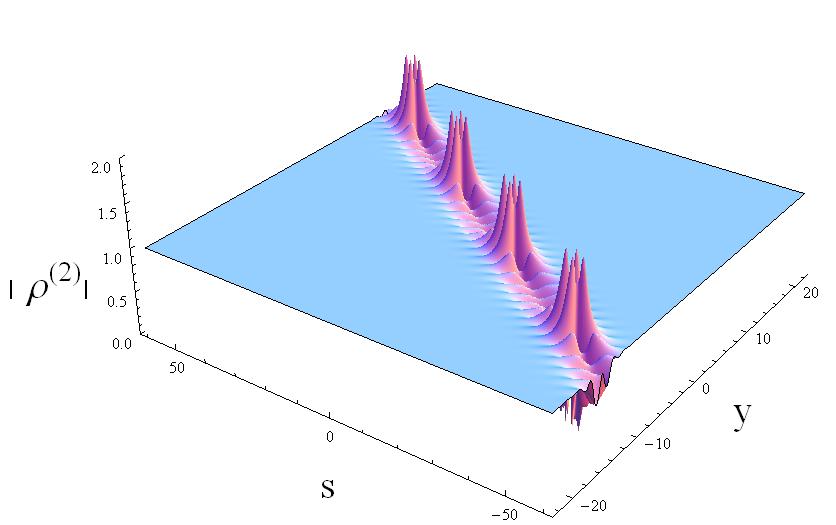}\\
{$(d)$ $|u^{(2)}|$ \hspace{4cm}  $(e)$ $|v^{(2)}|$ \hspace{4cm} $(f)$  $|\rho^{(2)}|$ }\\
\caption{Periodic-like wave solution for the f-def nonlocal cm-CID equation with $\alpha_1=\frac{3}{5},\beta_1=\frac{4}{5},\alpha_2=0,\beta_2=-1$:
$(a)$-$(c)$ $c_3=0,c_7=2$, $(d)$-$(f)$ $c_3=-1,c_7=2$, $(a)(d)$ periodic-like waves $|u^{(2)}|$, $(b)(e)$ periodic-like waves $|v^{(2)}|$, $(c)(f)$
breather-like wave $|\rho^{(2)}|$.}
\label{Fig.4}
\end{figure*}
\begin{figure*}
\centering
\includegraphics[height=2.8cm]{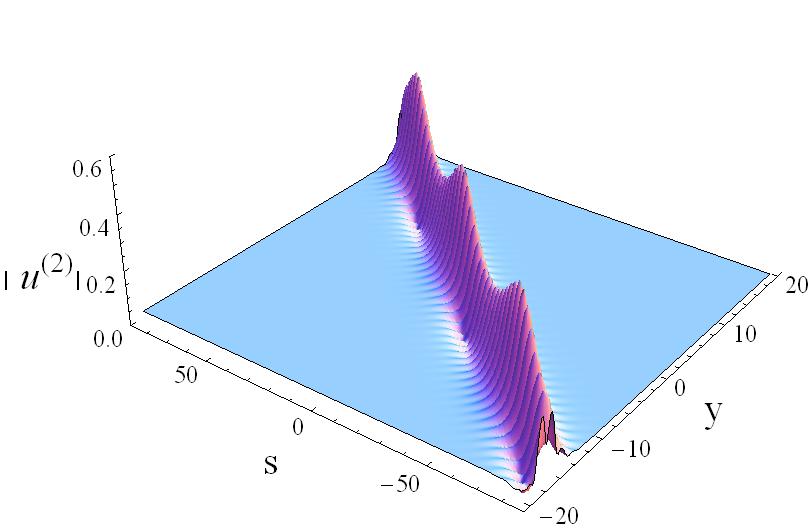}
~~\includegraphics[height=2.8cm]{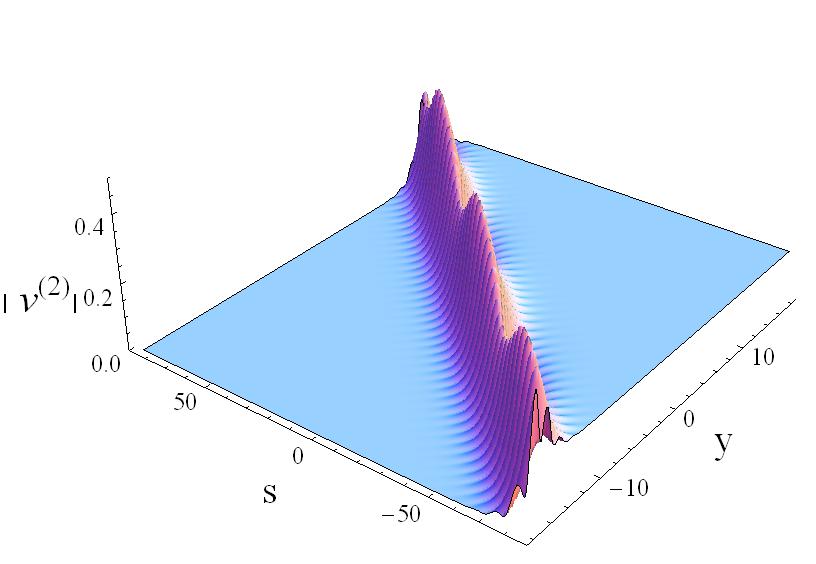}
~~\includegraphics[height=2.7cm]{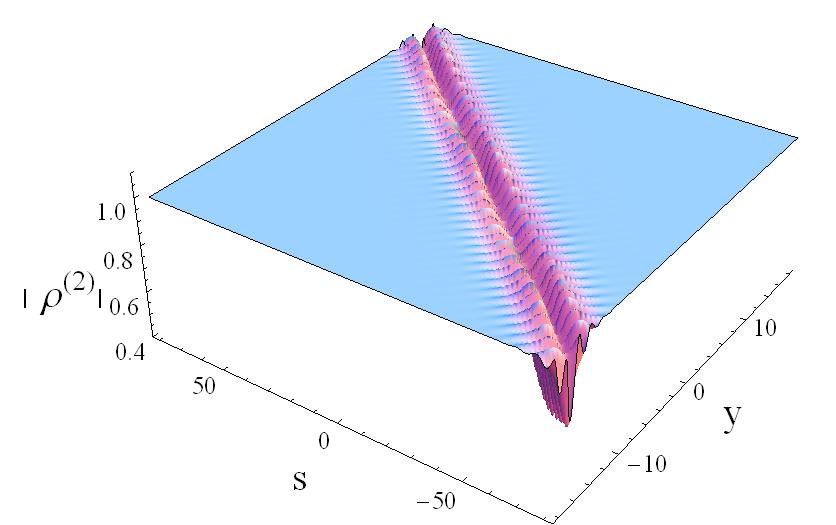}\\
{$(a)$ $|u^{(2)}|$ \hspace{3cm}  $(b)$ $|v^{(2)}|$ \hspace{3cm} $(c)$  $|\rho^{(2)}|$ }\\
\includegraphics[height=3cm]{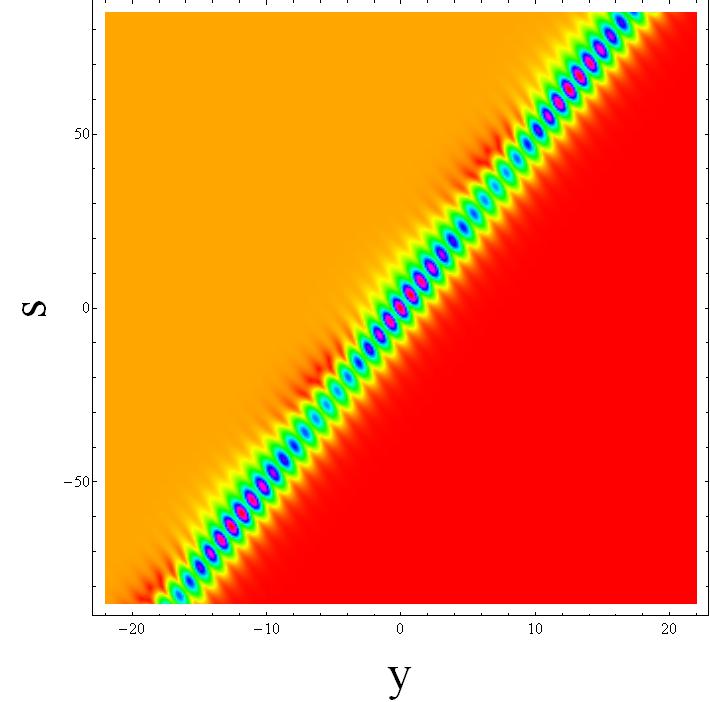}~~\qquad
~~\includegraphics[height=3cm]{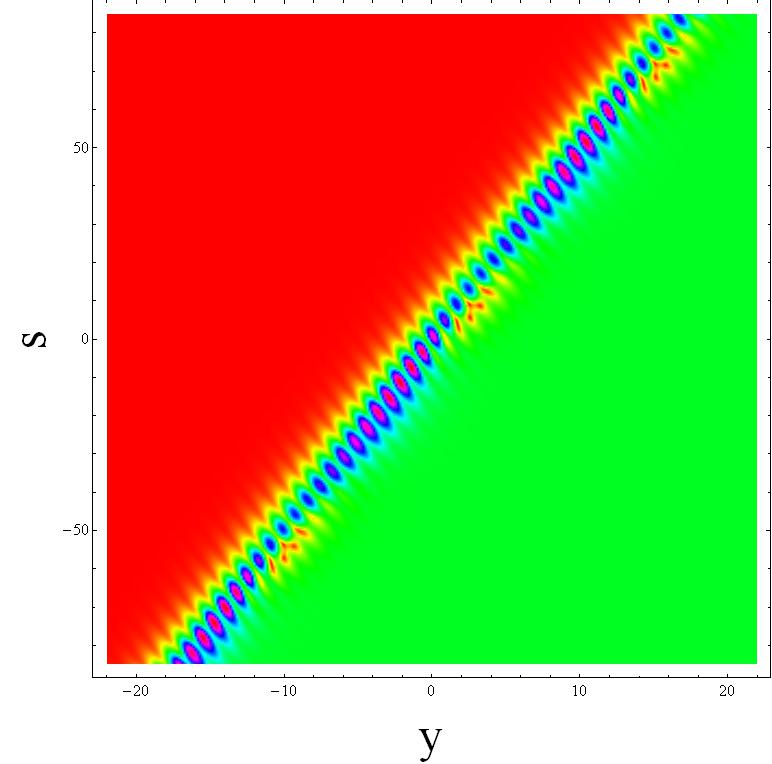}~~\qquad
~~\includegraphics[height=3cm]{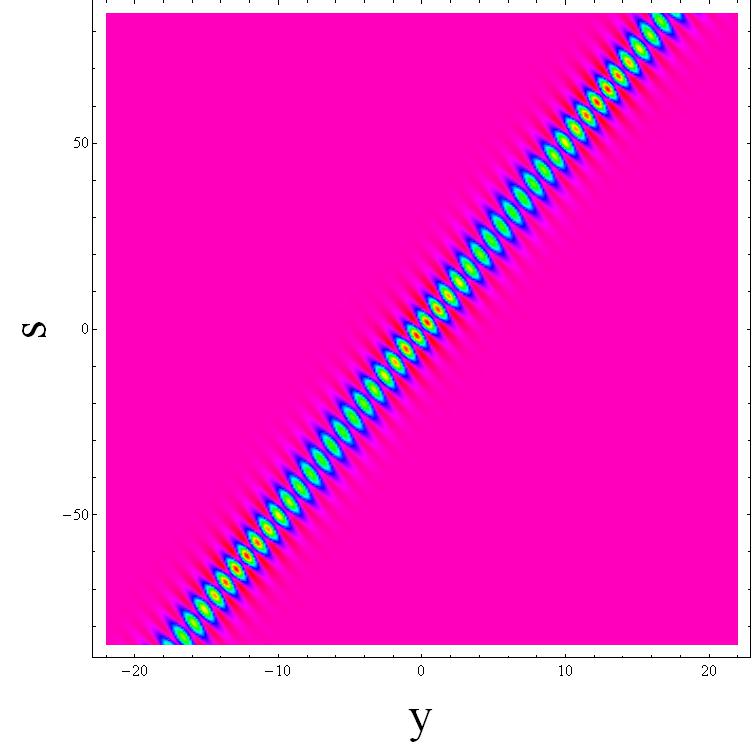}\\
{$(d)$ $|u^{(2)}|$ \hspace{3cm}  $(e)$ $|v^{(2)}|$ \hspace{3cm} $(f)$  $|\rho^{(2)}|$ }\\
\caption{Breather-like solution for the f-def nonlocal cm-CID equation with
$c_3=\frac{3}{2},c_7=1,\alpha_1=\frac{1}{2},\beta_1=1,\alpha_2=0,\beta_2=-1$: $(a)(d)$ bright breather-like soliton $|u^{(2)}|$, $(b)(e)$ bright
breather-like soliton $|v^{(2)}|$, $(c)(f)$ dark breather-like soliton $|\rho^{(2)}|$.}
\label{Fig.5}
\end{figure*}

\textbf{Case 2. periodic-like wave}

When only one of $\alpha_1$ and $\alpha_2$ is zero(e.g. $\alpha_2=0$), this solution is periodic-like wave or breather-like solution. The properties
of periodic-like solution for the f-f, def-def equation and that of the f-def equation are similar, here we only discuss the solution of the f-def
case. As example, with the parameters
$\gamma=1,c_1=c_5=c_8=1,c_2=c_6=\frac{1}{3},c_4=\frac{1}{2},\alpha_1=\frac{3}{5},\beta_1=\frac{4}{5},\alpha_2=0,\beta_2=-1$, we obtain the fomula of
periodic-like wave as
\begin{align*}
u^{(2)}=\frac{X_6}{5Y_2},~~v^{(2)}=\frac{X_7}{5Y_2},~~\rho^{(2)}=\frac{X_8}{25Y_2^2},~~
\end{align*}
with
\begin{align*}
&X_6=9\text{e}^{\xi_2}((4+3\text{i}+4(4-3\text{i})c_3^2)(1+3c_7)-16c_3(3+c_7))-64(6c_3\text{e}^{2\xi_1+4\xi_2}-\text{e}^{2\widetilde{\xi}_1^*+4\xi_2})\\
&~~~+72(1-c_7^2)(6c_3\text{e}^{2\xi_1}-\text{e}^{2\widetilde{\xi}_1^*})-32\text{e}^{2\xi_1+2\widetilde{\xi}_1^*+2\xi_2}(4-3\text{i}-(12+9\text{i})c_7),\\
&X_7=9\text{e}^{\xi_2}((4-3\text{i}+4(4+3\text{i})c_3^2)(3+c_7)-16c_3(1+3c_7))+64(3\text{e}^{2\xi_1+4\xi_2}-2c_3\text{e}^{2\widetilde{\xi}_1^*+4\xi_2})\\
&~~~-72(1-c_7^2)(3\text{e}^{2\xi_1}-2c_3\text{e}^{2\widetilde{\xi}_1^*})+32\text{e}^{2\xi_1+2\widetilde{\xi}_1^*+2\xi_2}((4-3\text{i})c_7-12-9\text{i}),\\
&X_8=1024(1087-863c_7^2+4c_3^2(1087c_7^2-863)-896c_3c_7)\text{e}^{2\xi_1+2\widetilde{\xi}_1^*+4\xi_2}\\
&~~~+25600((1-c_7^2)\text{e}^{2\xi_1+2\widetilde{\xi}_1^*}-8\text{e}^{2\xi_1+2\widetilde{\xi}_1^*+4\xi_2})^2+409600(1-2c_3c_7)^2(\text{e}^{4\widetilde{\xi}_1^*+4\xi_2}+\text{e}^{4\xi_1+4\xi_2})\\
&~~~-448(1-4c_3^2)(64\text{e}^{2\xi_1+2\widetilde{\xi}_1^*+8\xi_2}+81(1-c_7^2)^2\text{e}^{2\xi_1+2\widetilde{\xi}_1^*})\\
&~~~+1310720(1-2c_3c_7)(\text{e}^{2\xi_1+4\widetilde{\xi}_1^*+6\xi_2}+\text{e}^{4\xi_1+2\widetilde{\xi}_1^*+6\xi_2})+25(1-4c_3^2)^2(8\text{e}^{4\xi_2}-81(1-c_7^2))^2\\
&~~~+5120(1-2c_3c_7)(1-c_7^2)(81(1-4c_3^2)(\text{e}^{2\xi_1+2\xi_2}+\text{e}^{2\widetilde{\xi}_1^*+2\xi_2})-32(\text{e}^{4\xi_1+2\widetilde{\xi}_1^*+2\xi_2}+\text{e}^{2\xi_1+4\widetilde{\xi}_1^*+2\xi_2}))\\
&~~~-40960(1-4c_3^2)(1-2c_3c_7)(\text{e}^{2\xi_1+6\xi_2}+\text{e}^{2\widetilde{\xi}_1^*+6\xi_2}),\\
&Y_2=(1-4c_3^2)(81(1-c_7^2)+8\text{e}^{4\xi_2})+32(1-c_7^2)\text{e}^{2\xi_1+2\widetilde{\xi}_1^*}+256\text{e}^{2\xi_1+2\widetilde{\xi}_1^*+4\xi_2}\\
&~~~+128(\text{e}^{2\widetilde{\xi}_1^*+2\xi_2}+\text{e}^{2\xi_1+2\xi_2})(1-2c_3c_7),
\end{align*}
with $\xi_1=\frac{3+4\text{i}}{5}y-\frac{3-4\text{i}}{20}t$, $\xi_2=-\text{i}(y+\frac{1}{4}t)$.

In FIG.~\ref{Fig.4}, taking $c_3=0,c_7=2$, the periodic-like solution can be derived(see FIG.~\ref{Fig.4}$(a)$-$(c)$), where $|u^{(2)}|$ and
$|v^{(2)}|$ describes interaction of one periodic wave and one breather soliton, $|u^{(2)}|$ presents a periodic wave on the left of the breather
soliton and a zero plane on the right; $|v^{(2)}|$ shows that on the right side of the breather soliton are periodic waves and on the left side are
planes; $|\rho^{(2)}|$ displays the propagation of the breather-like solution. Setting the parameters $c_3=-1,c_7=2$, another type of periodic-like
wave can be obtained(see FIG.~\ref{Fig.4}$(d)$-$(f)$), where $|u^{(2)}|$ and $|v^{(2)}|$ shows interaction of two periodic waves and one breather
soliton, $|\rho^{(2)}|$ is bright breather-like wave. Let $c_3=\frac{3}{2},c_7=1$, we give the plots of breather-like solution for the f-def nonlocal
cm-CID equation(see FIG.~\ref{Fig.5}), where $|u^{(2)}|$ and $|v^{(2)}|$ are bright breather-like waves, $|\rho^{(2)}|$ is dark breather-like wave.

\textbf{Case 3. interaction of two-soliton wave}

When $\alpha_1\alpha_2\neq0$, this solution describes the collision of two breather solitons. Taking $\gamma=1,c_1=c_5=1$, and fixing $\xi_1\sim
O(1)$, $\xi_2\sim O(1)$ respectively, we analyze the asymptotic behavior of the two-soliton solution as follows:\\
i) if $\lambda_{2,R}(|\lambda_2|^2-|\lambda_1|^2)>0$ and $\lambda_{1,R}(|\lambda_2|^2-|\lambda_1|^2)<0$, we have
\begin{equation}\label{asymptotic-1}
\begin{aligned}
u^{(2)}\rightarrow\begin{cases}
u_{1}^{-}+u_{2}^{-},~ s\rightarrow-\infty,\\
u_{1}^{+}+u_{2}^{+},~ s\rightarrow+\infty,
\end{cases}v^{(2)}\rightarrow\begin{cases}
v_{1}^{-}+v_{2}^{-},~ s\rightarrow-\infty,\\
v_{1}^{+}+v_{2}^{+},~ s\rightarrow+\infty,
\end{cases}\rho^{(2)}\rightarrow\begin{cases}
\rho_{1}^{-}+\rho_{2}^{-},~ s\rightarrow-\infty,\\
\rho_{1}^{+}+\rho_{2}^{+},~ s\rightarrow+\infty,
\end{cases}
\end{aligned}
\end{equation}
ii) if $\lambda_{2,R}(|\lambda_2|^2-|\lambda_1|^2)<0$ and $\lambda_{1,R}(|\lambda_2|^2-|\lambda_1|^2)>0$, we get
\begin{equation}\label{asymptotic-2}
\begin{aligned}
u^{(2)}\rightarrow\begin{cases}
u_{1}^{+}+u_{2}^{+},~ s\rightarrow-\infty,\\
u_{1}^{-}+u_{2}^{-},~ s\rightarrow+\infty,
\end{cases}v^{(2)}\rightarrow\begin{cases}
v_{1}^{+}+v_{2}^{+},~ s\rightarrow-\infty,\\
v_{1}^{-}+v_{2}^{-},~ s\rightarrow+\infty,
\end{cases}\rho^{(2)}\rightarrow\begin{cases}
\rho_{1}^{+}+\rho_{2}^{+},~ s\rightarrow-\infty,\\
\rho_{1}^{-}+\rho_{2}^{-},~ s\rightarrow+\infty,
\end{cases}
\end{aligned}
\end{equation}
iii) if $\lambda_{2,R}(|\lambda_2|^2-|\lambda_1|^2)>0$ and $\lambda_{1,R}(|\lambda_2|^2-|\lambda_1|^2)>0$, we obtain
\begin{equation}\label{asymptotic-3}
\begin{aligned}
u^{(2)}\rightarrow\begin{cases}
u_{1}^{-}+u_{2}^{+},~ s\rightarrow-\infty,\\
u_{1}^{+}+u_{2}^{-},~ s\rightarrow+\infty,
\end{cases}v^{(2)}\rightarrow\begin{cases}
v_{1}^{-}+v_{2}^{+},~ s\rightarrow-\infty,\\
v_{1}^{+}+v_{2}^{-},~ s\rightarrow+\infty,
\end{cases}\rho^{(2)}\rightarrow\begin{cases}
\rho_{1}^{-}+\rho_{2}^{+},~ s\rightarrow-\infty,\\
\rho_{1}^{+}+\rho_{2}^{-},~ s\rightarrow+\infty,
\end{cases}
\end{aligned}
\end{equation}
iv) if $\lambda_{2,R}(|\lambda_2|^2-|\lambda_1|^2)<0$ and $\lambda_{1,R}(|\lambda_2|^2-|\lambda_1|^2)<0$, we have
\begin{equation}\label{asymptotic-4}
\begin{aligned}
u^{(2)}\rightarrow\begin{cases}
u_{1}^{+}+u_{2}^{-},~ s\rightarrow-\infty,\\
u_{1}^{-}+u_{2}^{+},~ s\rightarrow+\infty,
\end{cases}v^{(2)}\rightarrow\begin{cases}
v_{1}^{+}+v_{2}^{-},~ s\rightarrow-\infty,\\
v_{1}^{-}+v_{2}^{+},~ s\rightarrow+\infty,
\end{cases}\rho^{(2)}\rightarrow\begin{cases}
\rho_{1}^{+}+\rho_{2}^{-},~ s\rightarrow-\infty,\\
\rho_{1}^{-}+\rho_{2}^{+},~ s\rightarrow+\infty,
\end{cases}
\end{aligned}
\end{equation}
where
\begin{eqnarray*}
&&\rho_{1}^{-}=\gamma(1-\frac{(\lambda_1+\lambda_1^*)^2}{4|\lambda_1|^2}\text{sech}^2(2\xi_{1,R}+\frac{1}{2}\ln
A_1)),\rho_{1}^{+}=\gamma(1-\frac{(\lambda_1+\lambda_1^*)^2}{4|\lambda_1|^2}\text{sech}^2(2\xi_{1,R}-\frac{1}{2}\ln A_1^*)),\\
&&\rho_{2}^{-}=\gamma(1-\frac{(\lambda_2+\lambda_2^*)^2}{4|\lambda_2|^2}\text{sech}^2(2\xi_{2,R}+\frac{1}{2}\ln
A_2)),\rho_{2}^{+}=\gamma(1-\frac{(\lambda_2+\lambda_2^*)^2}{4|\lambda_2|^2}\text{sech}^2(2\xi_{2,R}-\frac{1}{2}\ln A_2^*)),\\
&&u_{1}^{-}=\frac{\sigma_2A_3(\sigma_1c_8f_1\text{e}^{2\text{i}\xi_{1,I}}+c_6^*g_1\text{e}^{-2\text{i}\xi_{1,I}})}{4|\lambda_1\lambda_2|^2}\text{sech}\left(2\xi_{1,R}+\ln\left(\frac{A_3(\sigma_1c_3^*c_7+\sigma_2c_4^*c_8)(1+\sigma_1\sigma_2c_2c_6^*)}{-|\lambda_1-\lambda_2^*|^2}\right)\right),\\
&&u_{2}^{+}=\frac{\sigma_2A_4^*(\sigma_1c_4f_4\text{e}^{2\text{i}\xi_{2,I}}+c_2^*g_7\text{e}^{-2\text{i}\xi_{2,I}})}{4|\lambda_1\lambda_2|^2}\text{sech}\left(2\xi_{2,R}+\ln\left(\frac{A_4^*(\sigma_1c_3c_7^*+\sigma_2c_4c_8^*)(1+\sigma_1\sigma_2c_2^*c_6)}{-|\lambda_1-\lambda_2^*|^2}\right)\right),\\
&&v_{1}^{-}=\frac{\sigma_2A_3(-\sigma_1c_7f_1\text{e}^{2\text{i}\xi_{1,I}}+c_6^*g_2\text{e}^{-2\text{i}\xi_{1,I}})}{4|\lambda_1\lambda_2|^2}\text{sech}\left(2\xi_{1,R}+\ln\left(\frac{A_3(\sigma_1c_3^*c_7+\sigma_2c_4^*c_8)(1+\sigma_1\sigma_2c_2c_6^*)}{-|\lambda_1-\lambda_2^*|^2}\right)\right),\\
&&v_{1}^{+}=\frac{\sigma_2A_3^*(c_8^*f_2\text{e}^{2\text{i}\xi_{1,I}}+\sigma_1g_4\text{e}^{-2\text{i}\xi_{1,I}})}{4|\lambda_1\lambda_2|^2}\text{sech}\left(2\xi_{1,R}+\ln\left(\frac{\sigma_1\sigma_2A_3^*(c_2-c_6)(c_4^*c_7^*-c_3^*c_8^*)}{|\lambda_1+\lambda_2|^2}\right)\right),\\
&&v_{2}^{-}=\frac{A_4(\sigma_2c_4^*f_3\text{e}^{2\text{i}\xi_{2,I}}+g_6\text{e}^{-2\text{i}\xi_{1,I}})}{4|\lambda_1\lambda_2|^2}\text{sech}\left(2\xi_{2,R}+\ln\left(\frac{\sigma_1\sigma_2A_4(c_2-c_6)(c_4^*c_7^*-c_3^*c_8^*)}{|\lambda_1+\lambda_2|^2}\right)\right),\\
&&v_{2}^{+}=\frac{\sigma_2A_4^*(-\sigma_1c_3f_4\text{e}^{2\text{i}\xi_{1,I}}-c_2^*g_8\text{e}^{-2\text{i}\xi_{2,I}})}{4|\lambda_1\lambda_2|^2}\text{sech}\left(2\xi_{2,R}+\ln\left(\frac{A_4^*(\sigma_1c_3c_7^*+\sigma_2c_4c_8^*)(1+\sigma_1\sigma_2c_2^*c_6)}{-|\lambda_1-\lambda_2^*|^2}\right)\right),\\
&&u_{1}^{+}=\frac{A_3^*(\sigma_1c_7^*f_2\text{e}^{2\text{i}\xi_{1,I}}+g_3\text{e}^{-2\text{i}\xi_{1,I}})}{4|\lambda_1\lambda_2|^2}\text{sech}\left(2\xi_{1,R}+\ln\left(\frac{\sigma_1\sigma_2A_3^*(c_2-c_6)(c_4^*c_7^*-c_3^*c_8^*)}{|\lambda_1+\lambda_2|^2}\right)\right),\\
&&u_{2}^{-}=\frac{A_4(\sigma_1c_3^*f_3\text{e}^{2\text{i}\xi_{2,I}}+g_5\text{e}^{-2\text{i}\xi_{1,I}})}{4|\lambda_1\lambda_2|^2}\text{sech}\left(2\xi_{2,R}+\ln\left(\frac{\sigma_1\sigma_2A_4(c_2-c_6)(c_4^*c_7^*-c_3^*c_8^*)}{|\lambda_1+\lambda_2|^2}\right)\right),
\end{eqnarray*}
with $\xi_{k,R}=\alpha_ky-\frac{\alpha_ks}{4|\lambda_k|^2}$, $\xi_{k,I}=\beta_ky+\frac{\beta_ks}{4|\lambda_k|^2}$ $(k=1,2)$, and
\begin{eqnarray*}
&&A_1=\frac{|\lambda_1-\lambda_2^*|^2(c_2^*-c_6^*)(c_3c_8-c_4c_7)}{|\lambda_1+\lambda_2|^2(1+\sigma_1\sigma_2c_2c_6^*)(\sigma_2c_3^*c_7+\sigma_1c_4^*c_8)},~A_2=\frac{|\lambda_1-\lambda_2^*|^2(c_2-c_6)(c_3^*c_8^*-c_4^*c_7^*)}{|\lambda_1+\lambda_2|^2(1+\sigma_1\sigma_2c_2c_6^*)(\sigma_2c_3^*c_7+\sigma_1c_4^*c_8)},\\
&&A_3=\frac{|(\lambda_1+\lambda_2)(\lambda_1-\lambda_2^*)|^2}{(c_2^*-c_6^*)(c_3c_8-c_4c_7)(1+\sigma_1\sigma_2c_2c_6^*)(\sigma_2c_3^*c_7+\sigma_1c_4^*c_8)},
\end{eqnarray*}
\begin{eqnarray*}
&&A_4=\frac{|(\lambda_1+\lambda_2)(\lambda_1-\lambda_2^*)|^2}{(c_2-c_6)(c_3^*c_8^*-c_4^*c_7^*)(1+\sigma_1\sigma_2c_2c_6^*)(\sigma_2c_3^*c_7+\sigma_1c_4^*c_8)},\\
&&f_1=\frac{\lambda_1^*\lambda_2^*(c_2^*-c_6^*)}{\lambda_1^*+\lambda_2^*}-\frac{\lambda_1\lambda_2^*c_2^*(1+\sigma_1\sigma_2c_2c_6^*)}{\lambda_1-\lambda_2^*}+\frac{|\lambda|^2c_6^*(1+\sigma_1\sigma_2|\lambda|^2)}{\lambda_1-\lambda_1^*},\\
&&f_2=\frac{\sigma_1\sigma_2\lambda_1\lambda_2c_2^*(c_2-c_6)}{\lambda_1+\lambda_2}+\frac{\lambda_1^*\lambda_2(1+\sigma_1\sigma_2c_2^*c_6)}{\lambda_1^*-\lambda_2}-\frac{|\lambda_1|^2(1+\sigma_1\sigma_2|c_2|^2)}{\lambda_1-\lambda_1^*},\\
&&f_3=\frac{\sigma_1\sigma_2\lambda_1\lambda_2c_6^*(c_2-c_6)}{\lambda_1+\lambda_2}+\frac{\lambda_1\lambda_2^*(1+\sigma_1\sigma_2c_2c_6^*)}{\lambda_1-\lambda_2^*}-\frac{|\lambda_2|^2(1+\sigma_1\sigma_2|c_6|^2)}{\lambda_2-\lambda_2^*},\\
&&f_4=\frac{-\lambda_1^*\lambda_2^*(c_2^*-c_6^*)}{\lambda_1^*+\lambda_2^*}+\frac{\lambda_1^*\lambda_2c_6^*(1+\sigma_1\sigma_2c_2^*c_6)}{\lambda_1^*-\lambda_2}+\frac{|\lambda_2|^2c_2^*(1+\sigma_1\sigma_2|c_6|^2)}{\lambda_2-\lambda_2^*},\\
&&g_1=\frac{\lambda_1\lambda_2c_3^*(c_4c_7-c_3c_8)}{\lambda_1+\lambda_2}-\frac{|\lambda_1|^2c_8(|c_3|^2+\sigma_1\sigma_2|c_4|^2)}{\lambda_1+\lambda_2}-\frac{\lambda_1^*\lambda_2c_6^*c_4(c_3^*c_7+\sigma_1\sigma_2c_4^*c_8)}{\lambda_1^*-\lambda_2},\\
&&g_2=\frac{\sigma_1\sigma_2\lambda_1\lambda_2c_4^*(c_4c_7-c_3c_8)}{\lambda_1+\lambda_2}+\frac{|\lambda_1|^2c_7(|c_3|^2+\sigma_1\sigma_2|c_4|^2)}{\lambda_1-\lambda_1^*}+\frac{\lambda_1^*\lambda_2c_3(c_3^*c_7+\sigma_1\sigma_2c_4^*c_8)}{\lambda_1^*-\lambda_2},\\
&&g_3=\frac{\sigma_1\sigma_2\lambda_1^*\lambda_2^*c_4(c_3^*c_8^*-c_4^*c_7^*)}{\lambda_1^*+\lambda_2^*}-\frac{|\lambda_1|^2c_7^*(|c_3|^2+\sigma_1\sigma_2|c_4|^2)}{\lambda_1-\lambda_1^*}-\frac{\lambda_1\lambda_2^*c_3^*(c_3c_7^*+\sigma_1\sigma_2c_4c_8^*)}{\lambda_1-\lambda_2^*},\\
&&g_4=\frac{\lambda_1^*\lambda_2^*c_3(c_4^*c_7^*-c_3^*c_8^*)}{\lambda_1^*+\lambda_2^*}+\frac{|\lambda_1|^2c_8^*(|c_3|^2+\sigma_1\sigma_2|c_4|^2)}{\lambda_1-\lambda_1^*}-\frac{\lambda_1\lambda_2^*c_4^*(c_3^*c_7+\sigma_1\sigma_2c_4c_8^*)}{\lambda_1-\lambda_2^*},\\
&&g_5=-\frac{\sigma_1\sigma_2\lambda_1^*\lambda_2^*c_8(c_3^*c_8^*-c_4^*c_7^*)}{\lambda_1^*+\lambda_2^*}+\frac{|\lambda_2|^2c_3^*(|c_7|^2+\sigma_1\sigma_2|c_8|^2)}{\lambda_2-\lambda_2^*}-\frac{\lambda_1^*\lambda_2c_7^*(c_3^*c_7+\sigma_1\sigma_2c_4^*c_8)}{\lambda_1^*-\lambda_2},\\
&&g_6=\frac{\sigma_1\sigma_2\lambda_1^*\lambda_2^*c_7(c_3^*c_8^*-c_4^*c_7^*)}{\lambda_1^*+\lambda_2^*}+\frac{|\lambda_2|^2c_4^*(\sigma_1\sigma_2|c_7|^2+|c_8|^2)}{\lambda_2-\lambda_2^*}+\frac{\lambda_1^*\lambda_2c_8^*(c_3^*c_7+\sigma_1\sigma_2c_4^*c_8)}{\lambda_1^*-\lambda_2},\\
&&g_7=\frac{\lambda_1\lambda_2c_7^*(c_3c_8-c_4c_7)}{\lambda_1+\lambda_2}-\frac{|\lambda_2|^2c_4(|c_7|^2+\sigma_1\sigma_2|c_8|^2)}{\lambda_2-\lambda_2^*}+\frac{\lambda_1\lambda_2^*c_8(c_3c_7^*+\sigma_1\sigma_2c_4c_8^*)}{\lambda_1-\lambda_2^*},\\
&&g_8=\frac{\sigma_1\sigma_2\lambda_1\lambda_2c_8^*(c_3c_8-c_4c_7)}{\lambda_1+\lambda_2}-\frac{|\lambda_2|^2c_3(|c_7|^2+\sigma_1\sigma_2|c_8|^2)}{\lambda_2-\lambda_2^*}+\frac{\lambda_1\lambda_2^*c_7(c_3c_7^*+\sigma_1\sigma_2c_4c_8^*)}{\lambda_1-\lambda_2^*}.
\end{eqnarray*}
\begin{figure*}
\centering
\includegraphics[height=2.8cm]{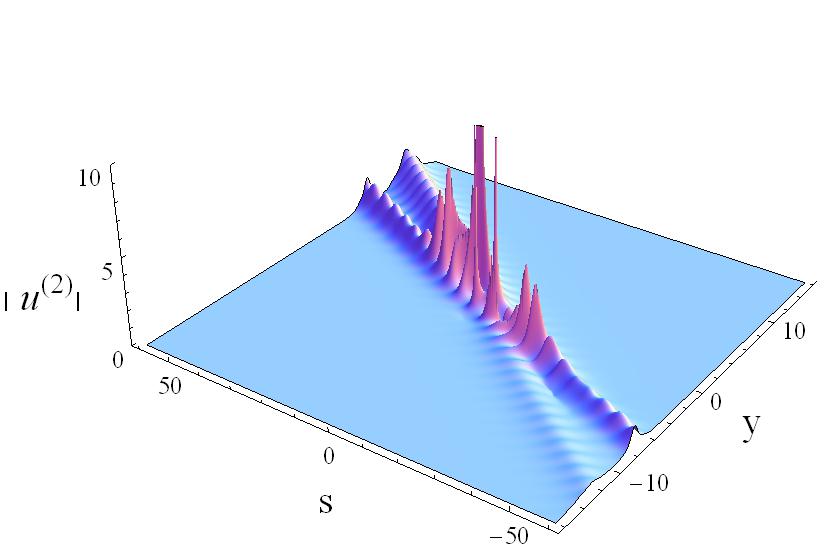}
~~\includegraphics[height=2.8cm]{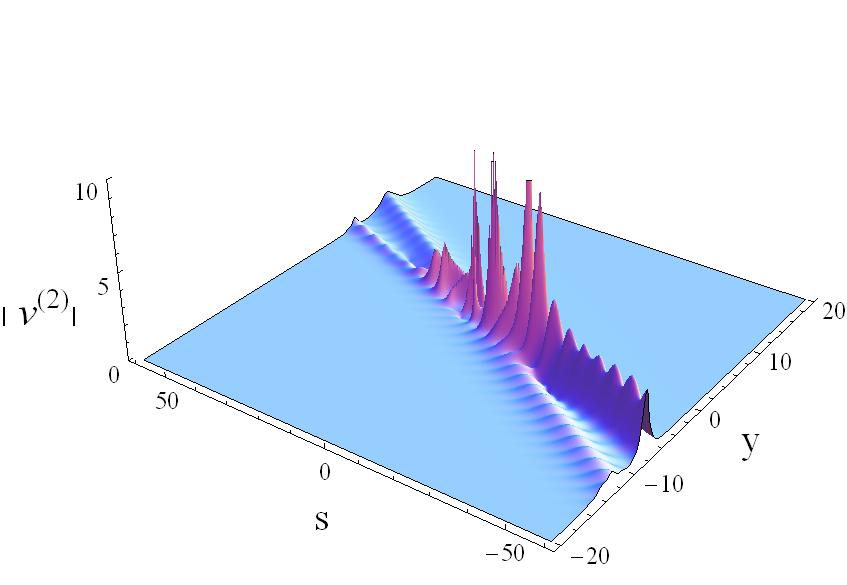}
~~\includegraphics[height=2.8cm]{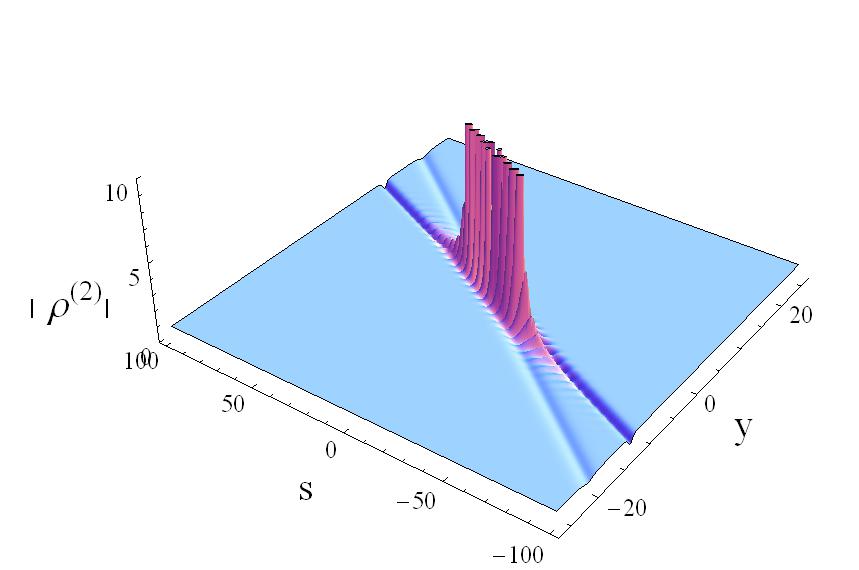}\\
{$(a)$ $|u^{(2)}|$ \hspace{3.2cm}  $(b)$ $|v^{(2)}|$ \hspace{3.2cm} $(c)$ $|\rho^{(2)}|$ }\\
\includegraphics[height=2.8cm]{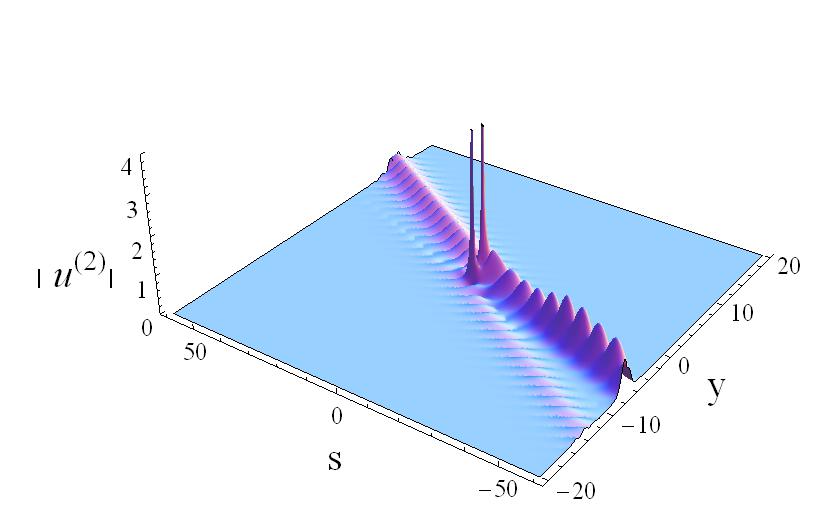}
~~\includegraphics[height=2.8cm]{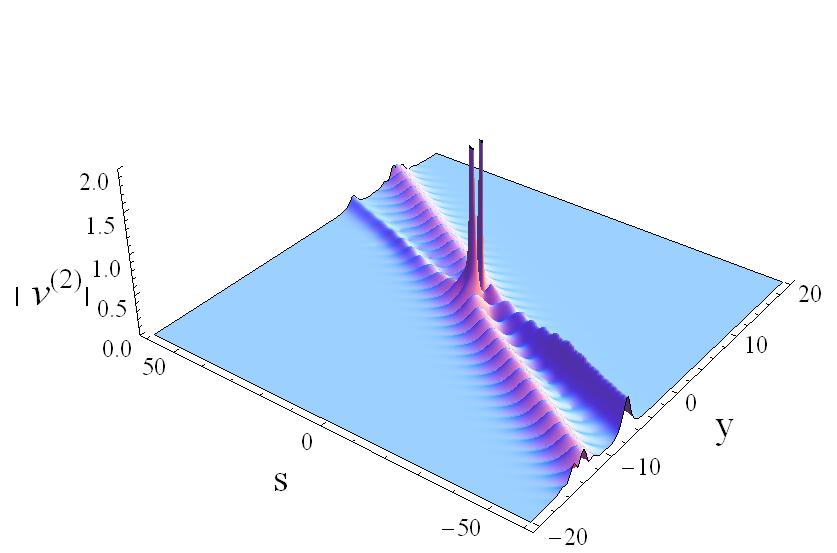}
~~\includegraphics[height=2.8cm]{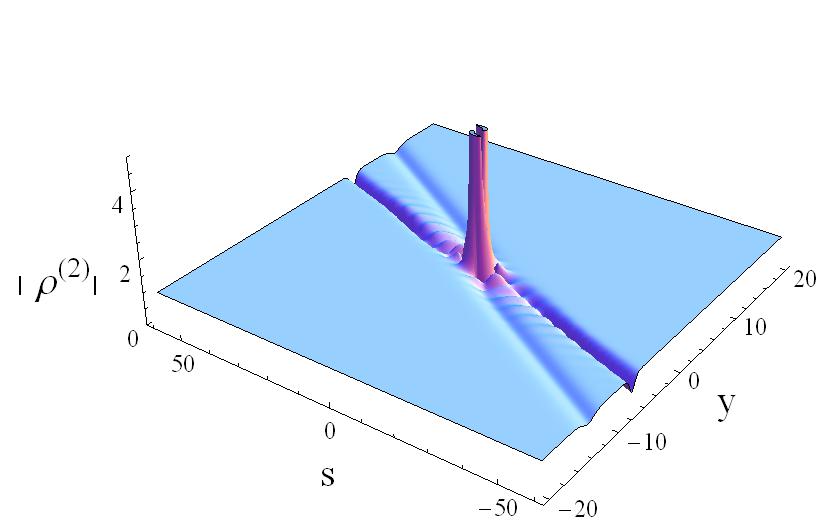}\\
{$(d)$ $|u^{(2)}|$ \hspace{3.2cm}  $(e)$ $|v^{(2)}|$ \hspace{3.2cm} $(f)$ $|\rho^{(2)}|$ }\\
\caption{Two soliton solution for the f-def nonlocal cm-CID equation, $(a)$-$(c)$ $c_6=\frac{1}{2}$, $c_8=2$, $(d)$-$(f)$ $c_6=c_8=0$. $(a)$
2-breather wave $|u^{(2)}|$, $(b)$ 2-breather wave $|v^{(2)}|$, $(c)$ 2-dark soliton wave $|\rho^{(2)}|$, $(d)$ fusion of two breather wave
$|u^{(2)}|$, $(e)$ collision of breather wave and soliton wave $|v^{(2)}|$, $(f)$ 2-dark soliton wave $|\rho^{(2)}|$.}
\label{Fig.45}
\end{figure*}

Taking $\alpha_1=1,\beta_1=1,\alpha_2=\frac{1}{2},\beta_1=1$,$c_7=1,c_2=\frac{1}{3},c_3=c_8=2,c_4=c_6=\frac{1}{2}$, the solution
($u^{(2)}$,$v^{(2)}$,$\rho^{(2)}$) display the interaction between soliton and breather-like wave(see FIG.~\ref{Fig.45}). The asymptotic behavior of
this interaction is as follows
\begin{equation}\label{asymptotic3}
\begin{aligned}
u^{(2)}\sim \begin{cases}
u_{1}^{+}+u_{2}^{-},~s\rightarrow-\infty,\\
u_{1}^{-}+u_{2}^{+},~s\rightarrow+\infty,
\end{cases}v^{(2)}\sim\begin{cases}
v_{1}^{+}+v_{2}^{-},~s\rightarrow-\infty,\\
v_{1}^{-}+v_{2}^{+},~s\rightarrow+\infty,
\end{cases}\rho^{(2)}\sim\begin{cases}
\rho_{1}^{+}+\rho_{2}^{-},~s\rightarrow-\infty,\\
\rho_{1}^{-}+\rho_{2}^{+},~s\rightarrow+\infty,
\end{cases}
\end{aligned}
\end{equation}
where $u_{1}^{\pm},u_{2}^{\pm},v_{1}^{\pm},v_{2}^{\pm},\rho_{1}^{\pm},\rho_{2}^{\pm}$ are defined by Eqs.\eqref{asymptotic-4}. It can be seen that
$|u^{(2)}|$ and $|v^{(2)}|$ describes the process of breather-like waves ($u_{1}^{+}$,$v_{1}^{+}$) and $(u_{2}^{-},v_{2}^{-})$ becoming breather-like
waves $(u_{1}^{-},v_{1}^{-})$ and $(u_{2}^{+},v_{2}^{+})$ after the collision(see FIG.~\ref{Fig.45}$(a)(b)$), note that these two waves occur blow-up
when they colliding. $|\rho^{(2)}|$ shows the collision of two dark soliton $\rho_{1}^{+}$ and $\rho_{2}^{-}$(see FIG.~\ref{Fig.45}$(c)$). Note that
from Eq.\eqref{asymptotic3}, we obtain the conclusion that for $|u^{(2)}|$ and $|v^{(2)}|$, the velocity of the waves before and after the collision
does not change, but the amplitude changes; for $|\rho^{(2)}|$, this collision is elastic collision, because the velocity and amplitude remain the
same before and after the collision, and only the phase changes. Taking $\alpha_1=1$, $\beta_1=1$, $\alpha_2=\frac{1}{2}$, $\beta_2=1$, $c_7=1$,
$c_2=-\frac{1}{3}$, $c_3=2,c_4=\frac{1}{2}$, $c_6=c_8=0$, we have $u_1^-=0$ and $v_1^{\pm}$ are solitons, this solution $u^{(2)}$ shows the process
of merging two breather-like waves into one breather-like wave; $v^{(2)}$ displays the collision of soliton $v_1^{+}$ and breather-like wave
$v_2^{-}$(see FIG.~\ref{Fig.45}$(d)(e)$); $\rho^{(2)}$ describes the elastic collision between two dark solitons(see FIG.~\ref{Fig.45}$(f)$).

\begin{figure*}
\centering
\includegraphics[height=3cm]{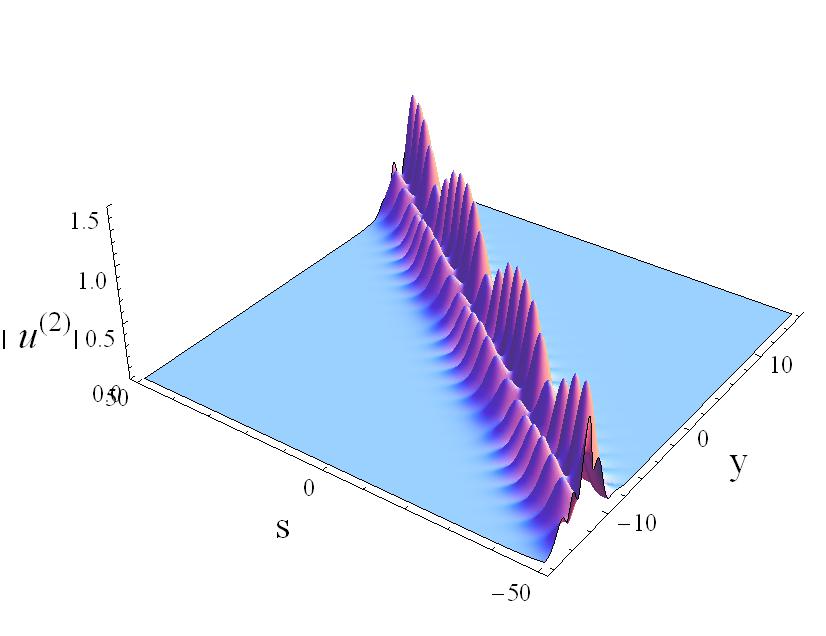}
~~\includegraphics[height=3cm]{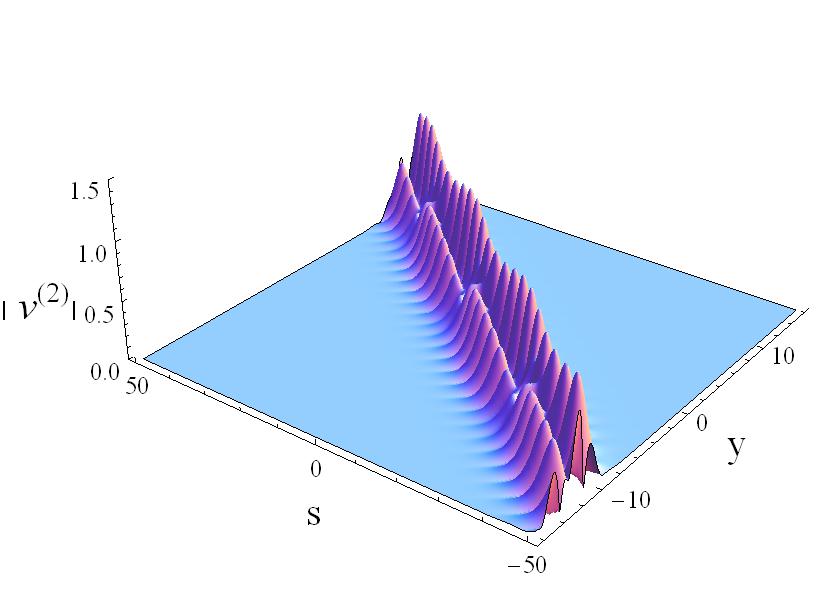}
~~\includegraphics[height=3cm]{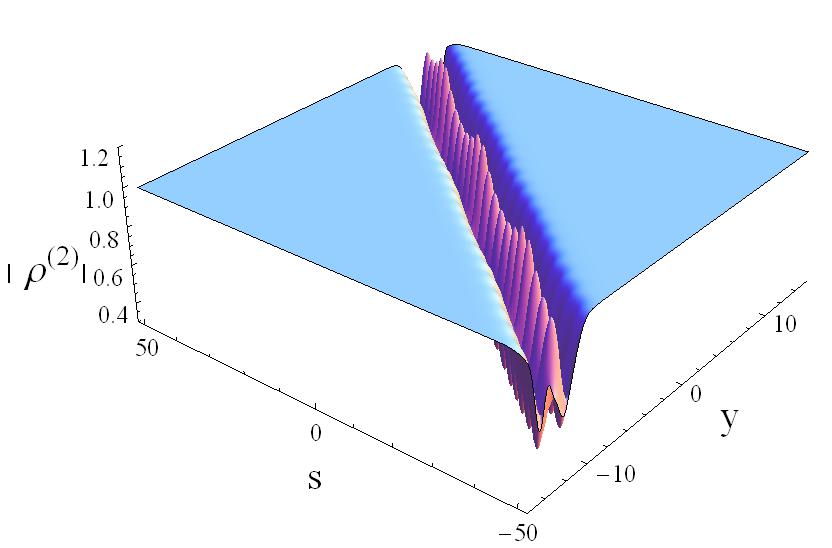}\\
{$(a)$ $|u^{(2)}|$ \hspace{3.2cm}  $(b)$ $|v^{(2)}|$ \hspace{3.2cm} $(c)$  $|\rho^{(2)}|$ }\\
\caption{Parallel breather-like wave for the nonlocal f-def cm-CID equation with $\alpha_1=\frac{3}{5}$, $\beta_1=-\frac{4}{5}$,
$\alpha_2=\frac{4}{5}$, $\beta_2=-\frac{3}{5}$, $(a)$$(b)$ parallel bright breather-like wave $|u^{(2)}|$, $|v^{(2)}|$, $(c)$ parallel dark
breather-like wave $|\rho^{(2)}|$.}
\label{Fig.47}
\end{figure*}

When $|\lambda_2|=|\lambda_1|$, this solution displays the propagation of two parallel breather-like waves. Taking the parameters $\sigma_1=1$,
$\sigma_2=-1$, $c_1=c_5=c_7=c_8=1$, $c_2=\frac{1}{3}$, $c_3=0$, $c_4=c_6=\frac{1}{2}$ and $\alpha_1=\frac{3}{5}$, $\beta_1=-\frac{4}{5}$,
$\alpha_2=\frac{4}{5}$, $\beta_2=-\frac{3}{5}$, the plots of parallel breather-like wave are displayed in FIG.~\ref{Fig.47}. It is can be seen that
$|u^{(2)}|$ and $|v^{(2)}|$ describe the propagation of two parallel bright breather-like waves, while $|\rho^{(2)}|$ shows the propagation of two
parallel dark breather-like waves.

Upon analysis as presented above, it becomes evident that starting from the zero seed, the properties of the solution derived via the quadratic DT
are more diverse than those obtained through the first DT.
In Case 1, we obtain the double periodic wave solution of the nonlocal f-def cm-CID equation. This solution is periodic in both the temporal and
spatial directions. Moreover, it contains two distinct peak values within each cycle.
In Case 2, we derive the periodic-like solution that is a hybrid of periodic wave and breather-like wave. In Case 3, the two soliton solution were
derived. This solution is the interaction between solitons and solitons or breathers, including catch up, fission, fusion and parallel propagation.

\section{\bf Soliton solution and rational solution with non-vanishing boundary condition}

In this section, first we choose a suitable non-zero seed solution for the nonlocal cm-CID equation. Then we obtain soliton solution and periodic
solution of the nonlocal f-f, f-def and def-def cm-CID equation, as well as the rational solution of the f-def nonlocal cm-CID equation.

With the seed solution $\rho=\gamma$, $u=b_1\text{e}^{\theta_1}$, $v=b_2\text{e}^{\theta_1}$, where $\theta_1=k_1y+w_1s$,
$w_1=\frac{-(\gamma^2+k_1^2(\sigma_1|b_1|^2+\sigma_2|b_2|^2))}{\gamma k_1}$, $b_1$, $b_2$ are complex constants, and $k_1$ is real constant.

\subsection{ Soliton solution with $\lambda_1\neq\frac{\pm
w_1\pm2\sqrt{\sigma_1|b_1|^2+\sigma_2|b_2|^2}}{-2w_1^2+8(\sigma_1|b_1|^2+\sigma_2|b_2|^2)}$}
Supposing the eigenfunction vector for the linear spectral problem \eqref{ncmCID-lax} at $\lambda=\lambda_1$ has the form as
\begin{equation}\label{parameter1}
\begin{aligned}
&|\zeta_{1}\rangle=(\psi_1,\psi_2,\psi_3,\psi_4)^T,~~\\
&\psi_1=d_1\text{e}^{\chi_1}+d_2\text{e}^{\chi_2},~~\psi_2=d_3\text{e}^{\chi_3-2\theta_1}+d_4\text{e}^{\chi_4-2\theta_1},\\
&\psi_3=\text{e}^{-\theta_1}(\sigma_1b_1^*(d_1h_2\text{e}^{\chi_1}+d_2h_1\text{e}^{\chi_2})-b_2d_1h_2(d_3h_3\text{e}^{\chi_3}+d_4h_4\text{e}^{\chi_4})),\\
&\psi_4=\text{e}^{-\theta_1}(\sigma_2b_2^*(d_1h_2\text{e}^{\chi_1}+d_2h_1\text{e}^{\chi_2})+b_1d_1h_2(d_3h_3\text{e}^{\chi_3}+d_4h_4\text{e}^{\chi_4})),
\end{aligned}
\end{equation}
where $\chi_{k}=\kappa_ky+\tau_ks$, and $d_k$ ($k=1,2,3,4$) are complex constants.
Substituting the eigenfunction \eqref{parameter1} to the temporal part of Eq.\eqref{ncmCID-lax}, we get
\begin{equation}\label{parameter0}
\begin{aligned}
&h_1=\frac{1+4\lambda_1\tau_2}{4\lambda_1(\sigma_1|b_1|^2+\sigma_2|b_2|^2)},~~
h_2=\frac{1+4\lambda_1\tau_1}{4\lambda_1(\sigma_1|b_1|^2+\sigma_2|b_2|^2)},~~\\
&h_3=\frac{1-8\lambda_1w_1+4\lambda_1\tau_3}{4\lambda_1(\sigma_1|b_1|^2+\sigma_2|b_2|^2)},~~
h_4=\frac{1-8\lambda_1w_1+4\lambda_1\tau_4}{4\lambda_1(\sigma_1|b_1|^2+\sigma_2|b_2|^2)},
\end{aligned}
\end{equation}
where $\tau_k(k=1,2,3,4)$ are the four roots of the equation
\begin{equation}\label{zero}
\begin{aligned}
&(4\lambda w_1+1+16\lambda^2(w_1\tau-\tau^2-\sigma_1|b_1|^2-\sigma_2|b_2|^2))\\
&\times(4\lambda w_1-1+16\lambda^2(2w_1^2-3w_1\tau+\tau^2+\sigma_1|b_1|^2+\sigma_2|b_2|^2))=0.
\end{aligned}\end{equation}
Here we choose
\begin{equation}\label{parameter}
\begin{aligned}
&\tau_1=\frac{w_1}{2}-\sqrt{\frac{(1+2\lambda_1w_1)^2}{16}-(\sigma_1|b_1|^2+\sigma_2|b_2|^2)},~~\\
&\tau_2=\frac{w_1}{2}+\sqrt{\frac{(1+2\lambda_1w_1)^2}{16}-(\sigma_1|b_1|^2+\sigma_2|b_2|^2)},\\
&\tau_3=\frac{3w_1}{2}-\sqrt{\frac{(1-2\lambda_1w_1)^2}{16}-(\sigma_1|b_1|^2+\sigma_2|b_2|^2)},~~\\
&\tau_4=\frac{3w_1}{2}+\sqrt{\frac{(1-2\lambda_1w_1)^2}{16}-(\sigma_1|b_1|^2+\sigma_2|b_2|^2)},
\end{aligned}
\end{equation}
where $\lambda_1\neq\frac{\pm w_1\pm2\sqrt{\sigma_1|b_1|^2+\sigma_2|b_2|^2}}{-2w_1^2+8(\sigma_1|b_1|^2+\sigma_2|b_2|^2)}$, and $w_1=0$, i.e.
$\tau_k(k=1,2,3,4)$ are simple root of Eq.\eqref{zero}.
Solving the spatial-part of Eq.\eqref{ncmCID-lax}, we have
\begin{align*}
&\kappa_1=\gamma\lambda_1+\frac{\lambda_1k_1^2}{\gamma}(\sigma_1|b_1|^2+\sigma_2|b_2|^2)+\frac{k_1}{2}(1+4\lambda_1\tau_1),\\
&\kappa_2=\gamma\lambda_1+\frac{\lambda_1k_1^2}{\gamma}(\sigma_1|b_1|^2+\sigma_2|b_2|^2)+\frac{k_1}{2}(1+4\lambda_1\tau_2),\\
&\kappa_3=\gamma\lambda_1+\frac{\lambda_1k_1^2}{\gamma}(\sigma_1|b_1|^2+\sigma_2|b_2|^2)+\frac{k_1}{2}(3+8\lambda_1w_1-4\lambda_1\tau_3),\\
&\kappa_4=\gamma\lambda_1+\frac{\lambda_1k_1^2}{\gamma}(\sigma_1|b_1|^2+\sigma_2|b_2|^2)+\frac{k_1}{2}(3+8\lambda_1w_1-4\lambda_1\tau_4).
\end{align*}
Substituting the eigenfunction \eqref{parameter1} into $\eqref{solution1}$, we obtain the soliton solutions of the nonlocal cm-CID equation
$\eqref{ncmCID}$.

\textbf{Case 1. soliton and periodic-like solution}

When $d_2=0$, $d_4=0$, with Darboux transformation, the solution of the nonlocal cm-CID equation \eqref{ncmCID} can be derived as follows
\begin{equation}\label{case1}
\begin{aligned}
&u^{(1)}=b_1\text{e}^{\theta_1}-\frac{\text{i}\sigma_1\lambda_{1,I}\text{e}^{\theta_1}}{|\lambda_1|^2A_1}\left(\sigma_1b_1\aleph_1-b_2^*d_1d_3^*\text{e}^{\chi_{1}-\chi_{3}^*}(h_2+h_3^*)\right),\\
&v^{(1)}=b_2\text{e}^{\theta_1}-\frac{\text{i}\sigma_2\lambda_{1,I}\text{e}^{\theta_1}}{|\lambda_1|^2A_1}\left(\sigma_2b_2\aleph_1+b_1^*d_1d_3^*\text{e}^{\chi_{1}-\chi_{3}^*}(h_2+h_3^*)\right),\\
&\rho^{(1)}=\gamma+\frac{2\text{i}\lambda_{1,I}\mu_1}{|\lambda_1|^2A_1^2}(|d_1|^4\text{e}^{4\text{i}\chi_{1,I}}\mu_2+|d_3|^4\text{e}^{4\text{i}\chi_{3,I}}\mu_3+\sigma_1\sigma_2|d_1d_3|^2\mu_4\text{e}^{-2\text{i}\chi_{1,I}-2\text{i}\chi_{3,I}}),
\end{aligned}
\end{equation}
where $h_k$ are defined by Eq.\eqref{parameter0}, and
\begin{align*}
&\mu_1=\sigma_1|b_1|^2+\sigma_2|b_2|^2,\aleph_1=|d_1|^2h_2^*\text{e}^{2\text{i}\chi_{1,I}}-\sigma_1\sigma_2|d_3|^2h_3\text{e}^{2\text{i}\chi_{3,I}},\\
&A_1=|d_1|^2\text{e}^{2\text{i}\chi_{1,I}}(1-\mu_1|h_2|^2)+\sigma_1\sigma_2|d_3|^2\text{e}^{2\text{i}\chi_{3,I}}(1-\mu_1|h_3|^2),\\
&\mu_2=-2\text{i} \text{Im}\big[\frac{\lambda_{1}k_1^2\mu_1}{\gamma}|h_2|^2+k_1\lambda_1h_2^*+k_1\lambda_1\mu_1h_2|h_2|^2\big],\\
&\mu_3=2\text{i}\text{Im}\big[k_1\lambda_1h_3^*(1+h_3^2\mu_1)-\frac{\lambda_{1}(\gamma^2+k_1^2\mu_1)}{\gamma}|h_3|^2\big],\\
&\mu_4=2\text{i} \text{Im}\big[k_1\lambda_1(h_2h_3\mu_1-1)(h_2^*-h_3^*)-\frac{\lambda_{1}(\gamma^2+k_1^2\mu_1)}{\gamma}(|h_2|^2+|h_3|^2)\big].
\end{align*}

\begin{figure*}
\centering
\includegraphics[height=3cm]{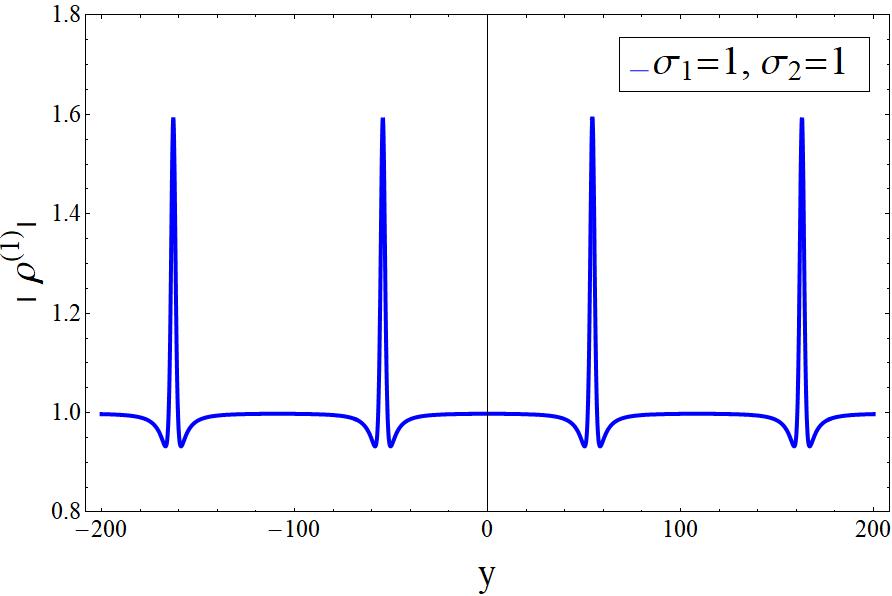}~~~~
\includegraphics[height=3cm]{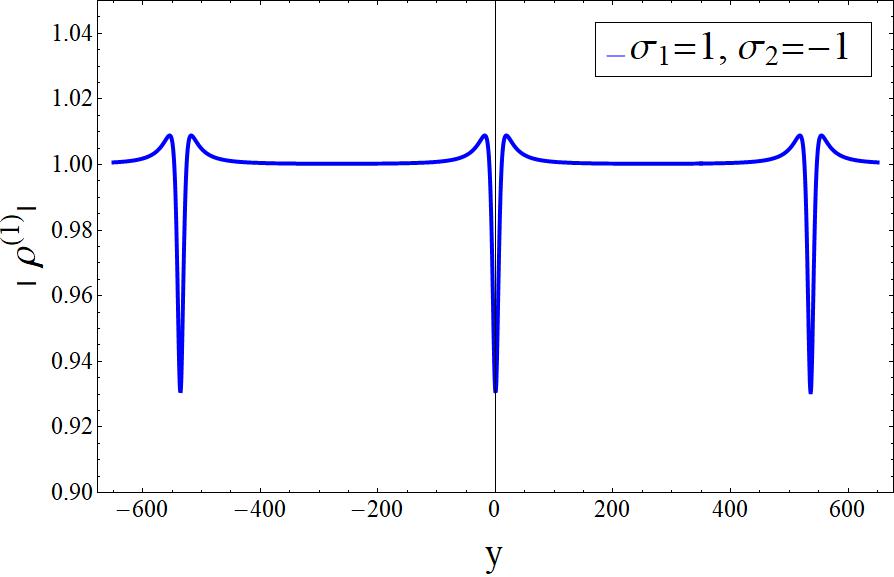}~~~~
\includegraphics[height=3cm]{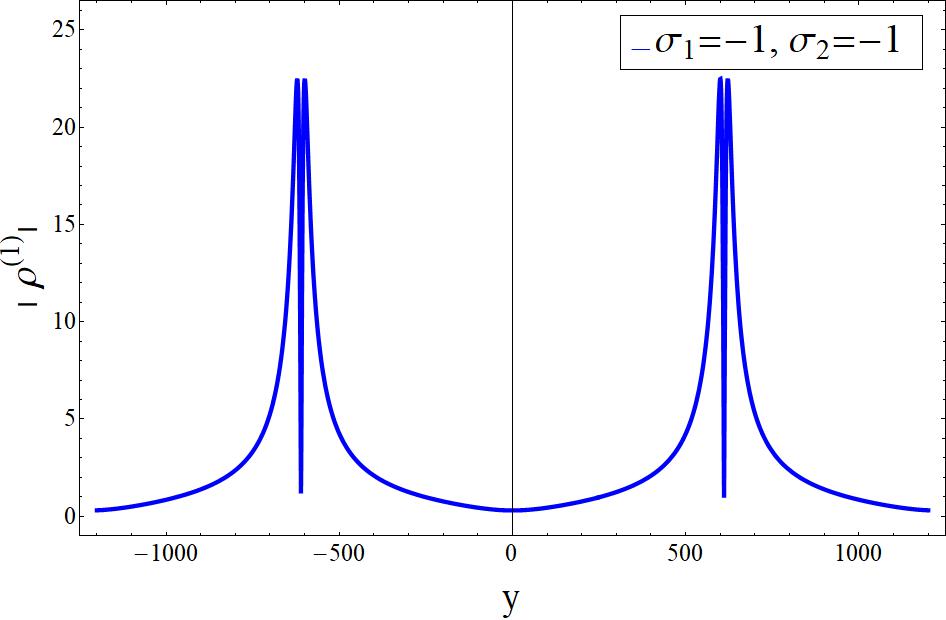}\\
{$(a)$ W-periodic wave~~~~~~~~~~~~~~~~~~~~~ $(b)$ M-periodic wave ~~~~~~~~~~~~~~~~~~~~~ $(c)$ M-periodic wave }\\
\includegraphics[height=3cm]{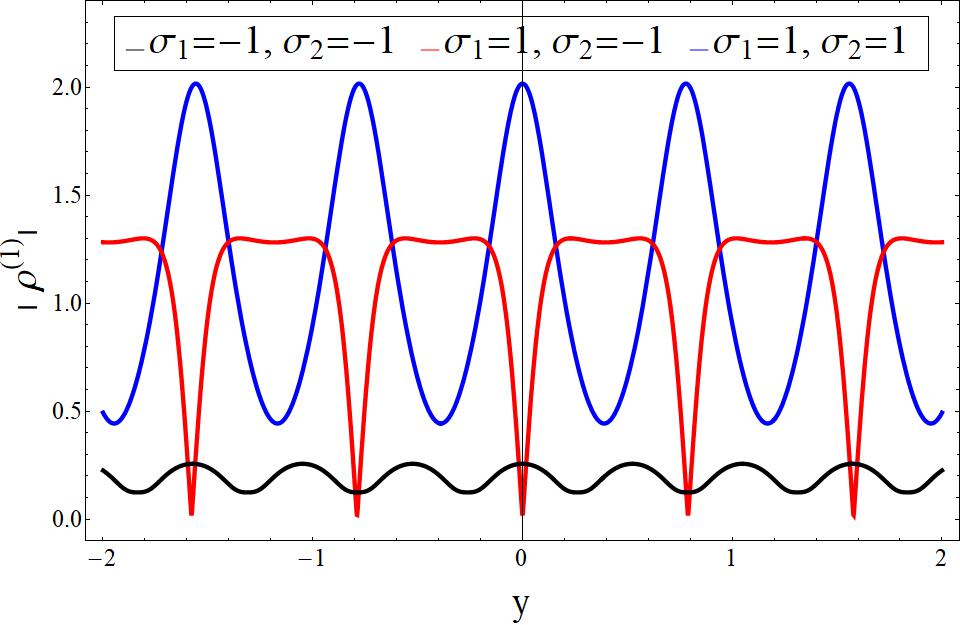}~~~~~~
\includegraphics[height=3cm]{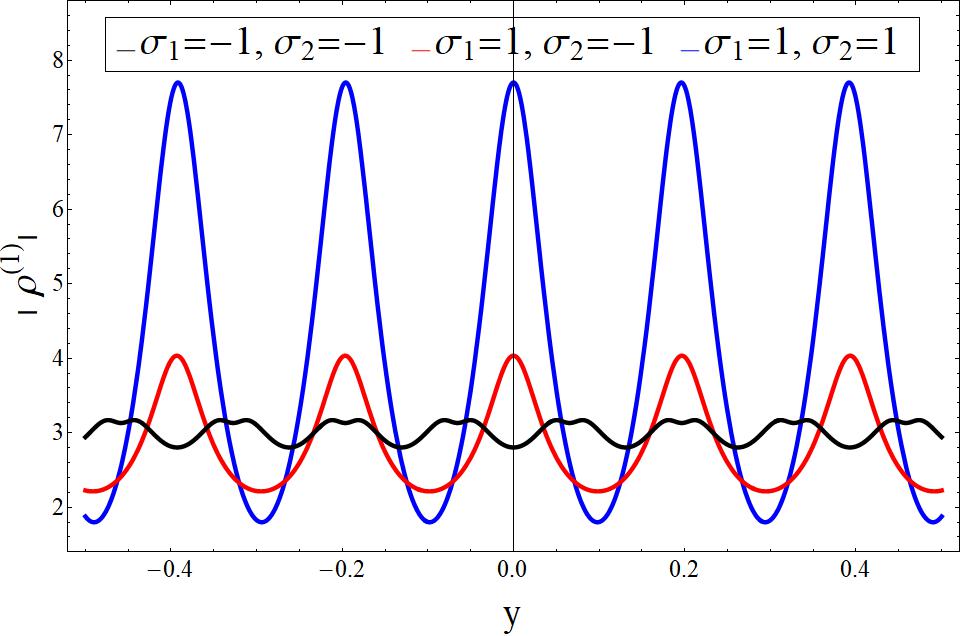}\\
{$(d)$ $\lambda_1=-1+\frac{1}{2}\text{i}$~~~~~~~~~~~~ $(e)$ $\lambda_1=-1+2\text{i}$}\\
\caption{Periodic wave solution $|\rho^{(1)}|$ for the nonlocal cm-CID equation: $(a)$ W-periodic wave with $\sigma_1=\sigma_2=1$,
$\lambda_1=\frac{1}{2}+\frac{\text{i}}{2}$, $(b)$ M-periodic wave with $\sigma_1=1,\sigma_2=-1$, $\lambda_1=\frac{1}{2}+\frac{\text{i}}{2}$, $(c)$
M-periodic wave with $\sigma_1=\sigma_2=-1$, $\lambda_1=\frac{1}{2}+\frac{\text{i}}{2}$, $(d)$ $\lambda_1=-1+\frac{1}{2}\text{i}$, $(e)$
$\lambda_1=-1+2\text{i}$.}
\label{Fig.41}
\end{figure*}

\begin{figure*}
\centering
\includegraphics[height=3cm]{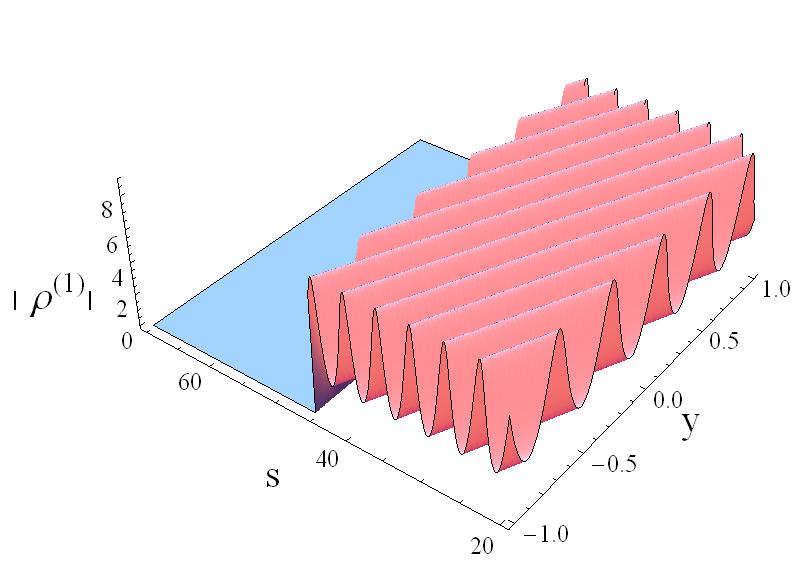}~~~~
\includegraphics[height=3cm]{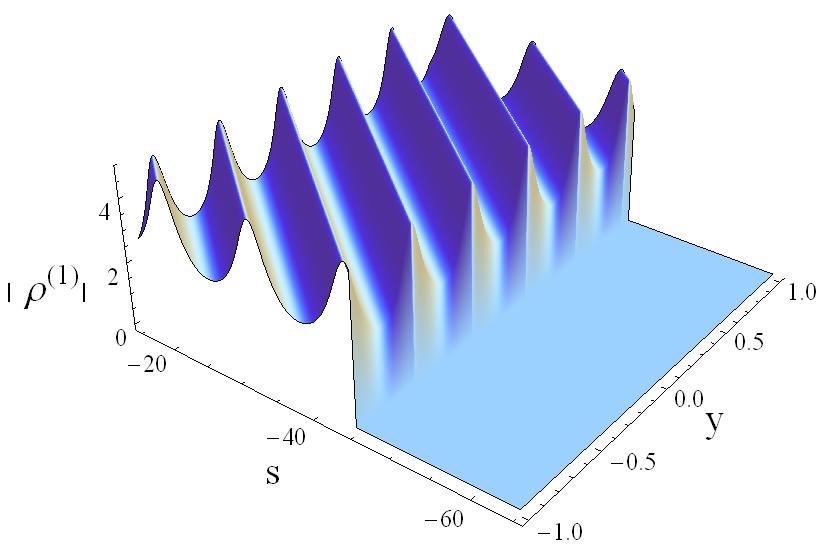}~~~~
\includegraphics[height=3cm]{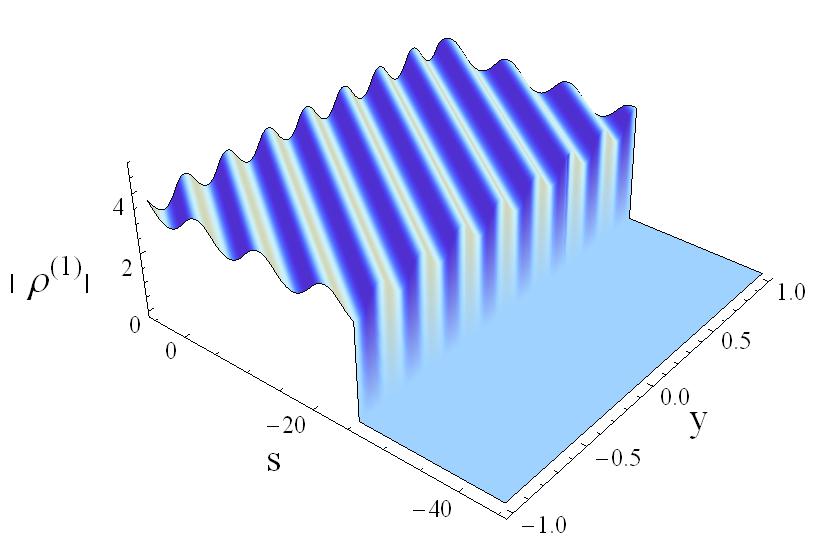}\\
{$(a)$ periodic-like wave\hspace{2cm}$(b)$ periodic-like wave \hspace{2cm} $(c)$ periodic-like wave }\\
\includegraphics[height=3cm]{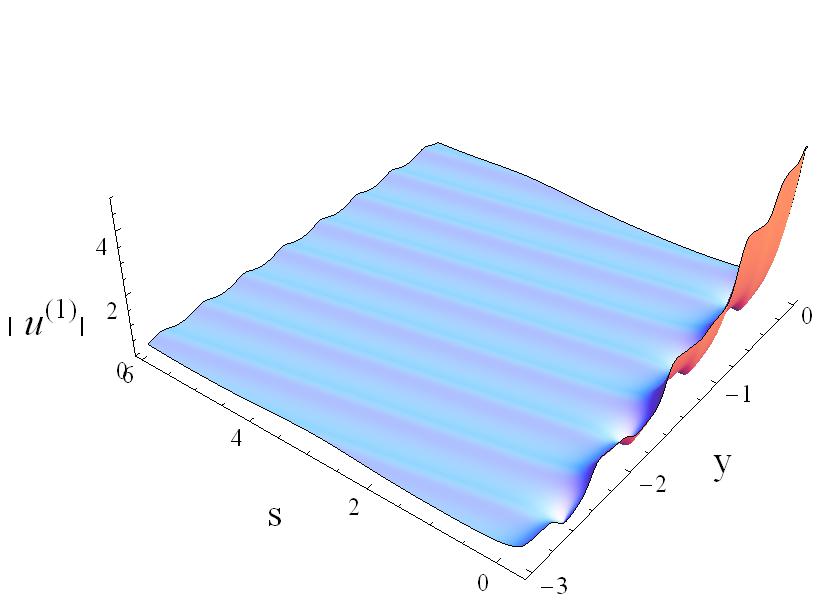}~~~~~~
\includegraphics[height=3cm]{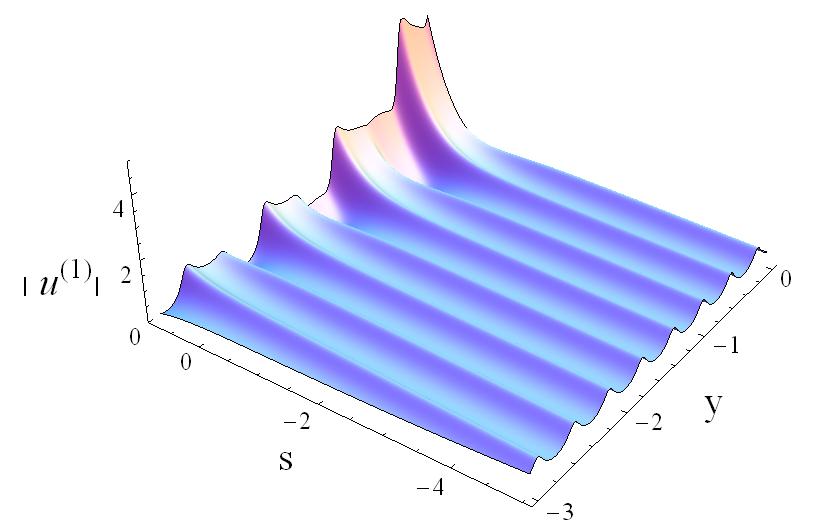}~~~~~~
\includegraphics[height=3cm]{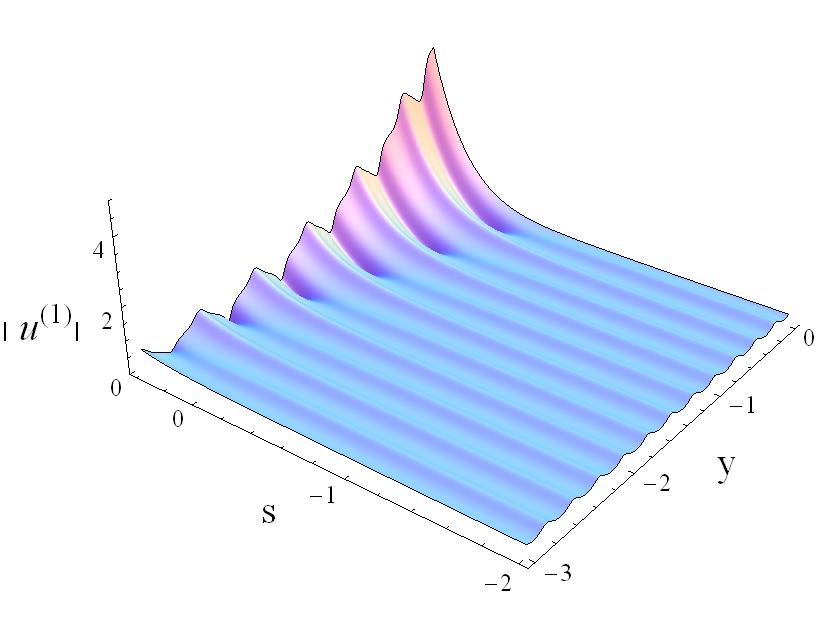}\\
{$(d)$ decaying-periodic-like\hspace{1.8cm} $(e)$ growing-periodic-like \hspace{1.8cm} $(f)$ growing-periodic-like }\\
\caption{Periodic-like wave solution for the nonlocal cm-CID equation with $\lambda_1=\text{i}$: $(a)$-$(c)$ periodic-like wave $|\rho^{(1)}|$,
$(d)$-$(f)$ decaying-, growing- periodic-like wave $|u^{(1)}|$. $(a)(d)$ $\sigma_1=\sigma_2=1$, $(b)(e)$ $\sigma_1=1,\sigma_2=-1$, $(c)(f)$
$\sigma_1=\sigma_2=-1$.}
\label{Fig.42}
\end{figure*}

Let $\gamma=1$, $d_1=d_3=1$, $b_1=1$, $b_2=2$, we give the plots of periodic wave and periodic-like wave of the nonlocal cm-CID equation. With
$\lambda_1=\frac{1}{2}+\frac{\text{i}}{2}$, the plots of periodic wave $\rho^{(1)}$ of the f-f, f-def and def-def equations are shown(see
FIG.~\ref{Fig.41} $(a)$-$(c)$). It is can be seen that solution $\rho^{(1)}$ of the f-f equation is W-periodic wave (which refers to a wave with one
wave crest and two troughs in each period), solution $\rho^{(1)}$ of the f-def equation is M-periodic wave (which refers to a wave with two wave
crests and one wave trough in each period), solution $\rho^{(1)}$ of the def-def equation is M-periodic wave. The plot of periodic-like wave
$u^{(1)}$ and $v^{(1)}$ are singular, we don't give it. Taking $\lambda_1=-1+\frac{\text{i}}{2}$, the plot of periodic wave solution for the f-f and
def-def equations, M-periodic wave for f-def equations is shown in FIG.~\ref{Fig.41}$(d)$. Taking $\lambda_1=-1+2\text{i}$, the plot of periodic wave
solution for the f-f and f-def equations, M-periodic wave for def-def equations is shown in FIG.~\ref{Fig.41}$(e)$.  With $\lambda_1=\text{i}$, the
solution $\rho^{(1)}$ of the nonlocal cm-CID equation is periodic-like solution, which is half planar and half periodic(see
FIG.~\ref{Fig.42}$(a)$-$(c)$), the plots of $u^{(1)}$ is growing- or decaying- periodic-like wave(see FIG.~\ref{Fig.42}$(d)$-$(f)$), the plot of
periodic-like wave $v^{(1)}$ are similar to $u^{(1)}$. Note that the plot of periodic-like wave seems to have been seen in the Sasa-Satsuma equation
equation\cite{Nimmo2015}. The growing-, decaying- periodic-like waves are different from growing-, decaying- periodic wave in FIG.~\ref{Fig.2}, which
has two non-equal peak values that increasing or decreasing exponentially in each cycle.

\textbf{Case 2. breather-like solution}

\begin{figure*}
\centering
\includegraphics[height=3.2cm]{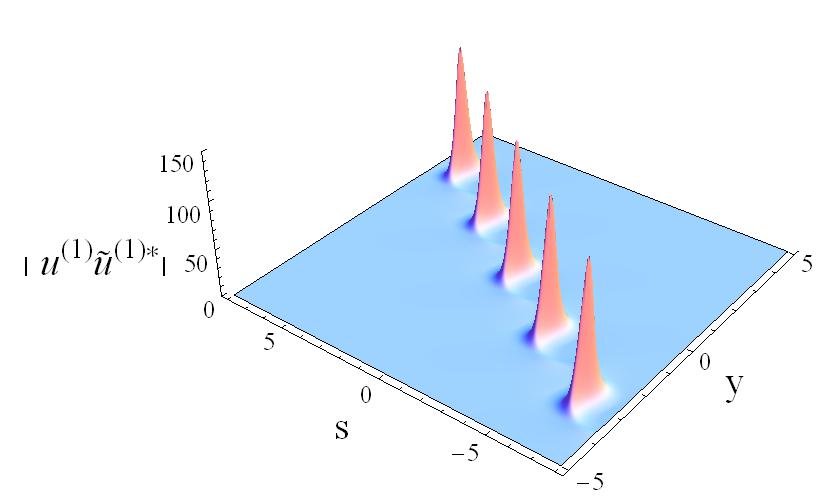}~~
\includegraphics[height=3.2cm]{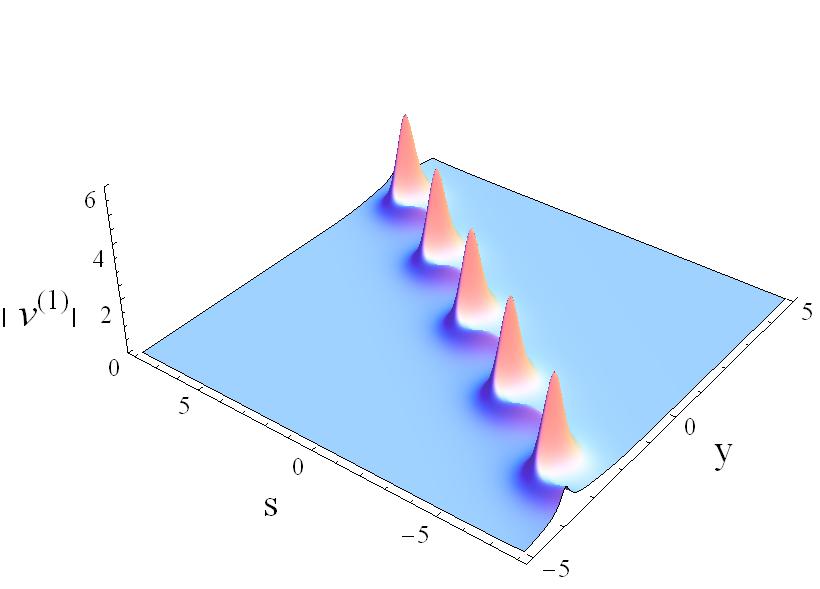}~~
\includegraphics[height=3.2cm]{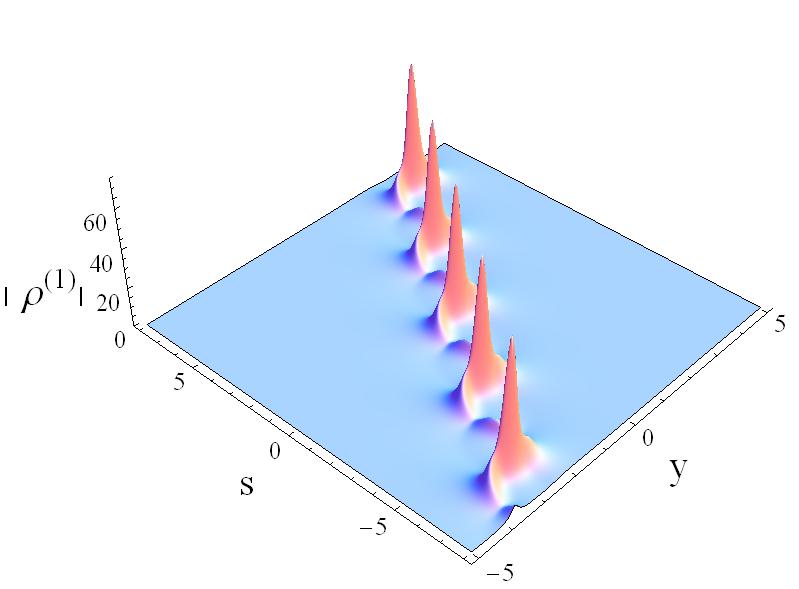}\\
{$(a)$ $|u^{(1)}\widetilde{u}^{(1)*}|$ \hspace{3.2cm} $(b)$ $|v^{(1)}|$\hspace{3.2cm} $(c)$ $|\rho^{(1)}|$ }\\
\includegraphics[height=3.5cm]{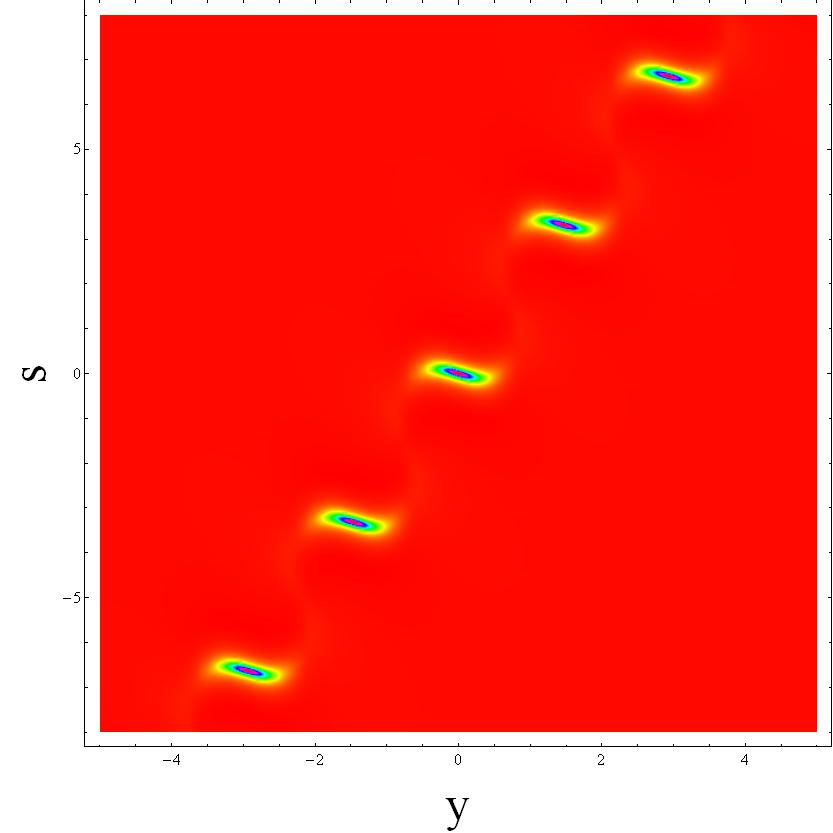}~~~~~~~~~~
\includegraphics[height=3.5cm]{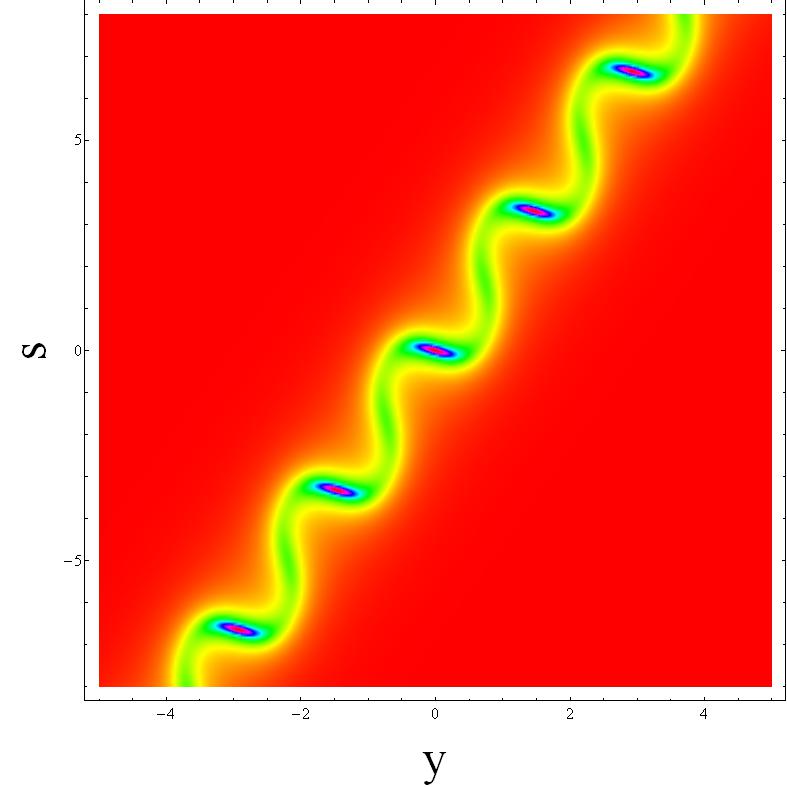}~~~~~~~~~~
\includegraphics[height=3.5cm]{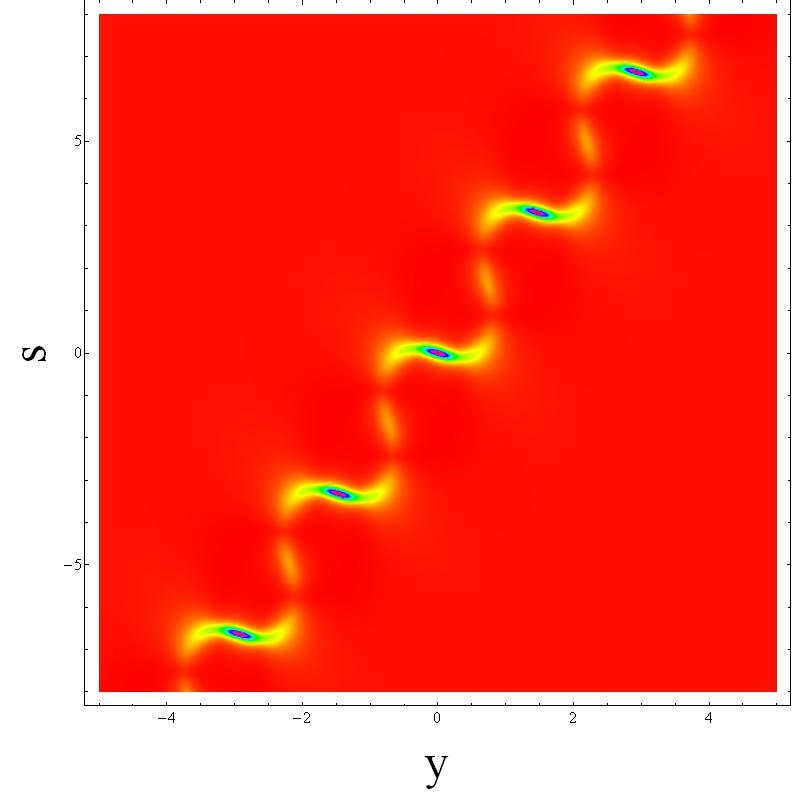}\\
{$(d)$ $|u^{(1)}\widetilde{u}^{(1)*}|$ \hspace{3.2cm} $(e)$ $|v^{(1)}|$\hspace{3.2cm} $(f)$ $|\rho^{(1)}|$ }\\
\caption{Breather-like solution for the nonlocal f-f cm-CID equation with $d_2=0$, $b_1=1$, $b_2=0$, $\lambda_1=\text{i}$.}
\label{Fig.50}
\end{figure*}

When $d_2$ and $d_4$ are not all zero, the solution $\rho^{(1)}$ of the nonlocal cm-CID equation is breather-like solution. Taking $\sigma_2=1$,
$\gamma=1$, $d_1=d_3=d_4=1$, $d_2=0$, $b_1=1$, $b_2=0$, $\lambda_1=\text{i}$, for the f-f nonlocal cm-CID equation, $u^{(1)}$ is the mixture of
breather wave and decaying-periodic wave, $|u^{(1)}\widetilde{u}^{(1)*}|$, $v^{(1)}$ and $\rho^{(1)}$ are breather-like wave(see FIG.~\ref{Fig.50}).
It is can be seen from the density diagram of this solution that $|u^{(1)}\widetilde{u}^{(1)*}|$, $v^{(1)}$ and $\rho^{(1)}$ describe the process of
one tall and one short breather propagating simultaneously under the influence of periodic function. For the f-def and def-def equations, those
solution usually are singular, we do not discuss it here.

\subsection{Rational and rational-soliton solutions with
$\lambda_1=\frac{-w_1-2\sqrt{\sigma_1|b_1|^2+\sigma_2|b_2|^2}}{-2w_1^2+8(\sigma_1|b_1|^2+\sigma_2|b_2|^2)}$}
When $\lambda_1\neq\frac{\pm w_1\pm2\sqrt{\sigma_1|b_1|^2+\sigma_2|b_2|^2}}{-2w_1^2+8(\sigma_1|b_1|^2+\sigma_2|b_2|^2)}$, we have growing-,
decaying-, breather soliton and periodic wave solution of the nonlocal cm-CID equation; when $\lambda_1=\frac{\pm
w_1\pm2\sqrt{\sigma_1|b_1|^2+\sigma_2|b_2|^2}}{-2w_1^2+8(\sigma_1|b_1|^2+\sigma_2|b_2|^2)}$ (e.g.
$\lambda_1=\frac{-w_1-2\sqrt{\sigma_1|b_1|^2+\sigma_2|b_2|^2}}{-2w_1^2+8(\sigma_1|b_1|^2+\sigma_2|b_2|^2)}$), we will obtain rational and
rational-soliton solution of the nonlocal cm-CID equation. Note that $\lambda_1$ can not be real number, so that the f-f equation has not exist the
solution of this case. Due to solution of the def-def equation is singular, we only discuss the f-def case, i.e. $\sigma_1=1,\sigma_2=-1$.

Taking $\lambda_1=\frac{-w_1-2\sqrt{|b_1|^2-|b_2|^2}}{-2w_1^2+8(|b_1|^2-|b_2|^2)}$ ($|b_1|<|b_2|$), the eigenfunction vector for the linear spectral
problem \eqref{ncmCID-lax} can be assumed as
\begin{eqnarray*}
&&|\zeta_{1}\rangle=(\psi_{1}^{(1)},\psi_{2}^{(1)},\psi_{3}^{(1)},\psi_{4}^{(1)})^T,\\
&&\psi_{1}^{(1)}=d_1\text{e}^{\chi_{1}}+d_2\text{e}^{\chi_{2}},~~~\psi_{2}^{(1)}=d_3(1+\iota_3y+t)\text{e}^{\chi_{3}-2\theta_1},\\
&&\psi_{3}^{(1)}=d_3h_3(h_3\iota_1-b_2(\iota_3y+t))\text{e}^{\chi_{3}-\theta_1}+\sigma_1b_1^*(d_2h_1\text{e}^{\chi_{2}-\theta_1}+d_1h_2\text{e}^{\chi_{1}-\theta_1}),\\
&&\psi_{4}^{(1)}=d_3h_3(h_3\iota_2+b_1(\iota_3y+t))\text{e}^{\chi_{3}-\theta_1}+\sigma_2b_2^*(d_2h_1\text{e}^{\chi_{2}-\theta_1}+d_1h_2\text{e}^{\chi_{1}-\theta_1}),
\end{eqnarray*}
where $\chi_j$, $h_j$ $(j=1,2,3)$ are defined by $\eqref{parameter}$ and $\eqref{parameter1}$. Solving the linear spectral problem
\eqref{ncmCID-lax}, we get
\begin{align*}
\iota_1=b_2(\sqrt{|b_1|^2-|b_2|^2}-1),~~\iota_2=b_1(1-\sqrt{|b_1|^2-|b_2|^2}),\iota_3=\frac{2\lambda_1k_1h_3(b_2+h_3\iota_1)(|b_1|^2-|b_2|^2)}{b_2}.
\end{align*}
By substituting $|\zeta_{1}\rangle$ into $\eqref{solution1}$, we obtain the rational-soliton solution of the nonlocal cm-CID equation \eqref{ncmCID}.

\textbf{Case 1. rational solution}

\begin{figure*}
\centering
\includegraphics[height=2.8cm]{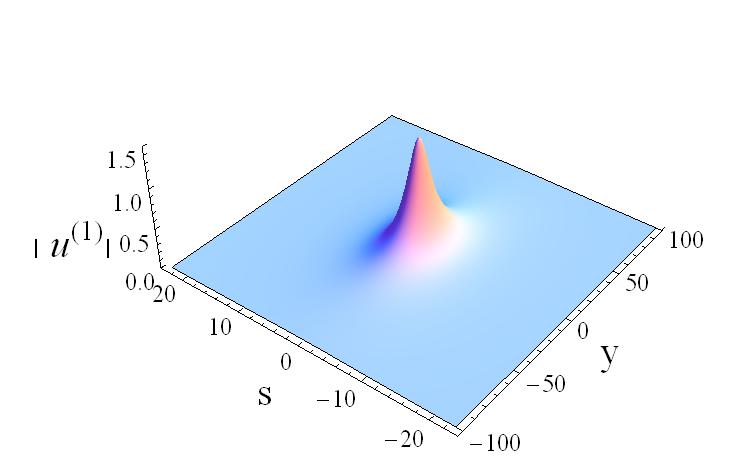}
\includegraphics[height=2.8cm]{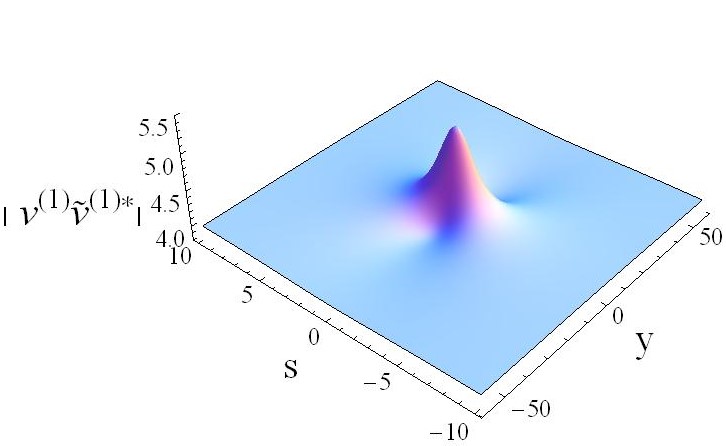}
\includegraphics[height=2.8cm]{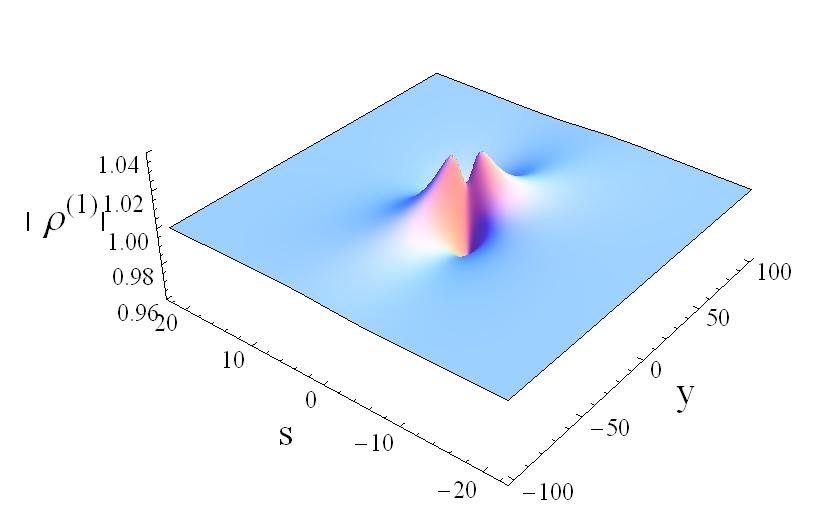}\\
{ $(a)$ $|u^{(1)}|$ \hspace{3.2cm} $(b)$ $|v^{(1)}\widetilde{v}^{(1)*}|$ \hspace{3.2cm} $(c)$ $|\rho^{(1)}|$}\\
\includegraphics[height=3cm]{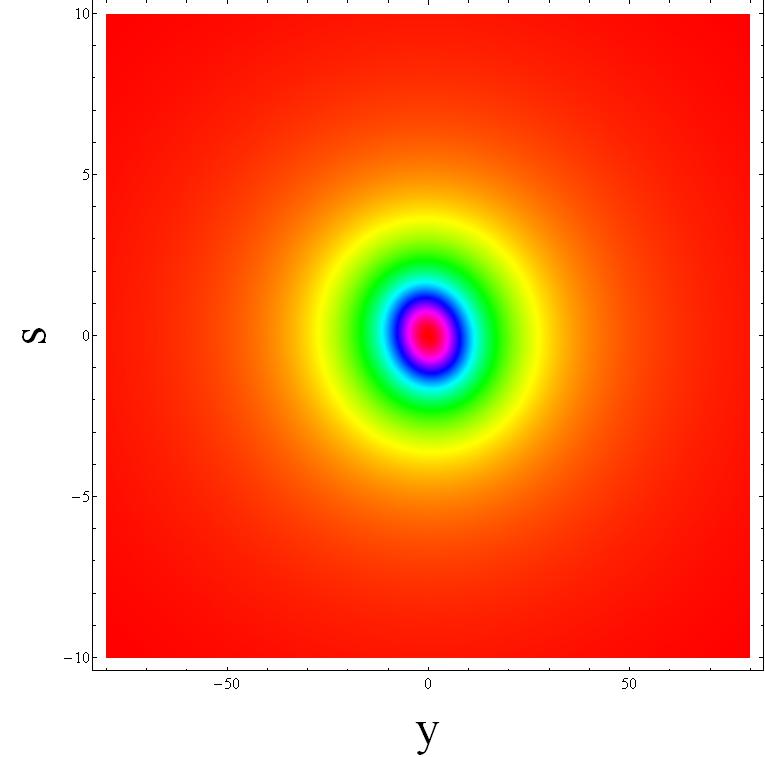}~~~~~~~~~~~~
\includegraphics[height=3cm]{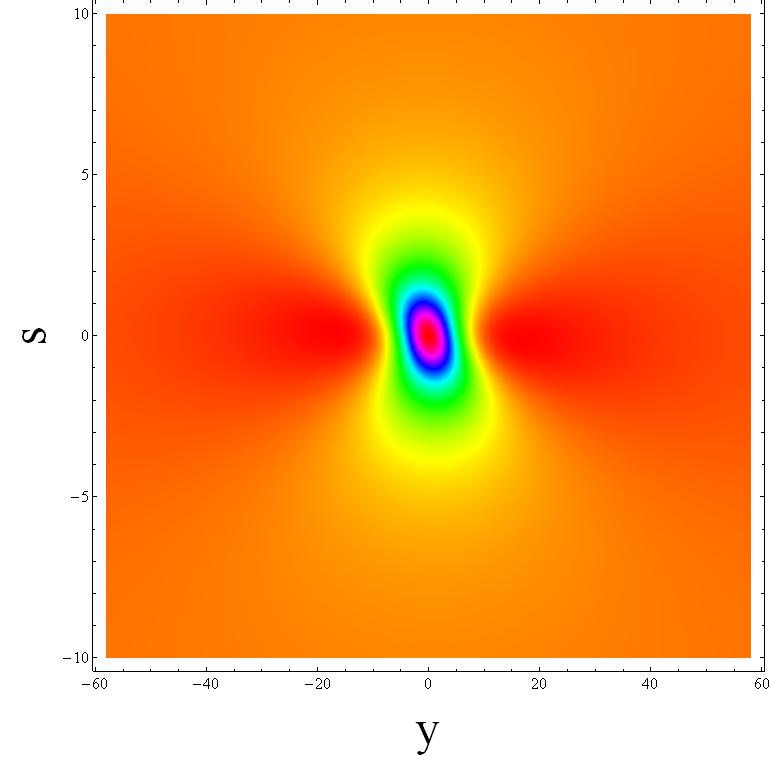}~~~~~~~~~~~~
\includegraphics[height=3cm]{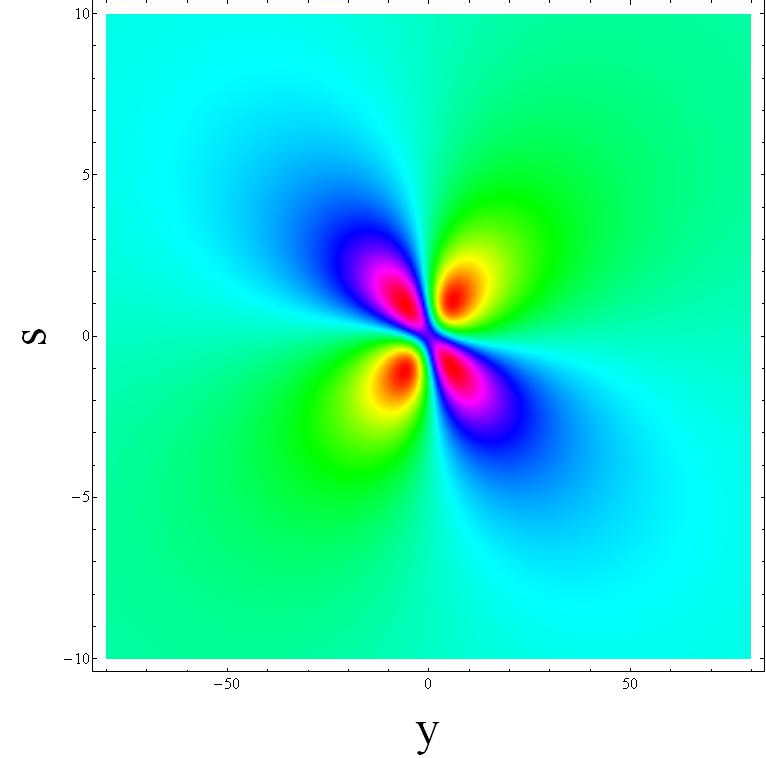}\\
{ $(d)$ $|u^{(1)}|$ \hspace{3.2cm} $(e)$ $|v^{(1)}\widetilde{v}^{(1)*}|$ \hspace{3.2cm} $(f)$ $|\rho^{(1)}|$}\\
\caption{Rational solutions for the nonlocal f-def cm-CID equation with $d_1=1$, $d_2=1$, $d_3=1$.}
\label{Fig.51}
\end{figure*}

When $w_1=0$, the solution \eqref{solution1} of the f-def cm CID equation could be non-singular rogue wave solution, which can be written as
\begin{equation}
\begin{aligned}
&u^{(1)}=\frac{8b_2^*\Xi_2+b_1\text{e}^{k_1y}(\Delta_3+4|d_3|^2(1-k_1y-2\sqrt{|b_1|^2-|b_2|^2}))}{\Delta_3},\\
&v^{(1)}=\frac{8b_1^*\Xi_2+b_2\text{e}^{k_1y}(\Delta_3+4|d_3|^2(1-k_1y-2\sqrt{|b_1|^2-|b_2|^2}))}{\Delta_3},\\
&\rho^{(1)}=\gamma+\left(\frac{2\sqrt{|b_1|^2-|b_2|^2}\Xi_3}{\Delta_3}\right)_y,
\end{aligned}
\end{equation}
where
\begin{align*}
&\Xi_2=d_3^*(d_1+d_2)\sqrt{|b_1|^2-|b_2|^2},\Xi_3=|d_3|^2(k_1y+2\sqrt{|b_1|^2-|b_2|^2})^2-4(|b_1|^2-|b_2|^2)(|d_1+d_2|^2+|d_3|^2t^2),\\
&\Delta_3=|d_3|^2(k_1y+2\sqrt{|b_1|^2-|b_2|^2})^2-4|d_1+d_2|^2(|b_1|^2-|b_2|^2)-|d_3|^2-|d_3|^2(2\sqrt{|b_1|^2-|b_2|^2}t-1)^2.
\end{align*}
Taking $b_1=0$, $b_2=2$, $\gamma=1$, $k_1=\frac{1}{2}$, we have $w_1=0$, $\tau_j=0(j=1,2,3)$, and then we obtain the rational solution $u^{(1)}$,
$\rho^{(1)}$ and $v^{(1)}$ of the nonlocal f-def cm-CID equation. When $8|d_1+d_2|^2-25|d_3|^2<0$, the rational solution $u^{(1)}$, $\rho^{(1)}$ is
singular, $v^{(1)}$ is mixture of rational wave and soliton; when $8|d_1+d_2|^2-25|d_3|^2\geq0$, the rational solution $u^{(1)}$, $\rho^{(1)}$ is
rogue wave, $v^{(1)}$ is mixture of rational wave and growing-soliton for $y$, while $|v^{(1)}\widetilde{v}^{(1)*}|$ is nonsingular rogue wave. With
$d_1=1,d_2=0,d_3=1$, the plots of rational solution are given in FIG.~\ref{Fig.51}.

\textbf{Case 2. interaction of a rational wave and a soliton }

\begin{figure*}
\centering
\includegraphics[height=3cm]{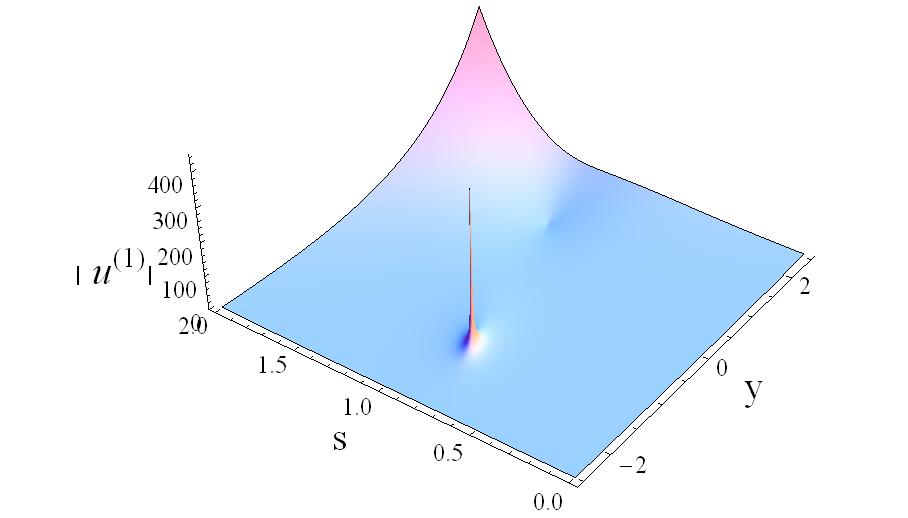}~
\includegraphics[height=3cm]{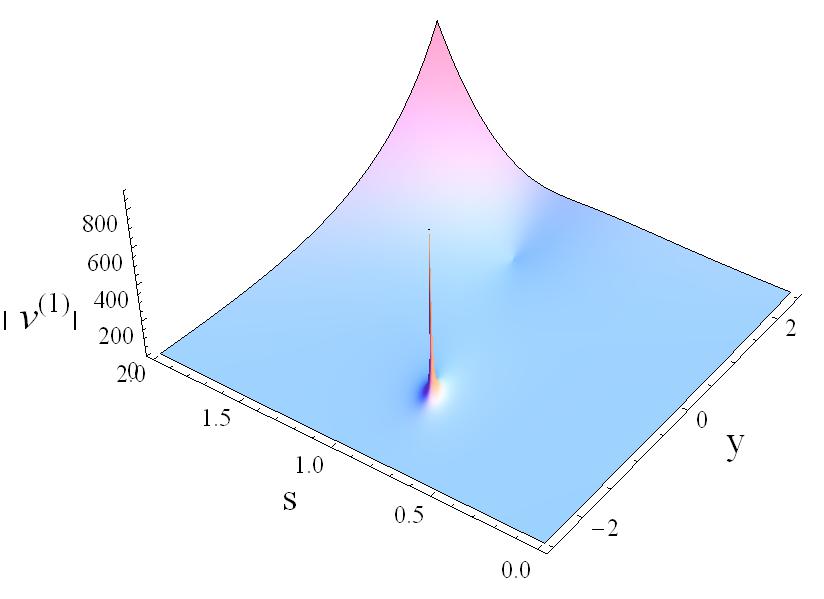}~
\includegraphics[height=3cm]{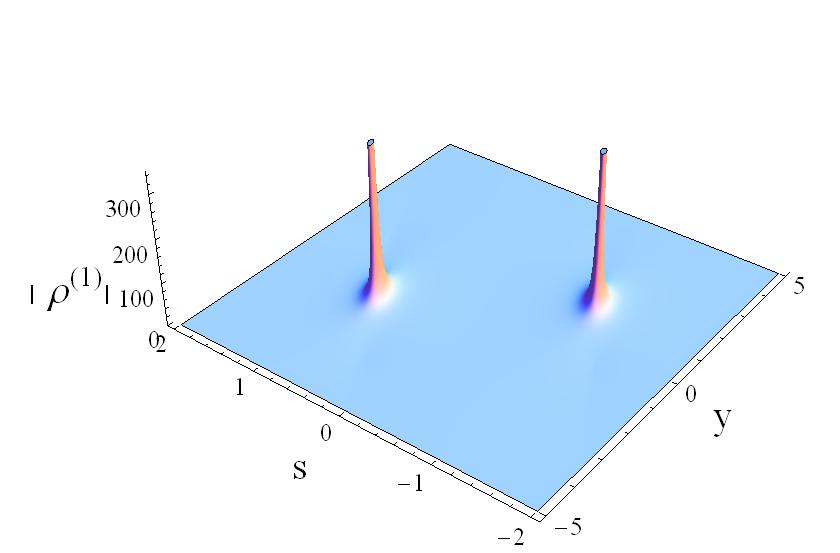}\\
{ $(a)$ rational-growing soliton \hspace{1cm}$(b)$ rational-growing soliton \hspace{1cm}$(c)$ rational solution   }\\
\includegraphics[height=3cm]{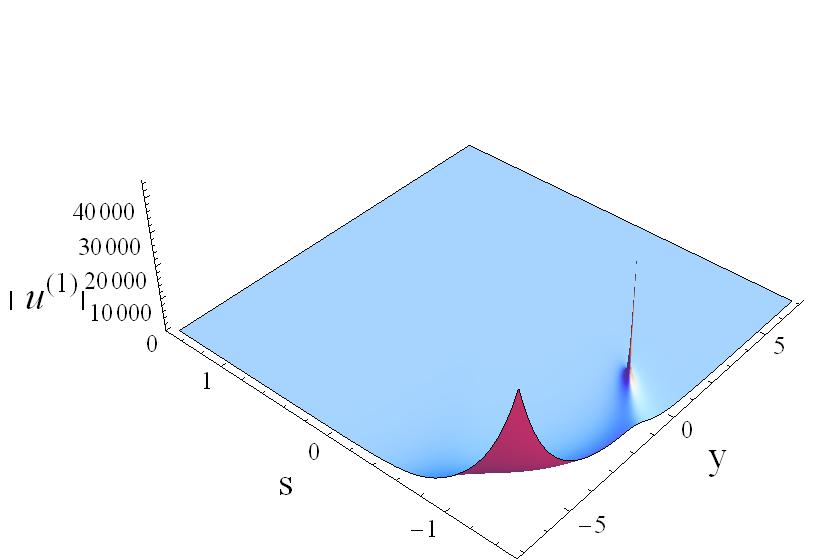}
\includegraphics[height=3cm]{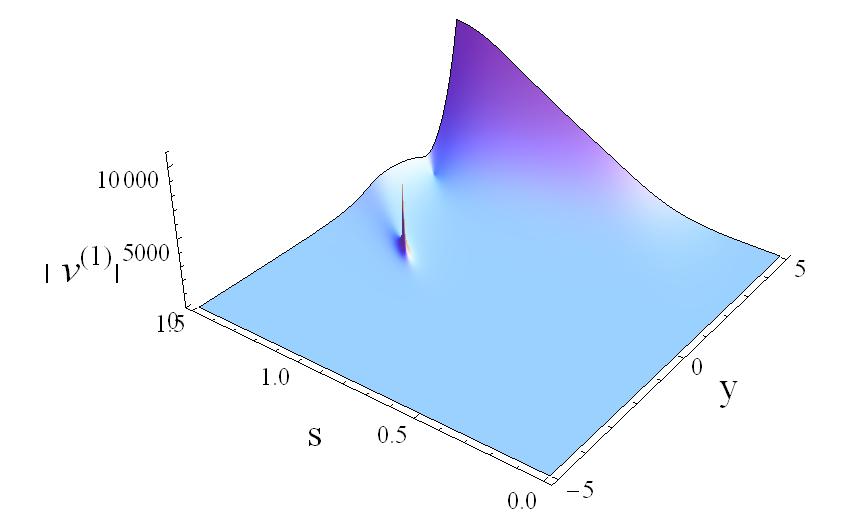}
\includegraphics[height=3cm]{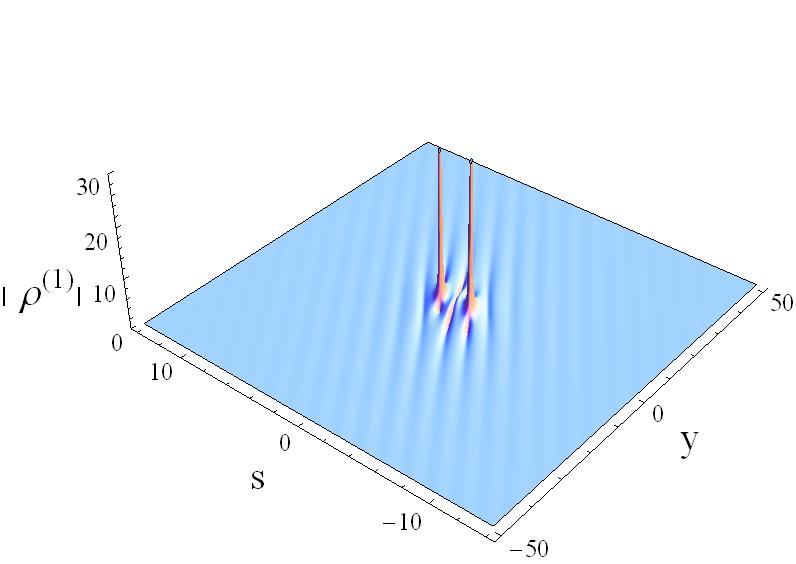}\\
{ $(d)$ rational-decaying soliton  \hspace{1cm} $(e)$ rational-growing soliton \hspace{1cm}~$(f)$ rational-periodic solution  }\\
\caption{Rational-soliton solution for the nonlocal f-def cm-CID equation: $(a)$-$(c)$ $b_1=1$, $b_2=2$, $d_1=0$, $(d)$-$(f)$ $b_1=0$, $b_2=2$,
$d_1=1$.}
\label{Fig.6}
\end{figure*}

When $w_1\neq0$, and at least one of $d_1$ and $d_2$ is zero, the solution \eqref{solution1} is a mixture of one rational wave and soliton or
periodic-like wave. If $d_1=d_2=0$, the rational solution of the f-def cm-CID equation can be derived as
\begin{equation}\label{rational}
\begin{aligned}
&u^{(1)}=b_1\text{e}^{\theta_1}(1+\frac{h_3(\lambda_1^*-\lambda_1)(1-t-\iota_3^*y)\Xi_4}{2|\lambda_1|^2\Delta_4}),~v^{(1)}=\frac{b_2}{b_1}u^{(1)},\\
&\rho^{(1)}=\gamma+\frac{(\lambda_1^*-\lambda_1)((1+t+\iota_3y)(1-t-\iota_3^*y)\Delta_5-\Xi_5\Delta_4)}{2|\lambda_1|^2\Delta_4^2},
\end{aligned}
\end{equation}
where
\begin{eqnarray*}
&&\Xi_4=h_3-h_3\sqrt{|b_1|^2-|b_2|^2}+t+\iota_3y,\Xi_5=((1+t+\iota_3y)\iota_3^*-\iota_3(1-t-\iota_3^*y)),\\
&&\Delta_4=|h_3|^2(h_3\iota_2+b_1t+b_1\iota_3y)(h_3^*\iota_2^*-b_1^*t-b_1^*\iota_3^*y)-(1+t+\iota_3y)(1-t-\iota_3^*y)\\
&&~~~~~-|h_3|^2(h_3\iota_1-b_2t-b_2\iota_3y)(h_3^*\iota_1^*+b_2^*t+b_2^*\iota_3^*y),\\
&&\Delta_5=\iota_3^*(1+t)-\iota_3(1-t-2\iota_3^*y)+\iota_3^*|h_3|^2(-b_2^*(h_3\iota_1-b_2t)-b_1^*(b_1t+\iota_2))\\
&&~~~~~+\iota_3|h_3|^2h_3^*(b_1\iota_2^*+b_2\iota_1^*)-\iota_3|h_3|^2(|b_1|^2-|b_2|^2)(t+2\iota_3^*y).
\end{eqnarray*}
Take parameters $\gamma=1,k_1=1,b_1=1,b_2=2$, we give the plots of rational-growing soliton solution $|u^{(1)}|$, $|v^{(1)}|$ and singular rational
solution $|\rho^{(1)}|$ in FIG.~\ref{Fig.6}$(a)$-$(c)$.

If $d_1=1$ and $d_2=0$, let $\gamma=1,k_1=1,d_3=1,b_1=0,b_2=2$, this solution $|u^{(1)}|$ is a mixture of decaying soliton and rational solution(see
FIG.~\ref{Fig.6}$(d)$); $|v^{(1)}|$ is a mixture of growing soliton and rational solution(see FIG.~\ref{Fig.6}$(e)$); $|\rho^{(1)}|$ is a mixture of
periodic-like wave and rational solution(see FIG.~\ref{Fig.6}$(f)$).

\textbf{Case 3. interaction solution of two-soliton waves }

\begin{figure*}
\centering
\includegraphics[height=3cm]{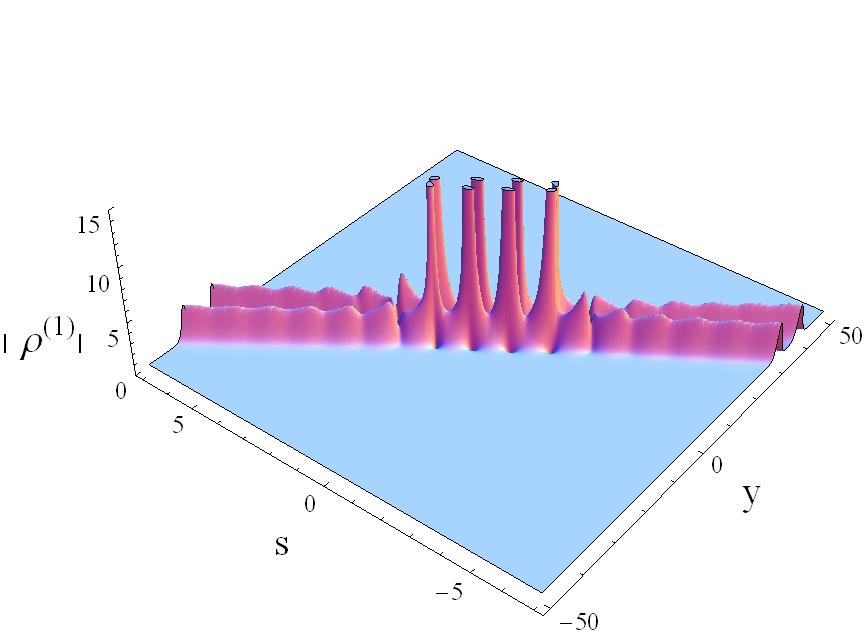}
\includegraphics[height=2.5cm]{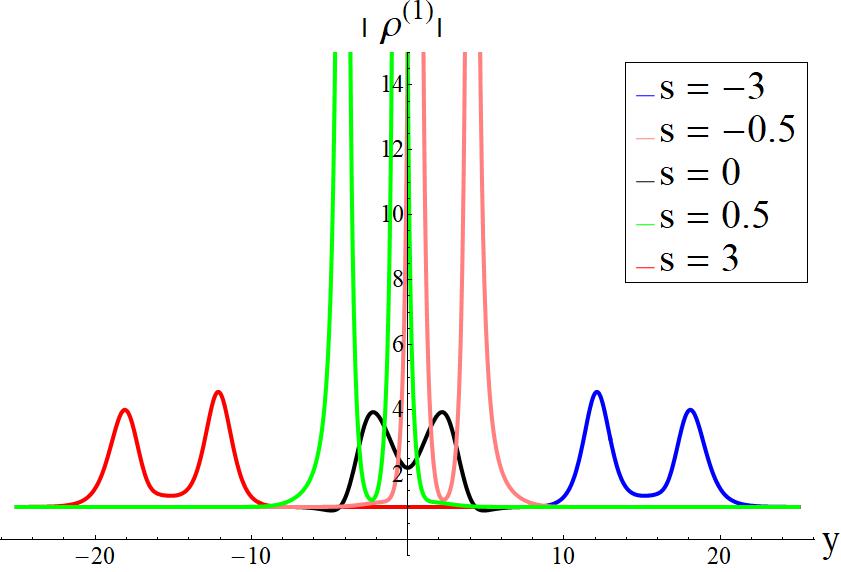}
\includegraphics[height=3cm]{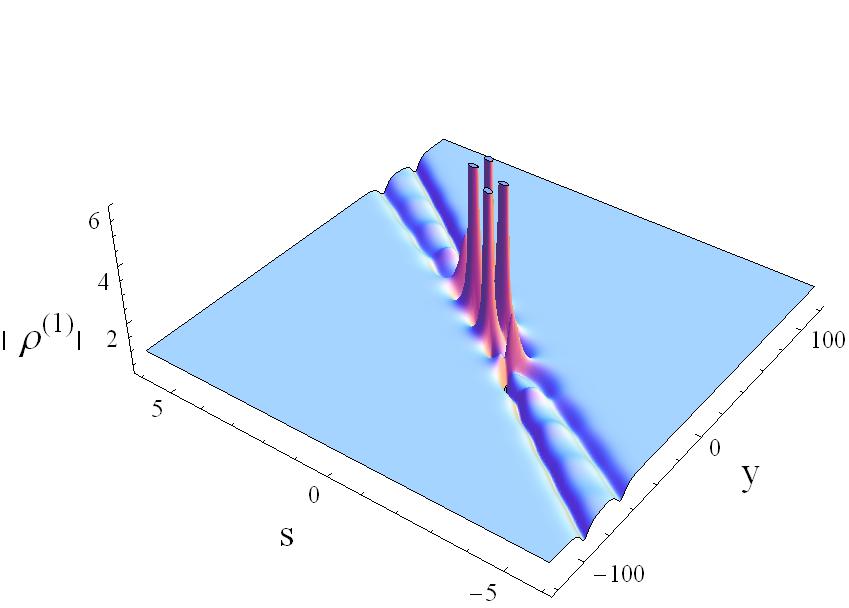}
\includegraphics[height=2.5cm]{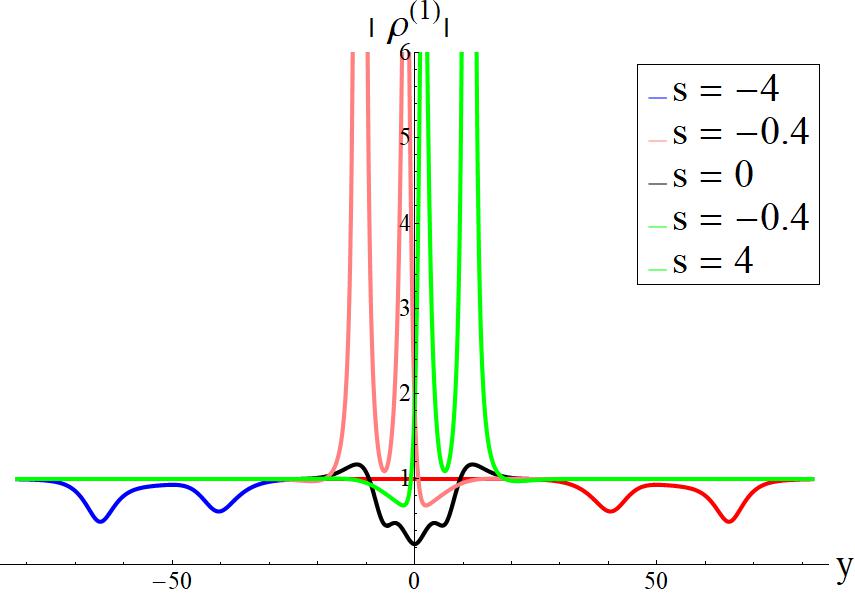}\\
{ $(a)$ $k_1=1$ \hspace{2.5cm} $(b)$ $k_1=1$ \hspace{2.5cm}$(c)$ $k_1=\frac{1}{3}$ \hspace{2.5cm}$(d)$ $k_1=\frac{1}{3}$ }\\
\caption{Interaction solution $|\rho^{(1)}|$ for the nonlocal f-def cm-CID equation: $(a)(b)$ collision of two bright solitons with $k_1=1$, $(c)(d)$
collision of two dark solitons with $k_1=\frac{1}{3}$.}
\label{Fig.7}
\end{figure*}

When $w_1\neq0$, and none of $d_1$ and $d_2$ is zero, the solution$\eqref{solution1}$ of the f-def cm-CID equation is a mixture of two soliton waves.
Taking parameters $b_1=0,b_2=2,\gamma=1,d_1=d_2=d_3=1$, we give the plots of interaction solution $\rho^{(1)}$ in FIG.~\ref{Fig.7}. With $k_1=1$,
this solution describes the interaction of two bright soliton waves(see FIG.~\ref{Fig.7}$(a)(b)$). With $k_1=\frac{1}{3}$, this solution displays the
interaction of two dark soliton waves(see FIG.~\ref{Fig.7}$(c)(d)$). Note that here $u^{(1)}$ shows the interaction of two bright waves and $v^{(1)}$
is the mixture of rational soliton and growing or decaying soliton, we do not discuss $u^{(1)}$ and $v^{(1)}$.

\section{Conclusions and discussions}

In this paper, we have presented the nonlocal cm-CID equation and the nonlocal cm-CSP equation, which can be transformed into each other through
hodograph transformation. We mainly studied the DT and various types of soliton solutions for the nonlocal cm-CID equation, including the f-f, f-def
and def-def cases. Under VBC, one soliton solutions of the nonlocal cm-CID equation were derived through the first DT, which includes periodic wave,
double periodic wave solution, decaying-, growing-soliton and decaying-, growing-periodic solution. By using quadratic DT, multiple types of exact
solutions, including periodic-like, breather-like solutions as well as two-soliton solutions, for the nonlocal cm-CID equation were obtained. We also
analyze the properties and asymptotic behavior of two-soliton solutions. Meanwhile, rational solution, M-periodic wave and breather-rational
solutions of the nonlocal cm-CID equation were obtained under NVBC. We would emphasize that the soliton solution of the nonlocal cm-CID equation has
different properties with those of the cm-CID equation, such as the nonlocal cm-CID equation has growing-, decaying-soliton, growing-,
decaying-periodic wave, periodic-like wave(which is the mixture of periodic wave and breather-like wave). Compared with the f-f cm-CID equation, the
solutions of the f-def cm-CID equation have rich properties, for instance the f-def cm-CID equation has rational solutions and dark soliton
solutions.

We anticipate exploring the following issues in the future: discussing rogue periodic waves of the nolocal cm-CID equation under a periodic
background. Solving the Cauchy problem for the nonlocal cm-CID equation by inverse scattering transform method. Constructing discretization of the
cm-CID equation. Naturally, the study of the integrability (including soliton solution, infinite conservation laws, infinite symmetry) and their
continuous limit of the discrete cm-CID equation are also interesting.

\vskip 16pt \noindent {\bf Acknowledgements}\\
The work of HQS is supported by the National Natural Science Foundation of China (NNSFC) under Grant No.12401319, that of SFS is supported by NNSFC
under Grant No.11871336, and that of ZNZ is supported by NNSFC under Grant No.12071286.

\small{

}

\begin{thebibliography}{a}
\bibitem{Hasegawa1973a}A. Hasegawa, F. Tappert, Transmission of stationary nonlinear optical pulses in dispersive dielectric fibers. I. Anomalous
    dispersion, Appl. Phys. Lett. 23 (1973) 142--144.~
\bibitem{Hasegawa1973b}A. Hasegawa, F. Tappert, Transmission of stationary nonlinear optical pulses in dispersive dielectric fibers. II. Normal
    dispersion, Appl. Phys. Lett. 23 (1973) 171--172.~
\bibitem{Benney1967} D. Benney, A. Newell, The propagation of nonlinear wave envelopes, J. Math. Phys. 46 (1967) 133--139.~
\bibitem{Zakharov1972}V.E. Zakharov, A. Shabat, Exact theory of two-dimensional self-focusing and one-dimensional self-modulation of waves in
    nonlinear media, Sov. Phys. JETP. 34 (1972) 62--69.~
\bibitem{Manakov1974}S.V. Manakov, On the theory of two-dimensional stationary self-focusing of electromagnetic waves, Sov. Phys. JETP 38 (1974) 248.
\bibitem{Kang1996}J.U. Kang, G.I. Stegeman, J.S. Aitchison, N. Akhmediev, Observation of Manakov spatial solitons in AlGaAs planar waveguides, Phys.
    Rev. Lett. 76 (1996) 3699.
\bibitem{Chen1997} Z. Chen, M. Segev, T.H. Coskun, D.N. Christodoulides, Y.S. Kivshar, Coupled photorefractive spatial-soliton pairs, J. Opt. Soc.
    Am. B 14 (1997) 3066.
\bibitem{Evangelides1992} S.G. Evangelides, L.F. Mollenauer, J.P. Gordon, N.S. Bergano, Polarization multiplexing with solitons, J. Lightwave
    Technol. 10 (1992) 28.
\bibitem{Hoefer2011} M.A. Hoefer, J.J. Chang, C. Hamner, P. Engels, Dark-dark solitons and modulational instability in miscible two-component
    Bose-Einstein condensates, Phys. Rev. A 84 (2011) 041605(R).
\bibitem{Ablowitz2013} M.J. Ablowitz, Z.H. Musslimani, Integrable nonlocal nonlinear Schr\"{o}dinger equation, Phys. Rev. Lett. 110 (2013) 064105.
\bibitem{Ablowitz2016} M.J. Ablowitz, Z.H. Musslimani, Inverse scattering transform for the integrable nonlocal nonlinear Schr\"{o}dinger equation,
    Nonlinearity 29 (2016) 915--946.
\bibitem{Li2015} M. Li, T. Xu, Dark and antidark soliton interactions in the nonlocal nonlinear Schr\"{o}dinger equation with the self-induced
    parity-time-symmetric potential, Phys. Rev. E 91 (2015) 033202.
\bibitem{Ma2016} L.Y. Ma, Z.N. Zhu, Nonlocal nonlinear Schr\"{o}dinger equation and its discrete version: Soliton solutions and gauge equivalence, J.
    Math. Phys. 57 (2016) 083507.
\bibitem{Gadzhimuradov2016} T.A. Gadzhimuradov, A.M. Agalarov, Towards a gauge-equivalent magnetic structure of the nonlocal nonlinear
    Schr\"{o}dinger equation, Phys. Rev. A 93 (2016) 062124.
\bibitem{Khare2015}A. Khare, A. Saxena, Periodic and hyperbolic soliton solutions of a number of nonlocal nonlinear equations, J. Math Phys. 56
    (2015) 032104.

\bibitem{Sarma2014} A.K. Sarma, M-A. Miri, Z.H. Musslimani, D.N. Christodoulides, Continuous and discrete Schr\"{o}dinger systems with
    parity-time-symmetric nonlinearities, Phys. Rev. E 89 (2014) 052918.
\bibitem{Ji2017a} J.L. Ji, Z.N. Zhu, On a nonlocal modified Korteweg-de Vries equation: Integrability, Darboux transformation and soliton solutions,
    Commun. Nonlinear. Sci. Numer. Simulat. 42 (2017) 699--708.
\bibitem{Ma2017} L.Y. Ma, S.F. Shen, Z.N. Zhu, Soliton solution and gauge equivalence for an integrable nonlocal complex modified Korteweg-de Vries
    equation, J. Math. Phys. 58 (2017) 103501.
\bibitem{Fokas2016} A.S. Fokas, Integrable multidimensional versions of the nonlocal nonlinear Schr\"{o}dinger equation, Nonlinearity 29 (2016)
    319--324.
\bibitem{Rabelo1989}M.L. Rabelo, On equations which describe pseudospherical surfaces, Stud. Appl. Math. 81 (1989) 221--248.~
\bibitem{Beals1989}R. Beals, M. Rabelo, K. Tenenblat, B\"{a}cklund transformations and inverse scattering solutions for some pseudospherical surface
    equations, Stud. Appl. Math. 81 (1989) 125--151.~
\bibitem{Schafer2004}T. Sch\"{a}fer, C.E. Wayne, Propagation of ultra-short optical pulses in cubic nonlinear media, Phys. D 196 (2004) 90--105.
\bibitem{Chung2005}Y. Chung, C.K.R.T. Jones, T. Sch\"{a}fer, C.E. Wayne, Ultra-short pulses in linear and nonlinear media, Nonlinearity 18 (2005)
    1351--1374.~

\bibitem{Qiao2003}Z.J. Qiao, C.W. Cao, W. Strampp, Category of nonlinear evolution equations. algebraic structure, and r-matrix, J. Math. Phys. 44
    (2003) 701--722.~
\bibitem{Sakovich2005}A. Sakovich, S. Sakovich, The short pulse equation is integrable, J. Phys. Soc. Jpn. 74 (2005) 239--241.~
\bibitem{Brunelli2005}J.C. Brunelli, The short pulse hierarchy, J. Math. Phys. 46 (2005) 123507.~
\bibitem{Brunelli2006}J.C. Brunelli, The bi-Hamiltonian structure of the short pulse equation, Phys. Lett. A 353 (2006) 475--478.~
\bibitem{Matsuno2007}Y. Matsuno, Multiloop and multibreather solutions of the short pulse model equation, J. Phys. Soc. Jpn. 76 (2007) 084003.~
\bibitem{Matsuno2008}Y. Matsuno, Periodic solutions of the short pulse model equation, J. Math. Phys. 49 (2008) 073508.~
\bibitem{Sakovich2006}A. Sakovich, S. Sakovich, Solitary wave solutions of the short pulse equation, J. Phys. A: Math. Theor. 39 (2006) L361.~
\bibitem{Kuetche2007}V.K. Kuetche, T.B. Bouetou, T.C. Kofane, On two-loop soliton solution of the Sch\"{a}fer-Wayne short-pulse equation using
    Hirota's method and Hodnett-Molony approach, J. Phys. Soc. Jpn. 76 (2007) 116001.~

\bibitem{Feng2010}B.F. Feng, K. Maruno, Y. Ohta, Integrable discretization of the short pulse equation, J. Phys. A: Math. Theor. 43 (2010) 085203.

\bibitem{Matsuno2011}Y. Matsuno, A novel multi-component generalization of the short pulse equation and its multisoliton solutions, J. Math. Phys. 52
    (2011) 123702.~
\bibitem{Feng2012}B.F. Feng, An integrable coupled short pulse equation, J. Phys. A: Math. Theor. 45 (2012) 085202.~
\bibitem{Kurt2013}L. Kurt, Y. Chung, T. Sch\"{a}fer, Higher-order corrections to the short pulse equation, J. Phys. A: Math. Theor. 46 (2013)
    285202.~

\bibitem{Feng2015}B.F. Feng, Complex short pulse and coupled complex short pulse equations, Phys. D 297 (2015) 62--75.~
\bibitem{Feng2016}B.F. Feng, L.M. Ling, Z.N. Zhu, A defocusing complex short pulse equation and its multi-dark-soliton solution by Darboux
    transformation, Phys. Rev. E 93 (2016) 052227.~
\bibitem{Shen2016}S.F. Shen, B.F. Feng, Y. Ohta, From the real and complex coupled dispersionless equations to the real and complex short pulse
    equations, Stud. Appl. Math. 136 (2016) 64--88.~
\bibitem{Ling2016}L.M. Ling, B.F. Feng, Z.N. Zhu, Multi-soliton, multi-breather and higher order rogue wave solutions to the complex short pulse
    equation, Phys. D 327 (2016) 13--29.~
\bibitem{Feng2022}B.F. Feng, L.M. Ling, Darboux transformation and solitonic solution to the coupled complex short pulse equation, Phys. D 437 (2022)
    133332.~

\bibitem{Gkogkou2022}A. Gkogkou, B. Prinari, B.F. Feng, A.D. Trubatch, Inverse scattering transform for the complex coupled short-pulse equation,
    Stud. Appl. Math. 148 (2022) 918--963.~
\bibitem{Xu2020}J. Xu, E.G. Fan, Long-time asymptotic behavior for the complex short pulse equation, J. Diff. Equ. 269 (2020) 10322--10349.~

\bibitem{Feng2021}B.F. Feng, L.M. Ling, Z.N. Zhu, A focusing and defocusing semi-discrete complex short pulse equation and its various soliton
    solutions, Proc. R. Soc. A 477 (2021) 20200853.

\bibitem{Sun2023}H.Q. Sun, Z.N. Zhu, Darboux transformation and soliton solutions of the spatial discrete coupled complex short pulse equation, Phys.
    D 436 (2022) 133312.

\bibitem{Matsuno2016}Y. Matsuno, Integrable multi-component generalization of a modified short pulse equation, J. Math. Phys. 57 (2016) 111507.~

\bibitem{Lv2022b}C. Lv, D.Q. Qiu, Q.P. Liu, Riemann-Hilbert approach to two-component modified short-pulse system and its nonlocal reductions, Chaos
    32 (2022) 093120.~

\bibitem{Li2022}X.Y. Li, Z.X. Zhang, Q.L. Zhao, C.Z. Li, Darboux transformation of two novel two-component generalized complex short pulse equations,
    Rep. Math. Phys. 90 (2022) 157--184.~

\bibitem{Lv2024}C. Lv, S.F. Shen, Q.P. Liu, Inverse scattering transform for the coupled modified complex short pulse equation: Riemann-Hilbert
    approach and soliton solutions, Phys. D 458 (2024) 133986.~

\bibitem{Sun2025}H.Q. Sun, S.F. Shen, Z.N. Zhu, Asymptotic analysis of soliton solutions for the coupled modified complex short pulse equation by
    binary Darboux transformation, J. Nonlinear Sci. 35 (2025) 57.~

\bibitem{Gu1987}C.H. Gu, Z.X. Zhou, On the Darboux matrices of B\"{a}cklund transformations for AKNS systems, Lett. Math. Phys. 13 (1987) 179--187.

\bibitem{Nimmo1997}J.J.C. Nimmo, Darboux transformations and the discrete KP equation, J. Phys. A: Math. Gen. 30 (1997) 8693--8704.

\bibitem{Bian2014} D.F. Bian, B.L. Guo, L.M. Ling, High-order soliton solution of Landau-Lifshitz equation, Stud. Appl. Math. 134 (2014) 181--214.
\bibitem{Ling2015} L.M. Ling, L.C. Zhao, B.L. Guo, Darboux transformation and multi-dark soliton for N-component nonlinear Schr\"{o}dinger equations,
    Nonlinearity 28 (2015) 3243--3261.~
\bibitem{Nimmo2015} J. Nimmo, H. Yilmaz, Binary Darboux transformation for the Sasa-Satsuma equation, J. Phys. A: Math. Theor. 48 (2015) 425202.~
\end{thebibliography}
\end{document}